\documentclass[aps,prx,showpacs,twocolumn,amsmath,amssymb,superscriptaddress]{revtex4-1}
\usepackage{tabularx}
\usepackage{bm}
\usepackage{graphicx}

\usepackage{color}

\usepackage{multirow}
\usepackage{dcolumn}
\usepackage{amssymb,amscd,xypic,bm,wasysym}
\usepackage{float}
\usepackage{cleveref}
\usepackage[caption=false,position=top,captionskip=0pt,farskip=0pt]{subfig}
\usepackage{soul}  

\begin{document}

\title{Emergent flat band lattices in spatially periodic magnetic fields}

\author{M. Tahir}
\affiliation{Department of Physics, Colorado State University, Fort Collins, CO 80523, USA}
\author{Olivier Pinaud}
\affiliation{Department of Mathematics, Colorado State University, Fort Collins, CO 80523, USA}
\author{Hua Chen}
\affiliation{Department of Physics, Colorado State University, Fort Collins, CO 80523, USA}
\affiliation{School of Advanced Materials Discovery, Colorado State University, Fort Collins, CO 80523, USA}
    
\begin{abstract}
Motivated by the recent discovery of Mott insulating phase and unconventional superconductivity due to the flat bands in twisted bilayer graphene, we propose more generic ways of getting two-dimensional (2D) emergent flat band lattices using either 2D Dirac materials or ordinary electron gas (2DEG) subject to moderate periodic orbital magnetic fields with zero spatial average. Employing both momentum-space and real-space numerical methods to solve the eigenvalue problems, we find stark contrast between Schr\"{o}dinger and Dirac electrons, i.e., the former show recurring ``magic'' values of the magnetic field when the lowest band becomes flat, while for the latter the zero-energy bands are asymptotically flat without magicness. By examining the Wannier functions localized by the smooth periodic magnetic fields, we are able to explain these nontrivial behaviors using minimal tight-binding models on a square lattice. The two cases can be interpolated by varying the $g$-factor or effective mass of a 2DEG and by taking into account the Zeeman coupling, which also leads to flat bands with nonzero Chern numbers for each spin. Our work provides flexible platforms for exploring interaction-driven phases in 2D systems with on-demand superlattice symmetries.
\end{abstract}

\maketitle

\section{Introduction}\label{sec:intro}
Moir\'{e} structures formed by stacking 2D crystals such as graphene, hexagonal boron nitride, transition metal dichalcogenides, etc. have attracted a lot of attention recently \cite{1_hass_2008,1_dean_2010,1_xue_2011,1_EWANG_2016,1_CZHANG_2017}. For incommensurate moir\'{e} structures, in-plane translation symmetry is broken, posing challenges to the paradigm of solid state physics based on Bloch's theorem. Nonetheless, in the long-wavelength limit and when the moir\'{e} potential is weak, one can still adopt a momentum-space description of the low-energy electronic states, and obtain ``moir\'{e} band structures" even in the case of incommensuration \cite{santos_2007,mele_2010,MD}. In this context, Bistritzer and MacDonald first found that the moir\'{e} structure formed by twisted bilayer graphene has flat bands at charge neutrality for certain ``magic angles" of twisting \cite{MD}. The strongly suppressed kinetic energy in these flat bands suggests potential for interaction-driven exotic phases, which were recently revealed experimentally in Refs.~\onlinecite{YCI,YCS,yankowitz_2018}, where both correlated insulating and unconventional superconducting ($T_c\sim 1$K) phases were found near charge neutrality in twisted bilayer graphene at the first magic angle $\theta \approx 1.05^\circ$. 

While the flat moir\'{e} bands in the family of twisted multilayer van der Waals materials \cite{DPH,BGY,volovik_2018} may host other interaction-driven phases, these phases will inevitably be restricted or selected by the symmetries of the moir\'{e} structures, which determine the form of interactions in the moir\'{e} bands \cite{22_xu_2018,2_dasilva_2016,2_baskaran_2018,2_dodaro_2018,2_gonzalez_2018,2_guo_2018, 2_laksono_2018,2_liu_2018,2_po_2018,2_ray_2018,2_thomson_2018,2_wu_2018,2_xu_2018,2_yuan_2018}. The spatial symmetry of a moir\'{e} structure, however, cannot be easily changed since it is dictated by the crystal symmetry of the constituent layers. For example, the moir\'{e} pattern of twisted bilayer graphene always has the form of triangular lattice with a 6-fold rotation symmetry. One main task of this paper is to provide practical ways of realizing 2D flat bands with different crystalline symmetries by design, \emph{not} relying on moir\'{e} structures, thus enabling exploration of exotic phases in a larger parameter space. This is made possible through a more generic understanding of the origin of moir\'{e} flat bands, which motivates us to replace the moir\'{e} potential \cite{PJF,LJW,LGZ} by periodic external magnetic fields or other artificial crystal potentials such as Zeeman or strain fields \cite{4_levy_2010,4_tang_2014,4_mueed_2018, Yuhang_Jiang_2019}, that can now be created and controlled experimentally. 

There has been a long effort of creating spatially periodic electric and magnetic fields and studying their influence on condensed matter systems. One of the earliest examples is the observation of Weiss oscillations in conventional two-dimensional electron gas (2DEG) in GaAs/AlGaAs subject to a one-dimensional periodic static electric potential, created by parallel fringes or metallic strip arrays, and a perpendicular homogeneous magnetic field \cite{DW}, which is due to the commensuration between the cyclotron radius and the period of the electric potential \cite{RRG,RWW,CWJB,PVF}. 2D periodic electric potentials on 2DEG \cite{RRD,CAJ,MCG,XFW,CAT}, with different symmetries \cite{SCC,SCA,YKA}, were also realized, which show Hofstadter butterfly spectra under moderate homogeneous magnetic fields. In parallel, spatially periodic (orbital) magnetic fields in 1D \cite{HAC,PDY,DPX,FMP}, 2D \cite{MCQ,PDM,PDR,EAS}, and Zeeman fields \cite{CBT} have been experimentally realized using periodic arrays of superconducting or ferromagnetic strips or dots. More recently, 1D \cite{MDJ} and 2D \cite{LAP,CRD,RKK} periodic electric potentials have also been realized in graphene.

In this work, we propose that 2D-periodic magnetic fields with zero average, applied on either 2D Dirac systems or ordinary 2DEG, are an effective and versatile way of creating flat bands with different superlattice symmetries in the low-energy electronic structure. Studies on 1D-periodic magnetic fields with zero average exist in literature \cite{5_ibrahim_1995,5_chiu_2008,5_luca_2009,5_masir_2009,5_tan_2010,5_taillefumier_2011,4_tang_2014}, but no general conclusions have been made on the existence and origin of 2D flatbands in non-quantizing 2D periodic magnetic fields. We find that for a simple 2D sinusoidal magnetic field forming a square Bravais lattice, Schr\"{o}dinger and Dirac electrons exhibit drastically different behaviors in the tendency of realizing flat, low-energy bands: The lowest band for the Schr\"{o}dinger electron (or 2DEG) becomes flat repeatedly at ``magic" values of the dimensionless parameter $\phi\equiv eB/\hbar K^2$, where $B$ is the amplitude of the periodic magic field and $K$ is the reciprocal lattice constant. In contrast, the two particle-hole-symmetric bands near zero energy of the Dirac electron only become asymptotically flat with increasing $\phi$ without ``magicness". The different behaviors of the two systems can be understood by looking into the Wannier functions of the low-energy bands and the accompanying tight-binding Hamiltonians. While in the Dirac case the lowest bands can be described by Gaussian-like Wannier functions localized around the centers of square plaquettes with a definite sign of the magnetic field, in the Schr\"{o}dinger case the lowest bands are best described by two Gaussian-like Wannier functions localized at the corners of a square plaquette. As a result, the nearest neighbor hopping for the Schr\"{o}dinger case is complex and varies with $\phi$ in an oscillatory way, and at special values of $\phi$ the kinetic energy vanishes due to destructive interference, which explains the magicness. Such a mechanism is reminiscent of the classic examples of flat band lattice models~\cite{3_sutherland_1986,3_lieb_1989,3_AME1,3_AME2,3_HTI1,3_HTI2}, and can also be captured by a minimal tight-binding model. On the other hand, in the Dirac case the nearest-neighbor hopping between Wannier functions at plaquettes centers is real and becomes monotonically smaller as $\phi$ increases. Moreover, by taking into account Zeeman coupling and spin degrees of freedom, one can naturally interpolate between Dirac and Schr\"{o}dinger electrons, by varying the $g$-factor or the effective mass of a 2DEG. In this case we find that it is common for the lowest flat band to have a nonzero Chern number for each spin species, despite the magnetic field having zero spatial average. Such a behavior can be qualitatively described by a three-band model. Our work thus provides flexible platforms for realizing 2D flat-band systems with different superlattice symmetries and nontrivial topology for exploring exotic interaction-driven phases.

The remainder of this paper is organized as follows: In Sec.~\ref{sec:flatbands} we solve the periodic magnetic field problem for Dirac and Schr\"{o}dinger electrons using momentum-space and real-space numerical methods and reveal the flat band behaviors. For the Dirac case we also provide an analytic solution which checks with the numerical results. In Sec.~\ref{sec:wannier} we obtain the maximally localized Wannier functions for the flat bands in both cases, based on which we construct Gaussian-like Wannier functions that can give physically intuitive real-space tight-binding Hamiltonians. In Sec.~\ref{sec:tb} we provide minimal nearest-neighbor tight-binding models based on the information of the Wannier functions obtained in Sec.~\ref{sec:wannier}, which can explain the contrasting behaviors of the two systems. In Sec.~\ref{sec:chern} we study the effect of Zeeman coupling of the periodic magnetic field, and show that the isolated low-energy flat band can quite often have a nonzero (spin) Chern number. Based on the knowledge of the Wannier functions of the low-energy bands we construct a minimal 3-band model that can describe this behavior. Brief discussions and conclusions are given in Sec.~\ref{sec:conclusion}.

\section{Band flattening for Dirac and Schr\"{o}dinger electrons in periodic magnetic fields}\label{sec:flatbands}
\subsection{Dirac electron}\label{subsec:Dirac}
We start by considering a generic 2D Dirac system subject to a perpendicular magnetic field having two cosinusoidal components along $x$ and $y$ directions, respectively: $\mathbf{B}=B[\cos (Kx) + \cos (Ky)]\hat{z}$, where $K\equiv 2\pi/a$ is the wave number with $a$ the period of the magnetic modulation. Specific material realizations and effects of more complex functional forms of fields will be discussed later. The single-particle Hamiltonian is
\begin{equation}\label{eq:HDirac}
H^D=v_{F} {\bm \sigma} \cdot \mathbf{ \Pi },
\end{equation}
where $v_{F}$ is the Fermi velocity of the Dirac electron, $\mathbf{\Pi} = -i\hbar\nabla + e\mathbf{A}$ is the kinetic momentum, with $e$ the absolute value of electron charge, and $\bm \sigma = \sigma_x \hat{x} + \sigma_y \hat{y}$. The vector potential $\mathbf{A}$ corresponding to the periodic magnetic field in the Coulomb gauge is
\begin{equation}\label{eq:Avec}
\mathbf{A}=\frac{B}{K}\left[-\sin (Ky)\hat{x} + \sin(Kx) \hat{y} \right].
\end{equation}
For such a simple vector potential it is convenient to use the plane wave expansion method to solve the eigenvalue problem \cite{MD,supp}. The momentum space Hamiltonian is an infinite-dimensional sparse matrix with the diagonal blocks being 
\begin{eqnarray}
H_0^D(\mathbf{k} + \mathbf{K}) = \mathbf{(k + K)}\cdot {\bm \sigma},
\end{eqnarray}  
where we have chosen $\hbar v_F K$ as the unit of energy, and $K$ as the unit of wave vectors. $\mathbf{K} = m \hat{x} + n \hat{y}$, $m,n\in \mathbb{Z}$, are the reciprocal lattice vectors, and $\mathbf{k}$ is restricted within the 1st Brillouin zone. There is coupling only between diagonal blocks with nearest-neighbor $\mathbf{K}$'s, i.e., separated by $\pm \hat{x}$ or $\pm \hat{y}$. These off-diagonal blocks are
\begin{eqnarray}\label{eq:VDirac}
V(\pm \hat{x}) = \pm \frac{\phi}{2i} \sigma_y, \,\, V(\pm \hat{y}) = \mp \frac{\phi}{2i} \sigma_x,
\end{eqnarray}
where $\phi \equiv eB/\hbar K^2$ is a single dimensionless parameter determining the strength of the magnetic potential. 

To obtain the band structure one has to truncate the momentum space Hamiltonian by choosing an appropriate bound of $\mathbf{K}$  for a given $\phi$ so that the low-energy band structure is converged. We have used a cutoff of the form 
\begin{eqnarray}
\max(|K_x|,|K_y|)\le K_c,
\end{eqnarray}
and found that convergence for moderate values of $\phi \sim 1$ can be well achieved with $K_c = 5$. As it has been noted previously \cite{MD,JNA} such a plane wave expansion method does not require $\phi$ to be small as long as $K_c$ is large enough.

\begin{figure}[ht]
	\subfloat[]{\includegraphics[width=1.8 in]{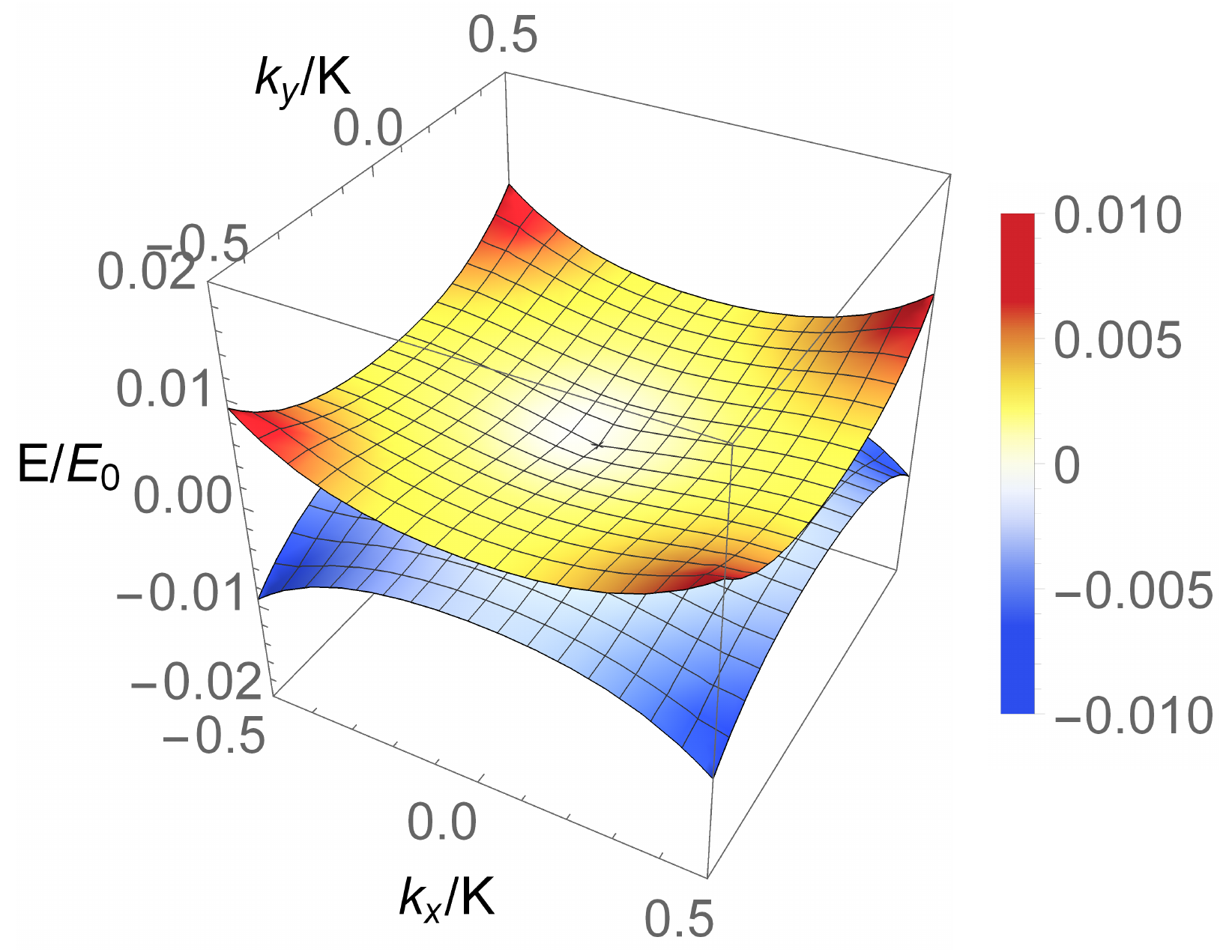}}
	\subfloat[]{\includegraphics[width=1.5 in]{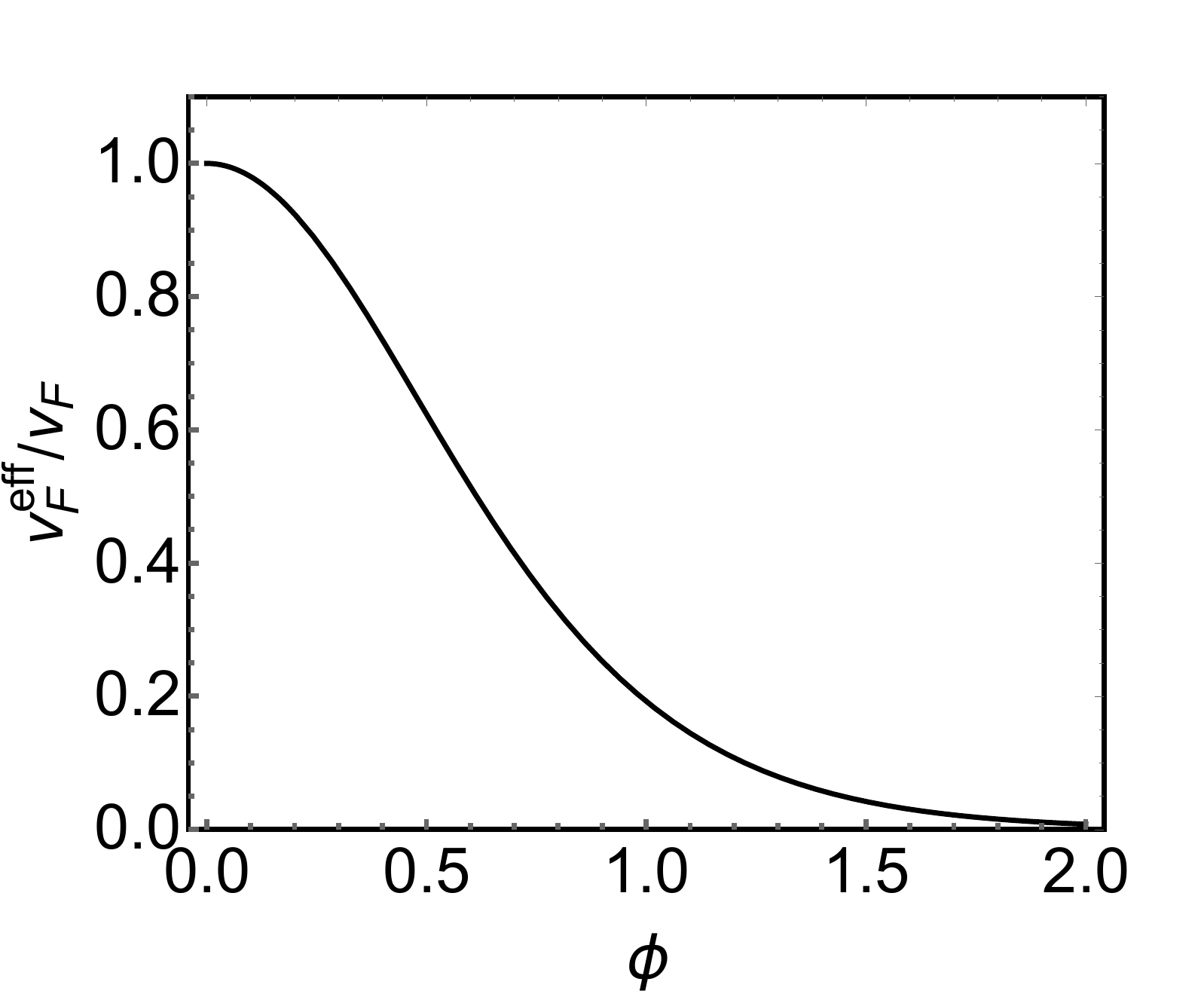}}
	\caption{Flat bands for Dirac electrons in periodic magnetic fields. (a) Band structure for the two particle-hole symmetric bands close to zero energy when $\phi = 2$. $E_0= \hbar v_F K$ is the energy unit. The color scale is the same as $E/E_0$. (b) Renormalized Fermi velocity $v_{F}^{\rm eff}$ vs. $\phi$. A plane wave cutoff of $K_c = 5K$ is used.} 
\label{fig:fbDirac}
\end{figure}

The Dirac Hamiltonian Eq.~\eqref{eq:HDirac} with the periodic vector potential Eq.~\eqref{eq:Avec} has a particle-hole symmetry: $\sigma_z H^D \sigma_z = - H^D$ and a zero energy solution (see below). By diagonalizing the truncated Hamiltonian and focusing on the two particle-hole symmetric bands near zero energy we found that the velocity at $\mathbf{k}=0$ monotonically decreases with increasing $\phi$, and approaches zero asymptotically, as shown in Fig.~\ref{fig:fbDirac}. The two low-energy bands are separated from other bands and their overall band width is also monotonically decreasing. Thus one can get as flat as possible low-energy bands by keeping increasing $\phi$, without fine-tuning which is needed for magic-angle twisted bilayer graphene. Moreover, the flatness is controlled by $\phi = eB/\hbar K^2$ instead of $B$ alone, and can thus be large by having a large period even with a relatively small $B$. Quantitative estimates will be given in Sec.~\ref{sec:conclusion}.

Such behavior of Dirac electrons in periodic magnetic fields can be obtained analytically by perturbing the zero-energy eigen solution of $H^D$ with $\hbar v_F \bm \sigma \cdot \mathbf{k}$, where $\mathbf{k}$ is a small wavevector \cite{JACKIW,snyman_2009,supp}. The effective Hamiltonian written in the two-fold subspace of the zero-energy eigenstates is
\begin{equation}
H_{\rm eff}^{D}=\hbar v_{F}^{\rm eff} \bm \sigma \cdot \mathbf{k},
\label{eq:Heff}
\end{equation}%
where the effective Fermi velocity $v_{F}^{\rm eff}$ for the simple sinusoidal vector potential Eq.~\eqref{eq:Avec} can be explicitly calculated as
\begin{equation}\label{eq:veff}
v_{F}^{\rm eff}=\frac{v_F}{[I_{0}(2\phi )]^2},  
\end{equation}%
where $I_{0}$ is the zeroth modified Bessel function of the first kind. Plotting Eq.~\eqref{eq:veff} vs. $\phi$ gives exactly the same curve as that in Fig.~\ref{fig:fbDirac}. At large $\phi$ one can use the asymptotic form of $I_0$ to get
\begin{equation}\label{eq:vfinal}
v_{F}^{\rm eff}=4\pi \phi e^{-4\phi } v_F.
\end{equation}
Therefore the renormalized Fermi velocity exponentially decreases with increasing $\phi$, but never becomes exactly zero. 

Above results can be easily generalized to (co)sinusoidal square lattice magic fields with unequal amplitudes $B_{1,2}$ and/or wave numbers $K_{1,2}$ along $x$ and $y$ directions, with $v_{F}^{\rm eff}=v_F/I_{0}(2\phi_1 )I_0(2\phi_2)$, where $\phi_{1,2} = eB_{1,2}/\hbar K_{1,2}^2$. The corresponding $\phi_{1,2}\gg 1$ asymptotic form is $v_{F}^{\rm eff}\approx 4\pi\sqrt{\phi_1\phi_2}e^{-2(\phi_1+\phi_2)}$. For a triangular lattice periodic magnetic field, we did not find an analytic expression of $v_{F}^{\rm eff}$, but numerical calculation shows that the band flattening behavior is qualitatively the same as the square lattice case \cite{supp}. Thus periodic magnetic fields can be used as an effective way of creating flat band Dirac systems with different superlattice symmetries. 

On the other hand, when $\phi\ll 1$ one can also obtain an effective $2\times 2$ Hamiltonian using perturbation theory and keeping the lowest order in $\phi$. Such a calculation \cite{supp} shows that $v^{\rm eff}_F\approx (1-\phi^2)v_F$ which describes the quadratic behavior of $v^{\rm eff}_F(\phi)$ at small $\phi$ in Fig.~\ref{fig:fbDirac}. When $\phi\gtrsim 1$ the perturbation theory obviously breaks down, but $\phi\sim 1$ can nevertheless be viewed as a critical scale of the magnetic field at which $v^{\rm eff}_F(\phi)$ starts to decay exponentially.  

We note that $v^{\rm eff}_F = 0$ does not necessarily mean the corresponding bands are flat throughout the Brillouin zone. In practice flat bands are interesting mainly because they lead to diverging density of states which makes correlation effects most pronounced. $v^{\rm eff}_F = 0$ at $\mathbf{k} = 0$ is not a sufficient condition for diverging density of states. However, for the simple form of the potential considered here, the overall flattening of the lowest band throughout the Brillouin zone is consistent with the behavior near $\mathbf{k} = 0$. This can be seen, for example, by looking at the momentum space Hamiltonian at the Brillouin zone boundary. The lowest bands at $\mathbf{k} = \frac{1}{2}\hat{x}$ are doubly degenerate in the absence of the magnetic field and have energies $\epsilon = \pm 1/2$. In each of the 2-fold degenerate subspaces, magnetic field induces a splitting proportional to $\phi/2$ according to Eq.~\eqref{eq:VDirac}. We note in passing that a periodic scalar potential does not split the two doublets, which is another reason why periodic magnetic fields are special in getting flat bands. Thus $\phi\sim 1$ is a crude estimate of when the lowest bands become very close to zero energy at the Brillouin zone boundary. (The estimate based on degenerate perturbation breaks down when $\phi \gtrsim 1$.) For a smooth vector potential such as Eq.~\eqref{eq:Avec} the lowest bands are not expected to vary strongly throughout the Brillouin zone. Thus the monotonic decrease of $v^{\rm eff}_F$ at $\mathbf{k} = 0$ together with the approaching of low-energy bands towards 0 at zone boundary suggest the overall flattening of the lowest band and the diverging density of states as $\phi$ increases. 

Another consequence of the flat band, at least near $\mathbf{k}=0$, is the immobility of the wavepacket centered around $\mathbf{k}=0$. Physically it means that particles described by such wavepackets will be easily trapped or localized by disorder. This is formally considered as the homogenization problem in PDE theory, which absorbs the effect of a periodic potential into an effective mass tensor by considering the dynamics at a much larger scale than the period. There is a large literature on the subject in the Schr\"odinger case, see e.g. \cite{allaire, naoufel} for some rigorous mathematical references. The situation is similar for the Dirac equation under appropriate assumptions, which will be addressed in a future work \cite{future}. In this context the vanishing $v^{\rm eff}_F$ directly corresponds to flat bands for the Dirac operator.

While the plane wave expansion method is generally applicable to any periodic potential, in reality it is sometimes more convenient to work in real space, especially when translational symmetry is broken. However, for the Dirac operator considered here, standard finite difference approximations are plagued by the so-called Fermion doubling problem: the obtained discrete dispersion relation is non-monotonic and, as a consequence, spurious unphysical modes are created by the numerical scheme. Some solutions, based on doubling the number of unknowns and introducing staggered grids, were proposed in \cite{hammer1,hammer2}. They result in schemes somewhat difficult to implement and we decided to follow a different approach: we used spectral methods, that have the advantage of providing a monotonic, high precision approximation of the linear dispersion relation of the free Dirac equation at a low computational and implementation cost. The method will be described in a forthcoming work \cite{future}. We have compared the band structures calculated with the spectral method to that from plane wave expansion and find they are in excellent agreement.

\subsection{Schr\"{o}dinger electron}\label{subsec:Schr}

We next show that periodic magnetic fields can lead to flat bands for 2D Schr\"{o}dinger electrons, but only at discrete values of the parameter $\phi$. Using the same vector potential Eq.~\eqref{eq:Avec}, the Hamiltonian is 
\begin{eqnarray}
H^S = \frac{1}{2m}\mathbf{\Pi}^2,
\end{eqnarray}
where $m$ is the effective mass of electrons in a given system. Using $\hbar^2 K^2/2m$ and $K$ as the units of energy and wave vector, respectively, the momentum space Hamiltonian matrix has the diagonal elements
\begin{eqnarray}\label{eq:HS0}
H_0^S(\mathbf{k+K}) = |\mathbf{k+K}|^2 + \phi^2.
\end{eqnarray}
The off-diagonal elements $V(\mathbf{K}')$ that couple $H_0^S(\mathbf{k+K})$ to $H_0^S(\mathbf{k+K-K'})$ are nonzero for the following values of $\mathbf{K'}$ \cite{supp}:
\begin{eqnarray}\label{eq:HSoffdiag}
&&V(\pm \hat{x}) = \mp i \phi (k_y+K_y),\\\nonumber
&&V(\pm \hat{y}) = \pm i\phi (k_x + K_x),\\\nonumber
&&V(\pm 2\hat{x}) = V(\pm 2\hat{y}) = -\frac{\phi^2}{4}.
\end{eqnarray}
Note that $V(\mathbf{K'})$ is also dependent on $\mathbf{k+K}$. 

\begin{figure}[ht]
		\subfloat[]{\includegraphics[width=1.7 in]{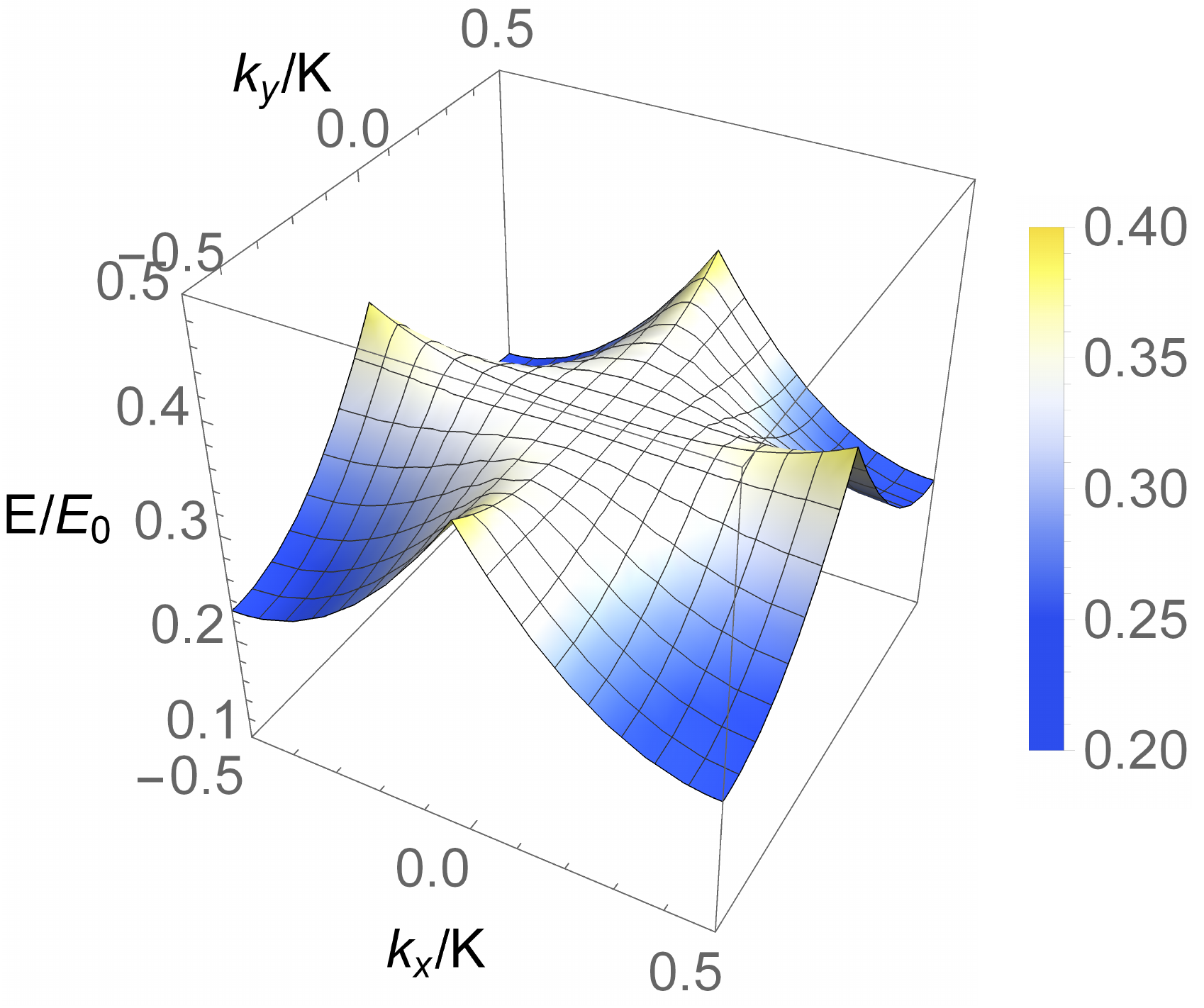}}
		\subfloat[]{\includegraphics[width=1.6 in]{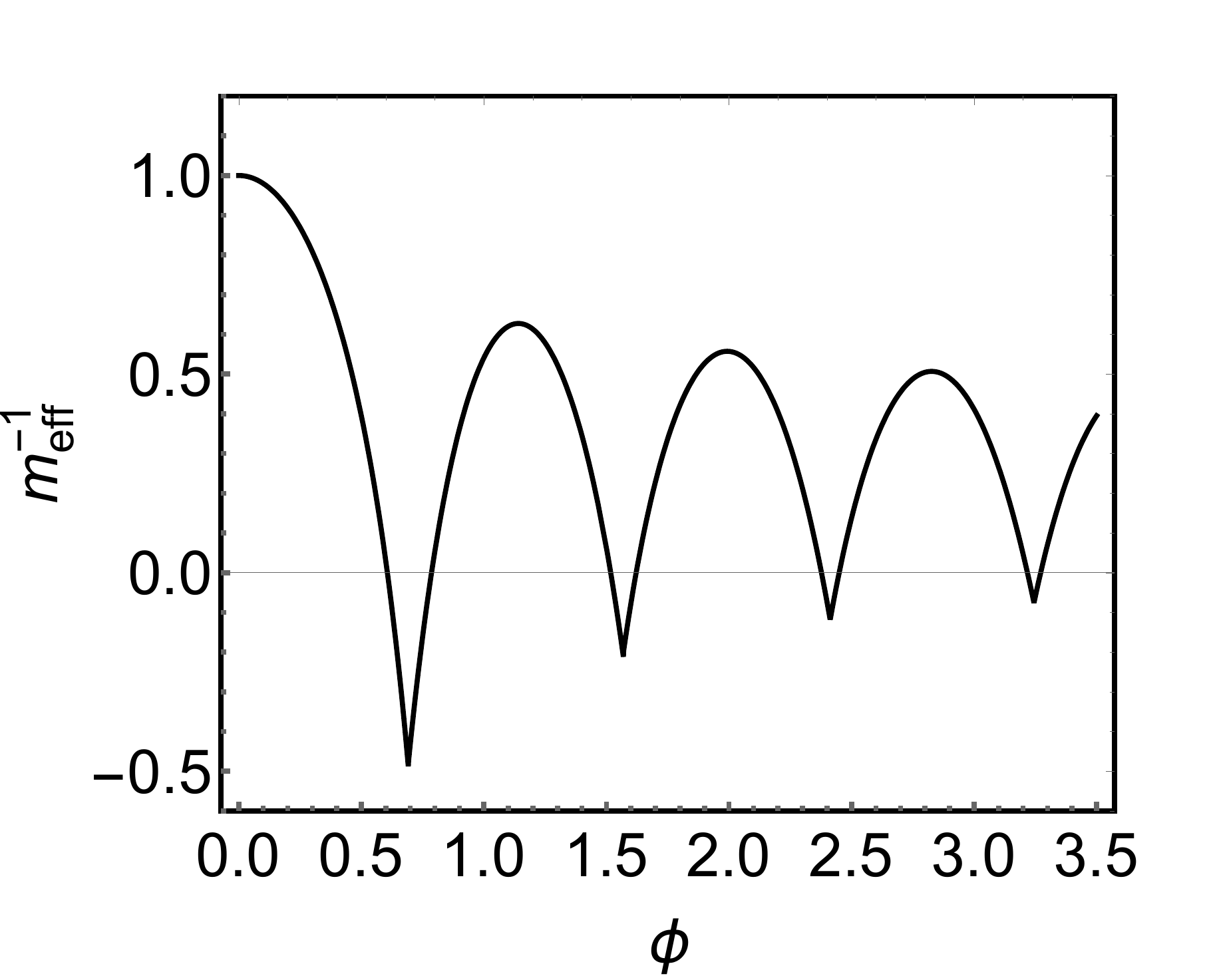}}
	\caption{\label{fig:fbSchr}
Flat bands for 2DEG in periodic magnetic fields. (a) Band structure for the lowest band when $\phi = 0.6$ near the first magic value. $E_0= \hbar^2 K^2/2m$ is the energy unit. The color scale is the same as $E/E_0$ with white corresponding to the energy at $\mathbf{k}=0$. (b) Renormalized inverse effective mass $m^{-1}_{\rm eff}$ (in units of $m^{-1}$) vs. $\phi$. A plane wave cutoff of $K_c = 9K$ is used.}
\end{figure}

By diagonalizing the momentum space Hamiltonian with a large enough cutoff, we calculate the inverse effective mass of the lowest band $m^{-1}_{\rm eff}$ at $\mathbf{k} = 0$ and plot it against $\phi$. Fig.~\ref{fig:fbSchr} (b) shows that $m^{-1}_{\rm eff}$ has an oscillatory dependence on $\phi$ and crosses zero repeatedly as $\phi$ increases. Our real space calculation using the spectral method gives the same result, although for the Schr\"{o}dinger equation a finite difference formula can also be used. Although for the smaller magic values of $\phi$ the width of the lowest band is not that small, the vanishing of $m^{-1}_{\rm eff}$ leads to a diverging density of states at the energy at $\mathbf{k}=0$. The vanishing $m^{-1}_{\rm eff}$ also leads to immobile wavepackets centralized at $\mathbf{k}=0$ in the homogenization sense. Our calculations for a triangular lattice periodic magnetic field also show similar oscillatory behavior \cite{supp}. Thus in contrast to Dirac electrons, 2DEG can have flat bands with exact vanishing of $m^{-1}_{\rm eff}$ at magic values of $\phi$. 

Unlike the Dirac case, for Schr\"{o}dinger electrons we are not able to find an analytic solution of the lowest band. However, since the smallest magic value $\phi\approx 0.6$ is less than 1, 2nd order perturbation may still be valid near this value \cite{supp}. The effective Hamiltonian thus obtained is
\begin{eqnarray}\label{eq:HSeff}
H^S_{\rm eff}(\mathbf{k}) = k^2(1-2\phi^2) + \phi^2.
\end{eqnarray}
Thus the inverse mass vanishes when
\begin{eqnarray}
\phi = \frac{1}{\sqrt{2}} \approx 0.707,
\end{eqnarray}
which is off by only about 15\%. That the 2nd order perturbation is approximately valid can also be seen from the exact result in Fig.~\ref{fig:fbSchr} (b), which shows that before reaching its first minimum $m^{-1}_{\rm eff}$ is roughly quadratic in $\phi$. Since the quadratic $\phi$ dependence in Eq.~\eqref{eq:HSeff} is accurate when $\phi\rightarrow 0$, it should serve as a good approximation until the behavior of $m^{-1}_{\rm eff} (\phi)$ significantly changes. However, to understand the origin of the recurring magic values in the Schr\"{o}dinger case and why there is no magicness in the Dirac case, we have to look into details of the wavefunctions associated with the flat bands.

\section{Wannier functions of the flat bands}\label{sec:wannier}

In this section we examine the Wannier functions associated with the lowest bands for both Dirac and Schr\"{o}dinger electrons, which sets the stage for our interpretation of the contrasting band flattening behaviors using minimal tight-binding models in the next section. We note that Wannier functions localized by periodic magnetic fields is by itself an interesting problem, as historically the discussion on the effect of magnetic fields on Wannier functions is mostly focused on slow-varying magnetic fields on the length scale of the Wannier functions or equivalently of the lattice constants \cite{luttinger_1951, wannier_1962, blount_1962} in crystalline solids. In this case the effect of magnetic fields can be approximately described as Peierls phase in the Hamiltonian written in the basis of Wannier functions, and the Wannier functions themselves are only slightly modified through a phase factor. In the present systems, however, the ``lattice constant" is set by the spatial period of the magnetic field, and the slow-variation assumption cannot be justified \emph{a priori}.

The Wannier function $\phi_n$ of an isolated band $n$ with Bloch eigenfunction $\psi_{n\mathbf{k}}$ is defined as
\begin{eqnarray}\label{eq:WFdef}
\phi_n(\mathbf{r-R}) = \frac{1}{V_{\rm BZ}} \int_{\rm BZ} d\mathbf{k} e^{-i\mathbf{k\cdot R}}\psi_{n\mathbf{k}}(\mathbf{r}),
\end{eqnarray}
where $\mathbf{R}$ is a lattice vector, BZ means Brillouin zone and $V_{\rm BZ}$ is its volume. While $\psi_{n\mathbf{k}}$ is determined up to a $\mathbf{k}$ dependent phase factor $e^{i\alpha_{n\mathbf{k}}}$ by the Hamiltonian, $\phi_n(\mathbf{r-R})$ is in general not unique or gauge invariant. It has been shown that for 1D systems Wannier functions are exponentially localized \cite{kohn_1959}, and for 2D and 3D systems Wannier functions are exponentially localized if the Chern numbers of the corresponding bands are zero \cite{ nenciu_1983, brouder_2007}. For the exponentially localized Wannier functions one can define a ``maximally localized" gauge  which minimizes the spread functional
\begin{eqnarray}
\Omega_n\equiv \langle r^2\rangle_n - \langle \mathbf{r}\rangle_n^2,
\end{eqnarray}
where $\langle \rangle_n$ means expectation value under the Wannier state $\phi_n$. The definition can be extended to a group of $N$ bands that are isolated from other bands, for which the Wannier functions have a gauge freedom of $U(N)$ and a maximally localized gauge is defined as that minimizes the sum of $\Omega_n$ over all $N$ Wannier functions. In the following we start from finding the maximally localized Wannier functions (MLWFs) of the lowest band (see below) of Dirac and Schr\"{o}dinger electrons in periodic magnetic fields.

We first introduce a trick which can help us describe Dirac and Schr\"{o}dinger electrons in a unified manner. Because of the particle-hole symmetry of the Dirac Hamiltonian $H^D$ in Eq.~\eqref{eq:HDirac}, one can get the eigenspectrum by considering $(H^{D})^2$, i.e. the Hamiltonian squared:
\begin{eqnarray}\label{eq:HD2}
(H^D)^2 = v_F^2 \mathbf{\Pi}^2 + e \hbar v_F^2 B(\mathbf{r}) \sigma_z,
\end{eqnarray}
which is identical to the Hamiltonian of a Schr\"{o}dinger electron of ``mass" $1/2v_F^2$ subject to the same vector potential $\mathbf{A}$ and a periodic ``Zeeman" potential $e\hbar v_F^2 B(\mathbf{r})$, despite the different dimensions. In the case of a uniform magnetic field this extra term shifts the 0th Landau level to zero energy and represents the $\pi$ Berry phase of Dirac electrons. Since there is no spin-orbit coupling in the present problem the periodic Zeeman field can be viewed as scalar potentials of opposite signs for opposite spin directions. Below we consider the branch corresponding to the positive eigenvalue of $\sigma_z$ in Eq.~\eqref{eq:HD2} unless otherwise noted.

In momentum space the diagonal elements of $(H^D)^2$ are the same as Eq.~\eqref{eq:HS0} in the dimensionless form (with $\hbar^2 v_F^2 K^2$ the ``energy" unit), and the extra Zeeman term modifies the off-diagonal elements by adding a $\phi/2$ to $V(\pm \hat{x})$ and $V(\pm \hat{y})$ in Eq.~\eqref{eq:HSoffdiag}. More generally, the Zeeman coupling for a 2DEG is 
\begin{eqnarray}
H_{\rm Zeeman} = \frac{g\mu_B}{2} \bm \sigma\cdot \mathbf{B(r)},
\end{eqnarray}
where $g$ is an effective $g$ factor and $\mu_B = e\hbar /2m_e$ is the Bohr magneton. Comparing it with the last term in $(H^D)^2$ in the dimensionless form, one can see that $(H^D)^2$ corresponds to the special case of $g m/m_e = 2$, i.e., free electron in vacuum, as expected. Conversely, the situation of a Dirac system in periodic magnetic fields can be captured by a 2DEG with $g m/m_e = 2$. We will consider the cases when $g m/m_e$ is different from 2 in Sec.~\ref{sec:chern}.

\begin{figure}[ht]
	\subfloat[]{\includegraphics[width=1.7 in]{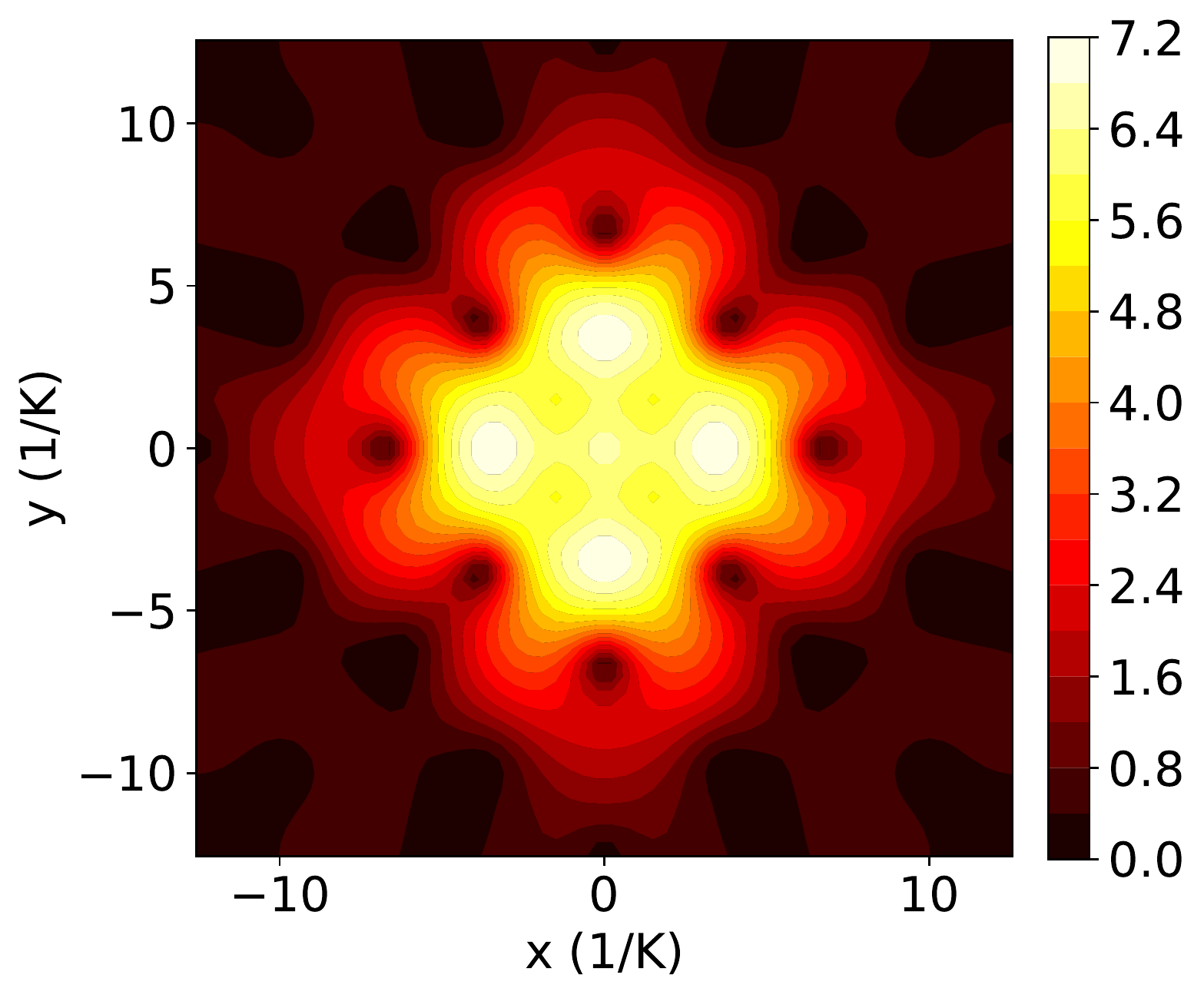}}
	\subfloat[]{\includegraphics[width=1.6 in]{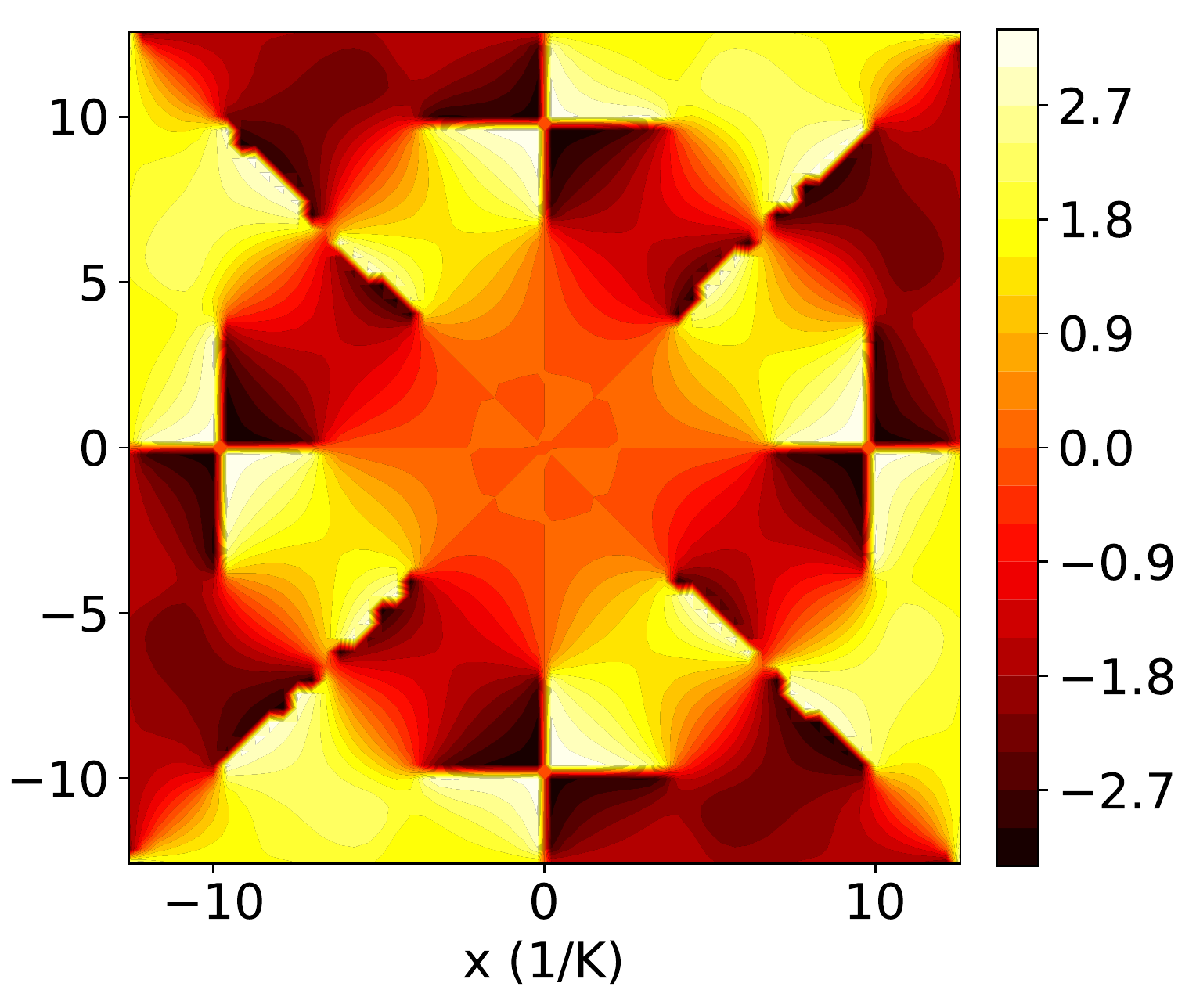}}
	\caption{\label{fig:MLWFSchr}
Norm (a) and phase (b) of the MLWF of the lowest band of a Schr\"{o}dinger electron in the periodic magnetic field near the first magic value of $\phi\approx 0.6$. A plane wave cutoff of $K_c = 5K$ and a Brillouin zone discretization of $11\times 11$ were used.}
\end{figure}

Using above trick we are able to get the same behavior of $v_{F}^{\rm eff} (\phi)$ in Fig.~\ref{fig:fbDirac} from the lowest band of $(H^D)^2$. For our purpose of getting the relevant Wannier functions for both Dirac and Schr\"{o}dinger electrons we now only need to minimize $\Omega$ for the lowest energy band with or without the Zeeman term. The minimization was done using the algorithm introduced in \cite{marzari_1997}. Because of the broken time-reversal symmetry the Wannier functions are in general complex and have a spatially dependent phase. The MLWF of the lowest band for the Schr\"{o}dinger case, obtained by starting from an initial guess of a Gaussian function located at the origin, is shown in Fig.~\ref{fig:MLWFSchr}. The absolute value of the Wannier function has four peaks at $\pm \frac{\pi}{K}\hat{x}$ and $\pm \frac{\pi}{K}\hat{y}$. 

To understand why peaks appear at these specific positions, we note that Eq.~\eqref{eq:WFdef} yields
\begin{eqnarray}\label{eq:unk0}
\psi_{n\mathbf{k}=0}(\mathbf{r}) = u_{n\mathbf{k}=0} (\mathbf{r}) = \sum_{\mathbf{R}} \phi_n(\mathbf{r-R}),
\end{eqnarray}
where $u_{n\mathbf{k}}(\mathbf{r})$ is the periodic part of $\psi_{n\mathbf{k}}(\mathbf{r})$. Thus $u_{n\mathbf{k}=0} (\mathbf{r})$ is a superposition of all Wannier functions shifted by different lattice vectors. Moreover, $u_{n\mathbf{k}=0}(\mathbf{r})$ is a solution of the original eigenvalue problem defined in the domain of a unit cell with periodic boundary condition. For such a problem the peaks of $u_{n\mathbf{k}=0}$ are determined by the minima of the potential $|\mathbf{A}|^2 = [\sin^2(Kx) + \sin^2(Ky)]B^2/K^2$, which are at $(x,y)=(0,0)$, $(\pi/K,\pi/K)$, $(\pi/K,0)$, and $(0,\pi/K)$ in the unit cell. Although Eq.~\eqref{eq:unk0} does not uniquely determine $\phi_n(\mathbf{r})$, when the Wannier function $\phi_n(\mathbf{r})$ is well localized within one unit cell the peaks of $u_{n\mathbf{k}=0}(\mathbf{r})$ should be the same as those of $\phi_n(\mathbf{r})$. However, Fig.~\ref{fig:MLWFSchr} indicates that this is not the case: The two peaks at $(\pi/K,0)$, and $(0,\pi/K)$ are more pronounced than that at $(0,0)$, while the one at $(\pi/K,\pi/K)$ is absent. That the two pairs of peaks have different behaviors can be partly understood in the following way. The periodic magnetic field divides the system into square plaquettes with either positive or negative fields along $z$, separated by lines with vanishing $\mathbf{B}$. The peak positions $(0,0)$ and $(\pi/K,\pi/K)$ are at the centers of plaquettes of opposite fields, while $(\pi/K,0)$, and $(0,\pi/K)$ are at the corners of a plaquette. Thus the two pairs of peaks do not have to have the same heights. 

\begin{figure}[ht]
	\subfloat[]{\includegraphics[width=1.7 in]{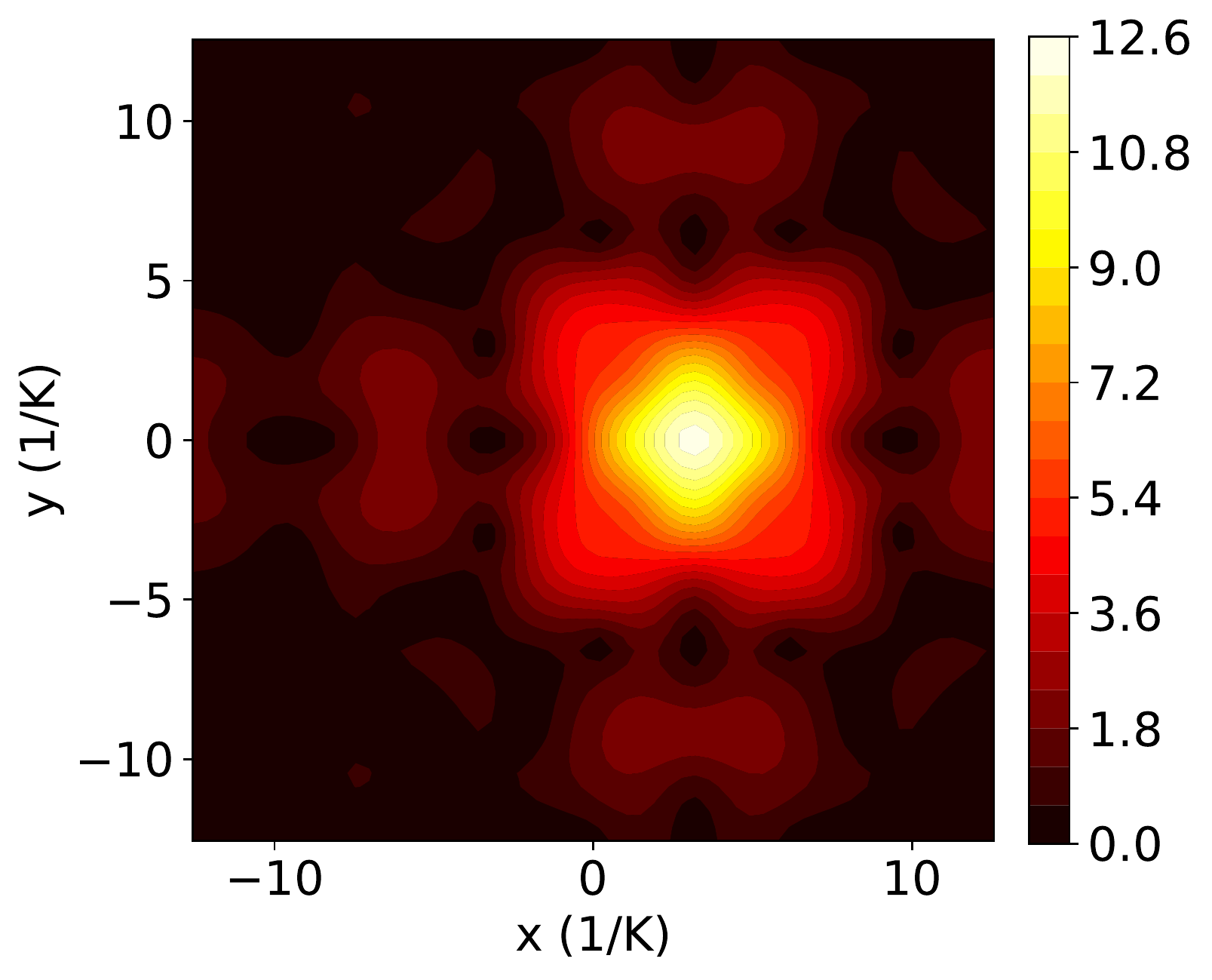}}
	\subfloat[]{\includegraphics[width=1.6 in]{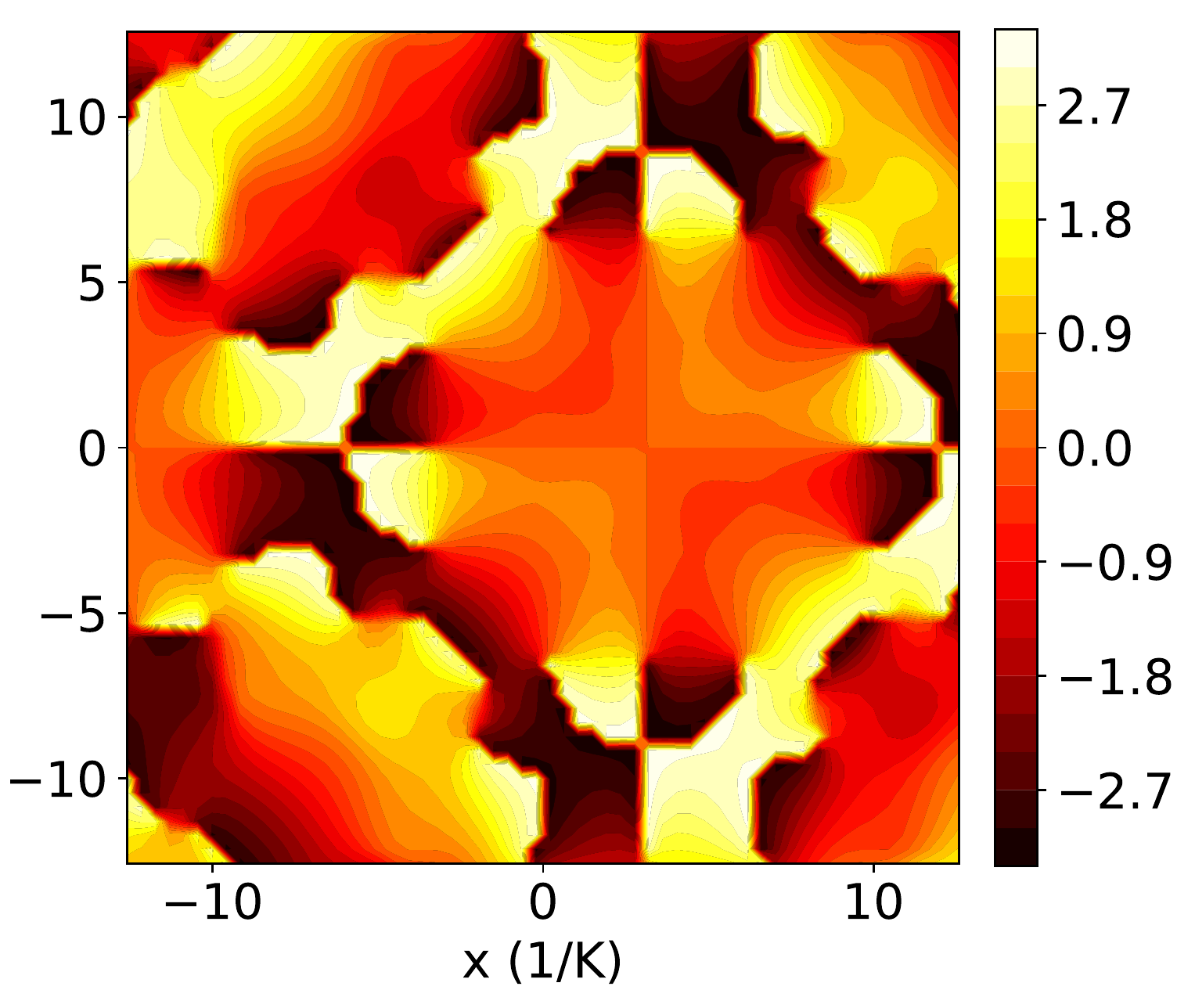}}\\
	\subfloat[]{\includegraphics[width=1.7 in]{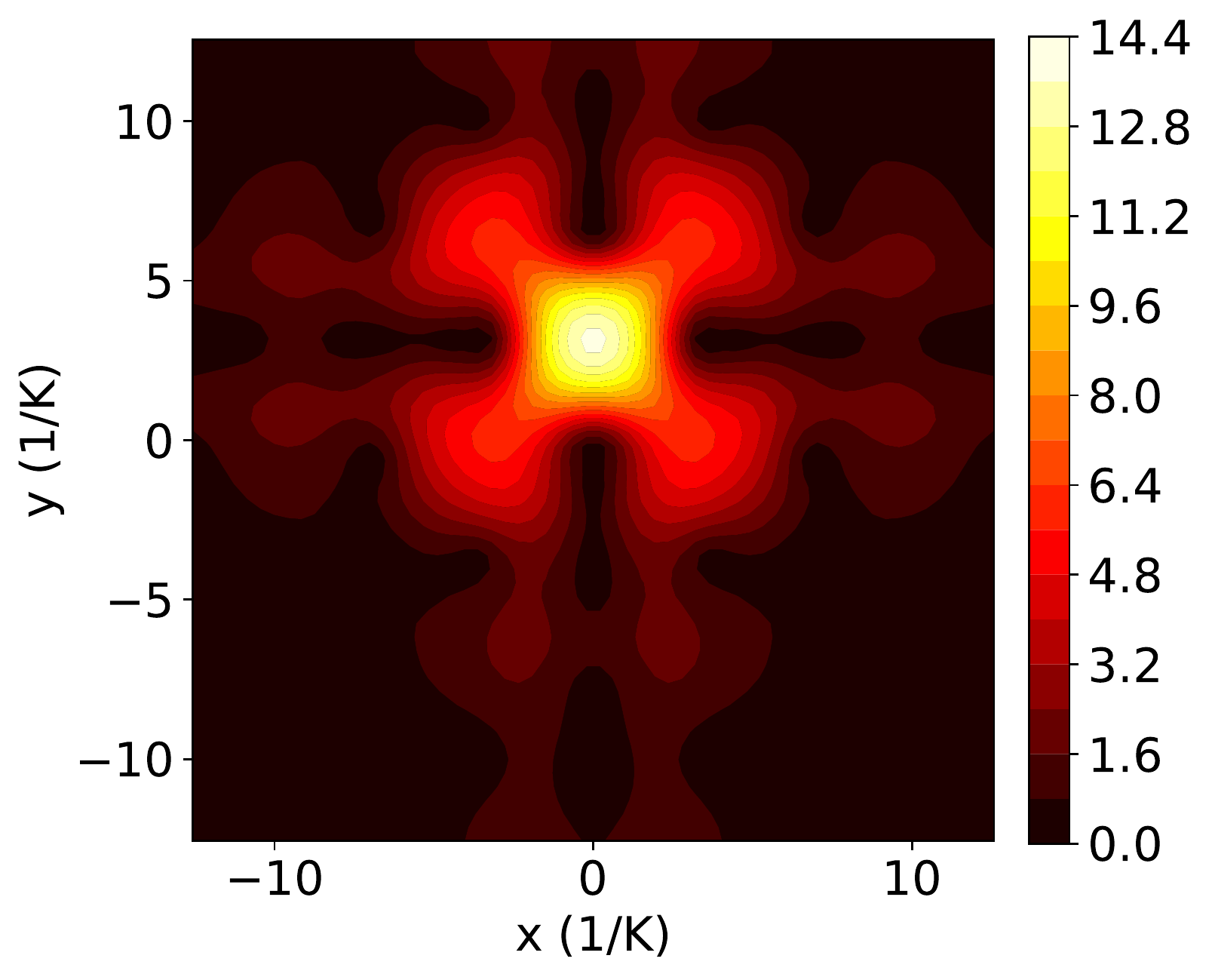}}
	\subfloat[]{\includegraphics[width=1.6 in]{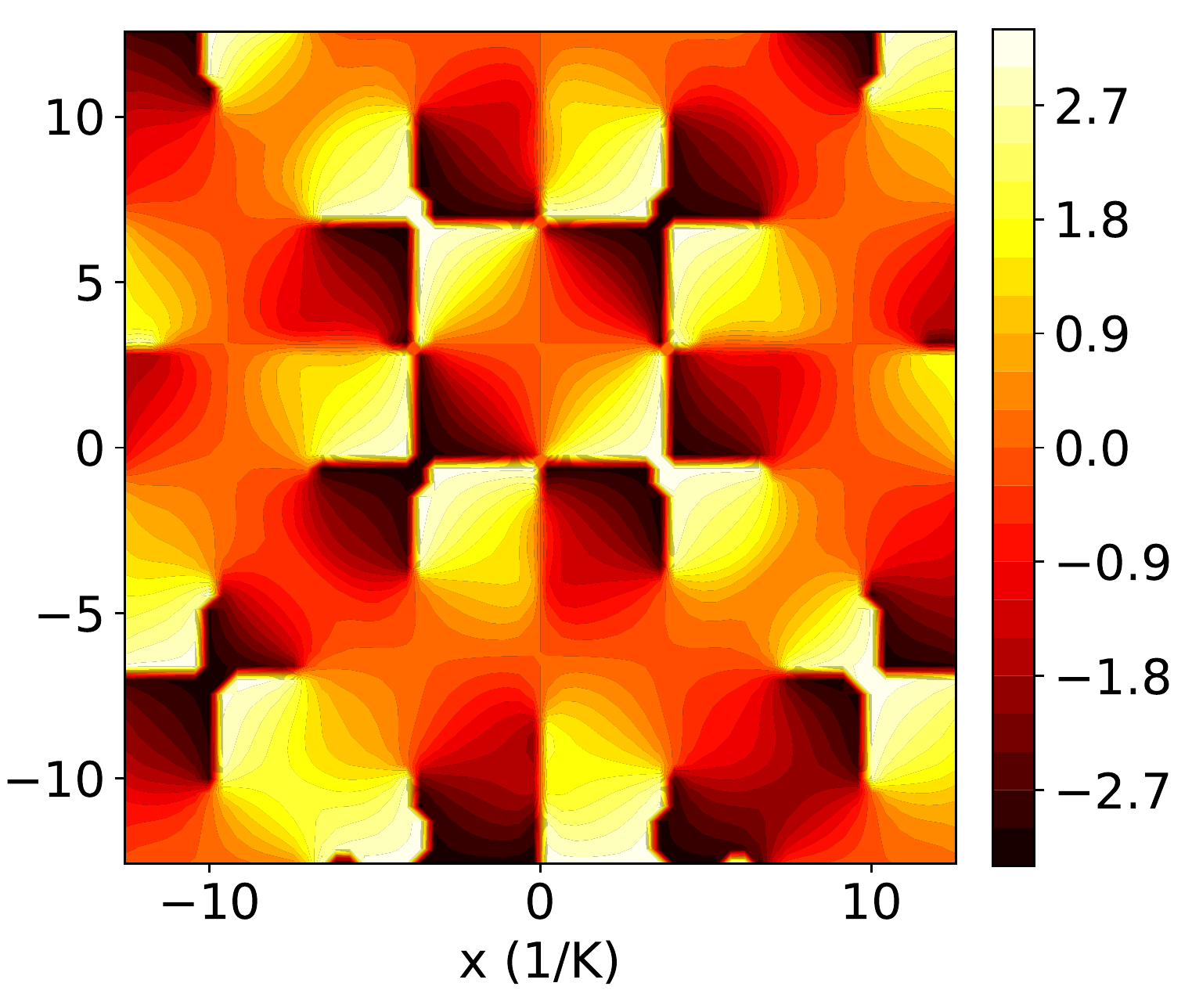}}\\
	\subfloat[]{\includegraphics[width=1.7 in]{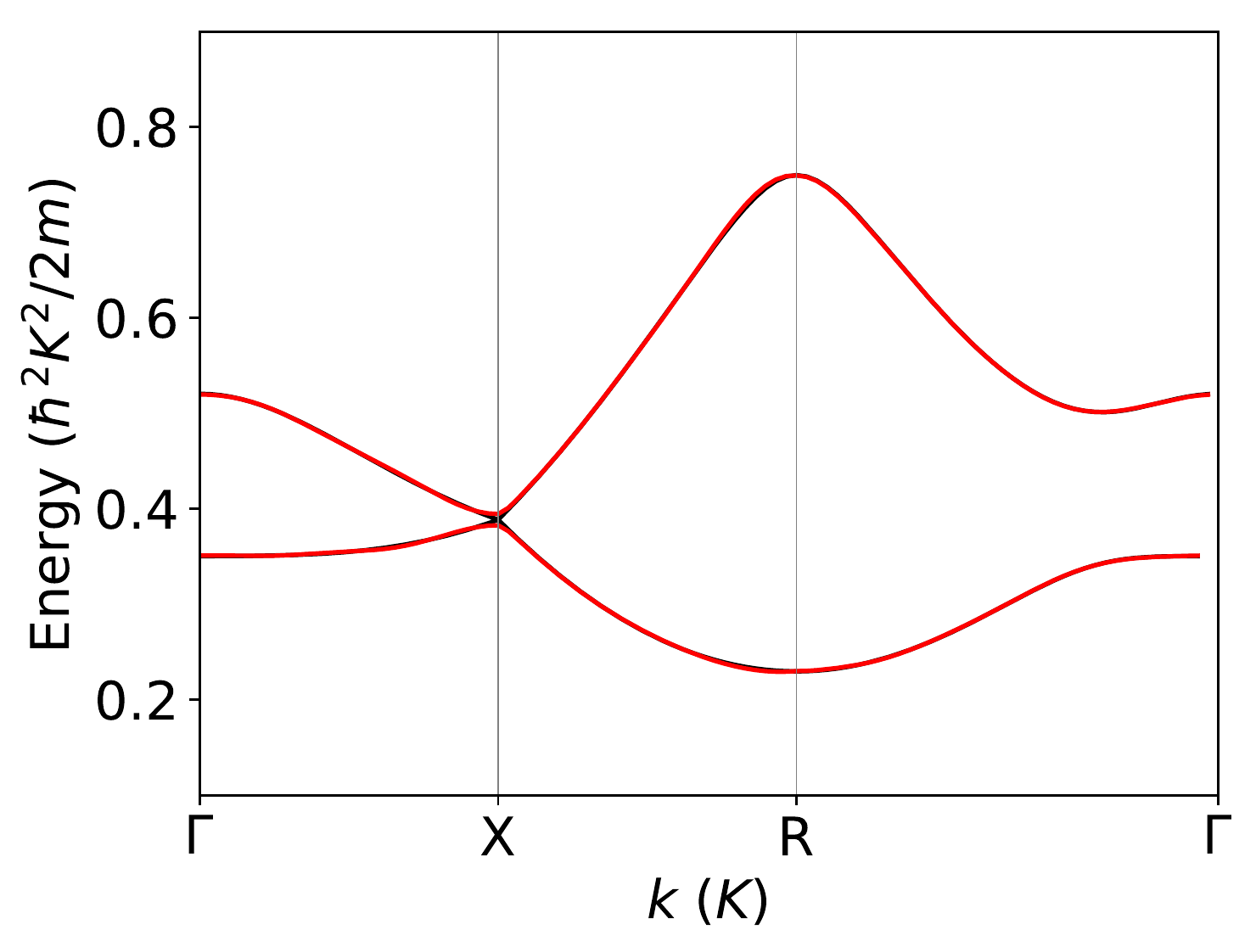}}
	\caption{\label{fig:WFSchrproj} Wannier functions of the two lowest bands of a Schr\"{o}dinger electron. (a) and (b): Norm and phase of the first Wannier function $\phi_A$ located at $(\pi/K,0)$. (c) and (d): Norm and phase of the 2nd Wannier function $\phi_B$ located at $(0,\pi/K)$. (e) Wannier-interpolated band structure (red solid lines) compared with the plane wave result (black solid lines). $\phi=0.6$. A plane wave cutoff of $K_c = 5K$ and a Brillouin zone discretization of $11\times 11$ were used. Width of the two Gaussians used for constructing the Wannier functions is set to $8/K$.}
\end{figure}

In the basis of this Wannier function (written as $\phi_1$ from now on) the lowest band of the Schr\"{o}dinger electron can be represented by a one-dimensional tight-binding Hamiltonian, with the hopping parameters
\begin{eqnarray}
t_{\mathbf{R}} \equiv \int d^2\mathbf{r} \phi^*_1(\mathbf{r}) H^S \phi_1(\mathbf{r-R}).
\end{eqnarray}
It is, however, not intuitive why such a Hamiltonian gives recurring flat band at magic values of $\phi$, since $t_{\mathbf{R}}$ depends on $\phi$ through $\phi_1$ and $H^S$ in a complicated way. To go further, we note that the peaks of $|\phi_1|$ suggest that it may be possible to use a basis of two Gaussian-like Wannier functions, located at the plaquette corners $(\pi/K,0)$ and $(0,\pi/K)$ to describe the lowest band. Moreover, the phase around these two peaks, as shown in Fig.~\ref{fig:MLWFSchr}, changes fastest along the plaquette boundaries, which is similar to the behavior in slow-varying magnetic fields described by the Peierls phase. We thus project $\psi_{1\mathbf{k}}$ and $\psi_{2\mathbf{k}}$, Bloch functions of the two lowest bands, onto two Gaussians $g_A$ and $g_B$ located at $(\pi/K,0)$ and $(0,\pi/K)$ respectively:
\begin{eqnarray}
&&\phi_{A\mathbf{k}}(\mathbf{r}) = \langle g_A | \psi_{1\mathbf{k}}\rangle \psi_{1\mathbf{k}} (\mathbf{r}) + \langle g_A | \psi_{2\mathbf{k}}\rangle \psi_{2\mathbf{k}}(\mathbf{r})\\\nonumber
&&\phi_{B\mathbf{k}}(\mathbf{r}) = \langle g_B | \psi_{1\mathbf{k}}\rangle \psi_{1\mathbf{k}} (\mathbf{r}) + \langle g_B | \psi_{2\mathbf{k}}\rangle \psi_{2\mathbf{k}}(\mathbf{r}),
\end{eqnarray}
which are then orthonormalized. Even though we did not run the maximal localization routine for the reason explained further below, the tight-binding Hamiltonian in this basis has fast decaying hopping parameters with increasing distance \cite{supp}, and the interpolated band structure from this Hamiltonian fits that obtained using the plane wave method very well [Fig.~\ref{fig:WFSchrproj} (e)].

Although the shapes of the two Wannier functions deviate from Gaussian-like after projection and orthonormalization, they are still localized at $(\pi/K,0)$ and $(0,\pi/K)$. Moreover, each of them has a phase distribution qualitatively consistent with the Peierls form, i.e., the phase increases fastest along the lines with large line integral of the vector potential. One would then wonder if the real space tight-binding Hamiltonian in the basis of these two Wannier functions also has complex hopping parameters with Peierls phases. We find that this is indeed the case. For example, the nearest neighbor hopping from $\phi_A$ to $\phi_B$ at $\phi = 0.6$ is about $0.057 i$ along $\pm(\hat{x}+\hat{y})$, and $- 0.057 i$ along $\pm(\hat{x}-\hat{y})$, which are mutually complex conjugate as expected from the behavior of $\exp(i\frac{e}{\hbar}\int \mathbf{A}\cdot d\mathbf{l})$. Moreover, it is surprising that the nearest-neighbor hopping is almost purely imaginary near the first magic value of $\phi$. This behavior motivates us to propose the minimal tight-binding model in Sec.~\ref{sec:tb}, based on which we explain the recurring magic values of Schr\"{o}dinger electrons. We have also tried to run the maximal localization routine for these two Wannier functions. However, the resulting MLWFs are of more complex shape with multiple peaks at $(\pm \pi/K ,0)$, $(0, \pm \pi/K)$, and $(0,0)$ \cite{supp}, which is somewhat expected based on the lowest-band MLWF in Fig.~\ref{fig:MLWFSchr}. Such a basis does not give as intuitive hopping parameters as that from the Gaussian-like Wannier functions without maximal localization. Therefore we will not discuss about them any further. 

\begin{figure}[ht]
	\subfloat[]{\includegraphics[width=1.7 in]{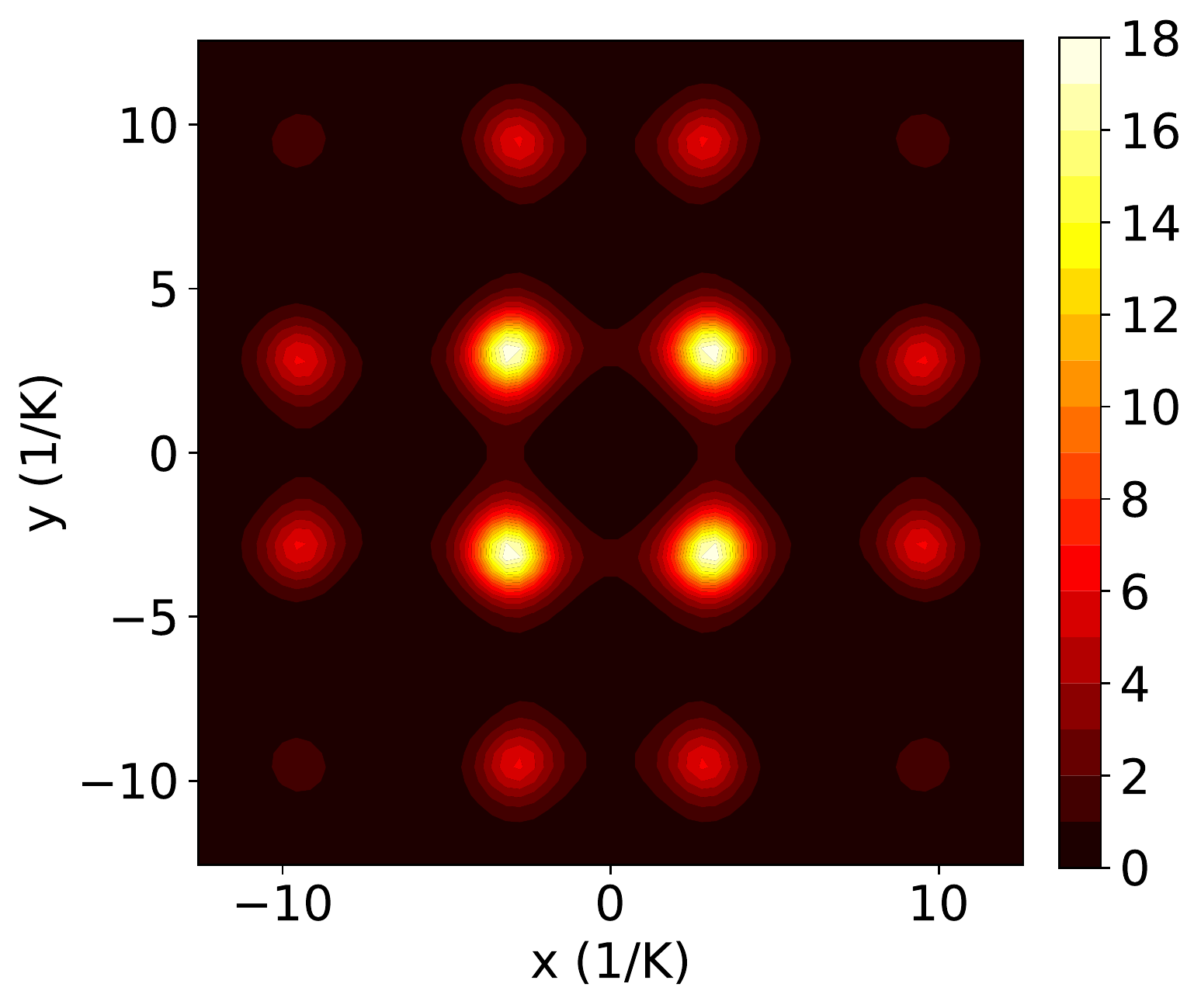}}
	\subfloat[]{\includegraphics[width=1.65 in]{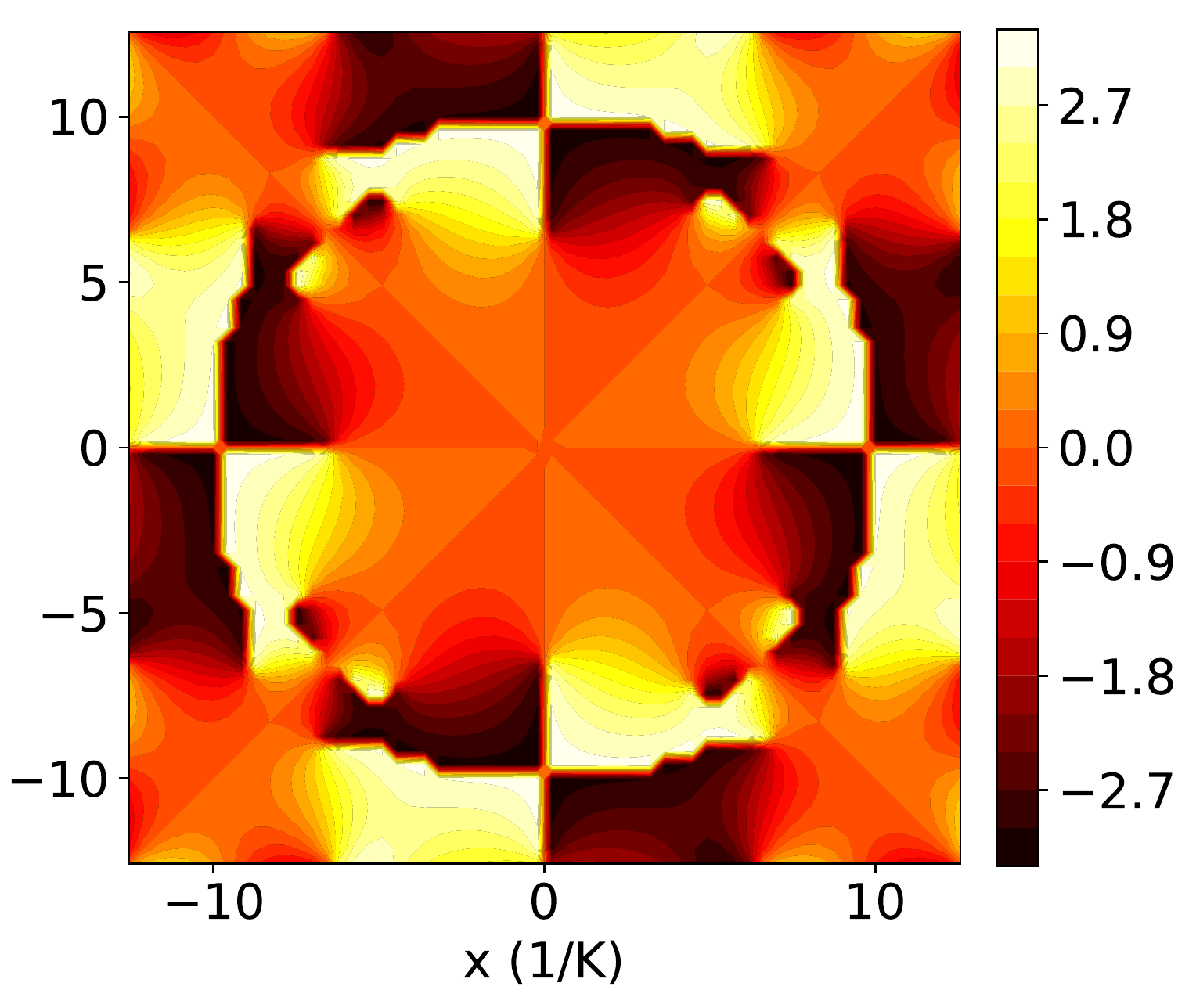}}\\
	\subfloat[]{\includegraphics[width=1.7 in]{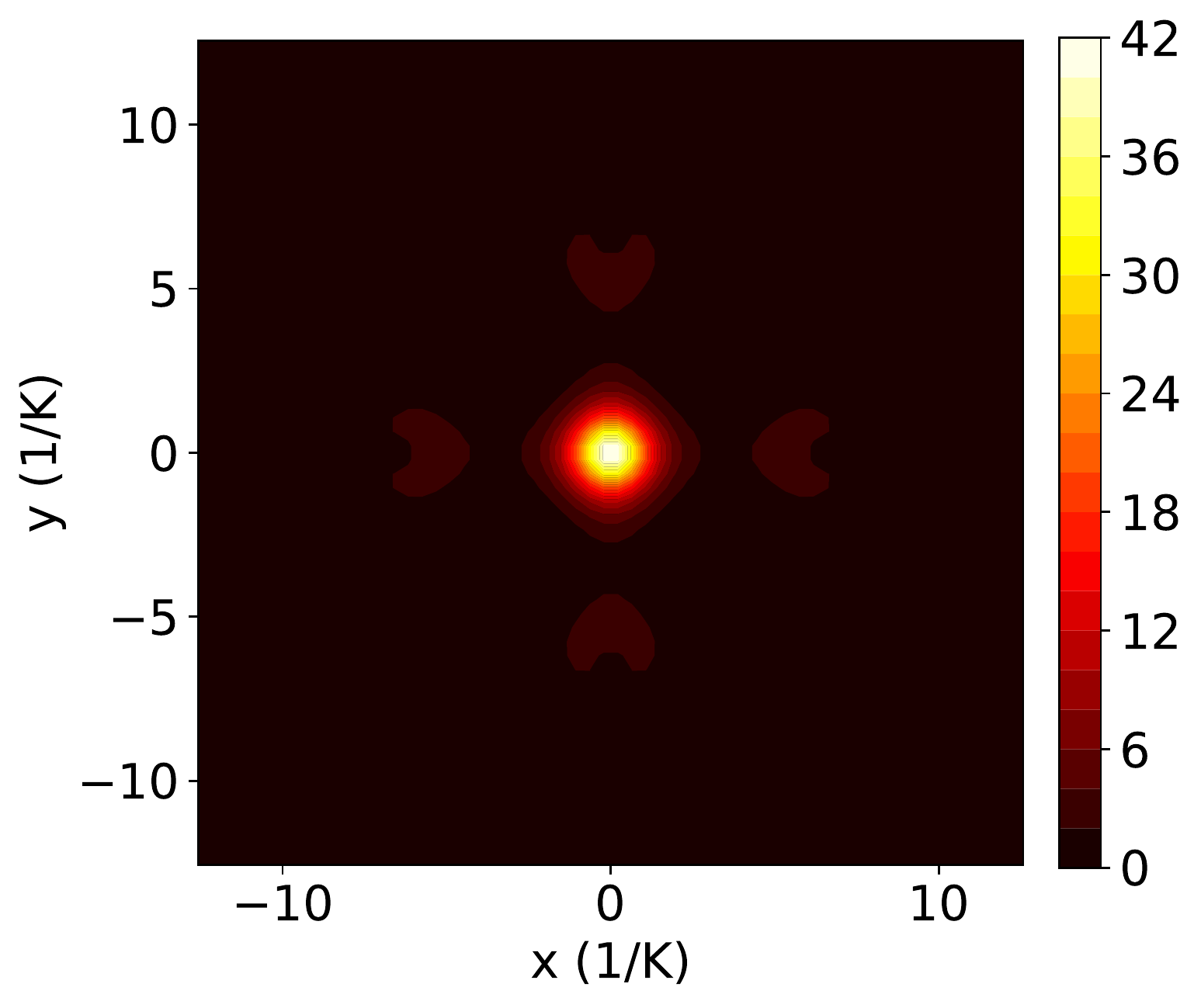}}
	\subfloat[]{\includegraphics[width=1.65 in]{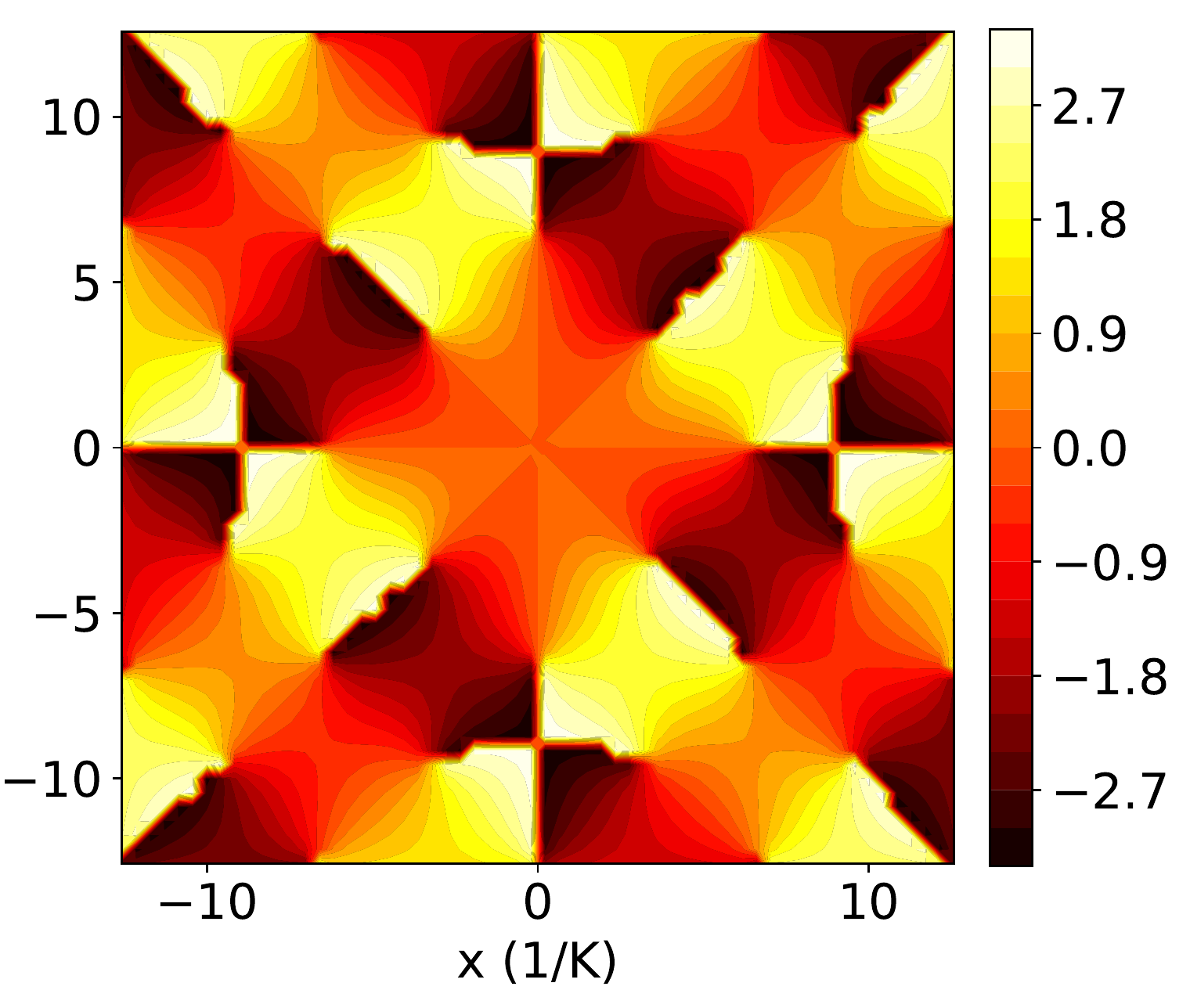}}
	\caption{\label{fig:MLWFDirac} Norm (a and c) and phase (b and d) of the MLWFs of the lowest band of a Dirac electron with spin up and down [Eq.~\eqref{eq:HD2}], respectively. $\phi = 1.5$. A plane wave cutoff of $K_c = 5K$ and a Brillouin zone discretization of $11\times 11$ were used.}
\end{figure}

We next turn to the Dirac case. The MLWFs of the lowest band, obtained at $\phi = 1.5$ for spin up and down in Eq.~\eqref{eq:HD2}, are shown in Fig.~\ref{fig:MLWFDirac}. The specific value of $\phi$ is chosen so that the lowest band is flat enough, but is not essential for the shapes of the MLWFs. MLWFs obtained when $\phi = 0.6$, i.e. same as that for the Schr\"{o}dinger case, also have the similar shapes. In stark contrast to the Schr\"{o}dinger case, the peaks are now located at $(\pm \pi/K, \pm \pi/K)$ (four equivalent points) and $(0,0)$, which are nothing but the minima of $\pm B(\mathbf{r})$ for spin up and down, respectively. This can also be understood from the behavior of $u_{n\mathbf{k}=0}$. Because at its minima the Zeeman potential is negative, it always dominates over the potential wells of $|\mathbf{A}|^2$ and thus defines the positions where $u_{n\mathbf{k}=0}$ should be localized at. Since the tight-binding Hamiltonians are one-dimensional now, all the hopping parameters are real and monotonically decrease as $\phi$ increases, since the wells of $\pm B(\mathbf{r})$ become monotonically deeper, which is the reason for the asymptotic band flattening. 

Before ending this section, we note that for both cases the lowest band is touching the next lowest one at Brillouin zone boundary. For the Schr\"{o}dinger case the band touching is at the X point or $(k_x,k_y)=(1/2,0)$ and its symmetry related points, while for the Dirac case [either $H^D$ or $(H^D)^2$], it is at the R point or $(1/2,1/2)$ and its symmetry related points. If such degeneracies are removed and the lowest band has a nonzero Chern number, which is possible because of the broken time-reversal symmetry in the present systems, exponentially localized Wannier functions for the lowest band cannot exist. We will discuss on the Chern number in more detail in Sec.~\ref{sec:chern}.

\section{Minimal tight-binding models for the flat band lattices}\label{sec:tb}

\begin{figure}[h]
\subfloat[]{\includegraphics[width=1.3in]{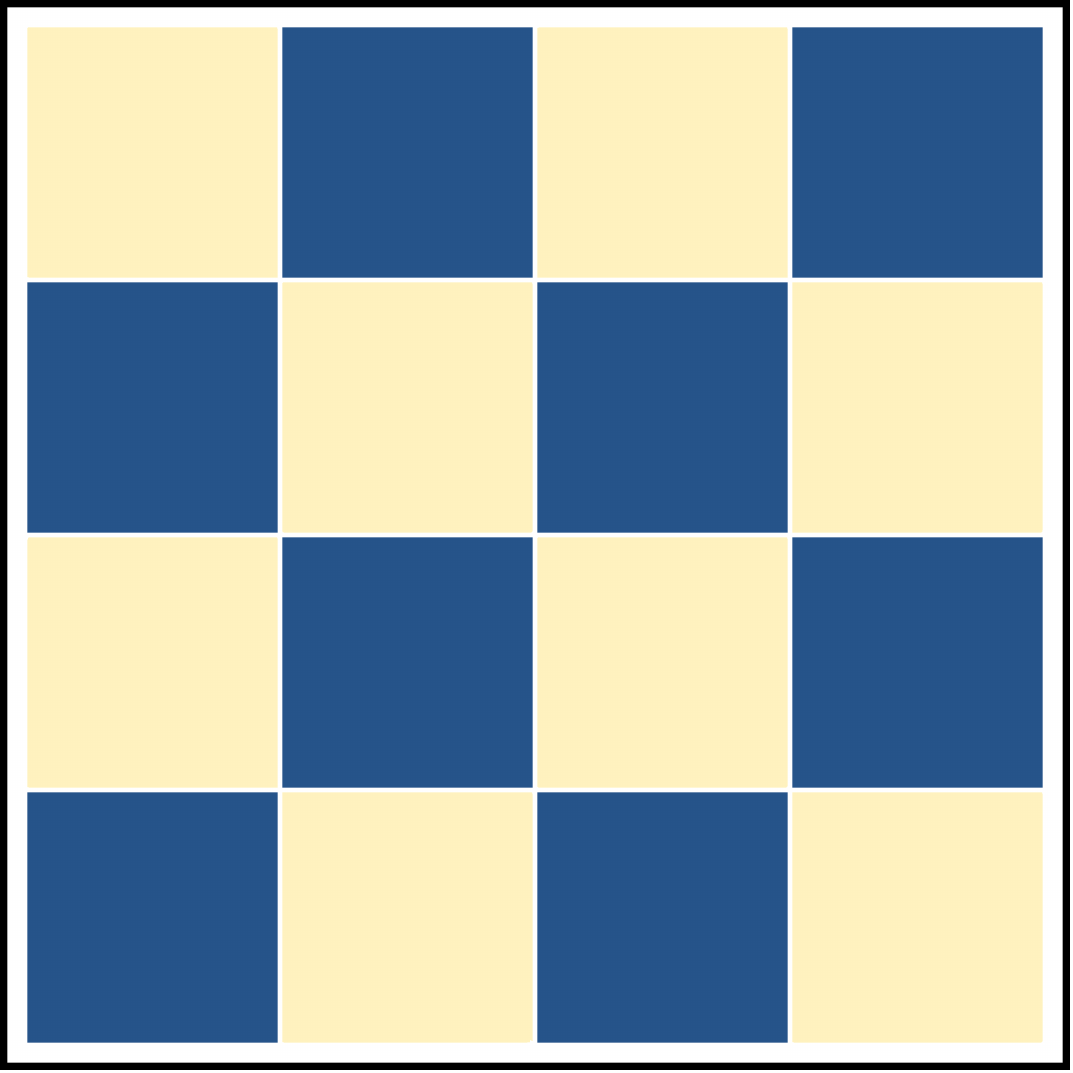}}\quad
\subfloat[]{\includegraphics[width=1.8in]{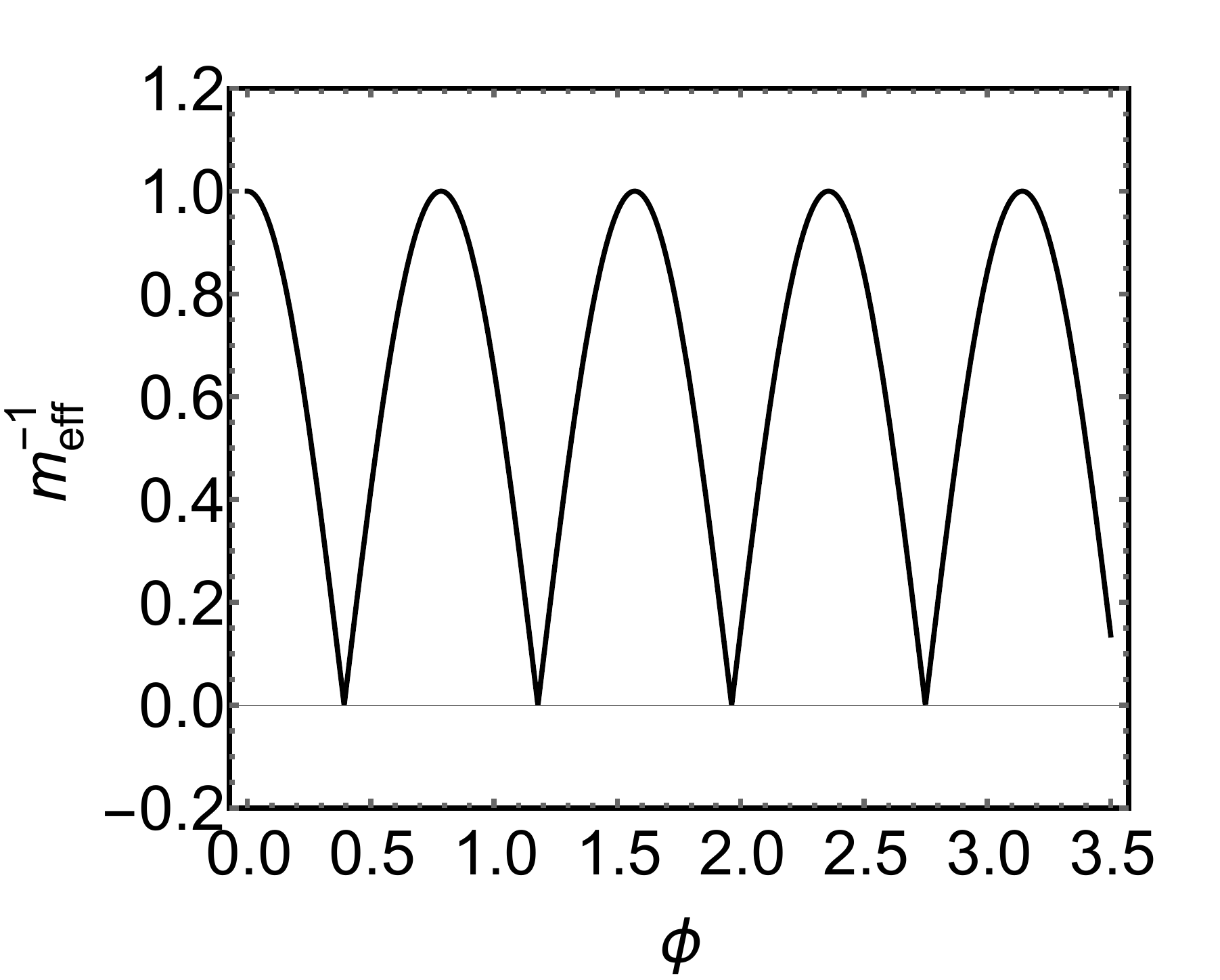}}
	\caption{(a) Tight-binding model on a square lattice with staggered magnetic fields for Schr\"{o}dinger electrons. The $xy$ axes are rotated by $\pi/4$ compared to that used for Eq.~\eqref{eq:Avec}. (b) Inverse effective mass versus $\phi$ based on Eq.~\eqref{eq:eigentb}.}
	\label{fig:tb}
\end{figure}

The Wannier functions given in the previous section motivate us to construct a minimal tight-binding model to explain the recurring magic values for the Schr\"{o}dinger case. Although usually the quantum effects of magnetic fields are treated in the Landau level basis, it is most convenient for slow-varying and strong magnetic fields. The Wannier function basis, which exploits the discrete translational symmetry, is more advantageous for the present problem of relatively weak and periodic magnetic fields. We thus consider the following spinless tight-binding model on a 2D square lattice with the lattice sites coinciding with the plaquette corners, i.e. positions of the Wannier functions in Fig.~\ref{fig:WFSchrproj}:
\begin{eqnarray}\label{eq:tbSchr}
H = -\sum_{\langle ij \rangle } t e^{i\varphi_{ij}} c_{i}^\dag c_j + 4 t,
\end{eqnarray}
where $t = \hbar^2/2m a^2$ is the hopping parameter between nearest neighbors, and the summation is over nearest neighbors. For convenience we have rotated the coordinate system by $\pi/4$ around the $z$ axis, compared to that used for Eq.~\eqref{eq:Avec}. The $4t$ is needed to shift the band bottom at zero magnetic field to zero energy. For the 2D-cosinusoidal magnetic field used above the absolute value of the flux through a plaquette is $\Phi = 16 B/K^2 =8B a^2/\pi^2$. All positive flux plaquettes only share edges with negative flux ones. The square lattice looks like a checkerboard, with two sites per unit cell, and the black and white squares correspond to positive and negative magnetic fluxes of the same size [Fig.~\ref{fig:tb} (a)]. Based on the spatial distribution of the phase of the Wannier functions in the previous section, we expect it to be qualitatively correct to include the magnetic field as a Peierls phase in the hopping parameter, which is the $e^{i\varphi_{ij}}$ in Eq.~\eqref{eq:tbSchr}. Integrating the vector potential in Eq.~\eqref{eq:Avec} along the bonds gives the phase $\varphi_{ij}$
\begin{eqnarray}
\varphi_{ij} = \pm \frac{4 e B }{\hbar K^2} = \pm \frac{\pi \Phi}{2 \Phi_0} = \pm 4\phi,
\end{eqnarray}
where positive sign means the plaquette on the left of the directional hopping path has positive flux, and $\Phi_0 = h/e$. The phase can also be obtained without choosing an explicit gauge, by considering symmetry and the value of the total flux through a plaquette \cite{haldane_1988}.

The Fourier-transformed Hamiltonian is written as a $2\times 2$ matrix
\begin{eqnarray}\label{eq:Htbk}
H (\mathbf{k}) = \left ( \begin{array}{cc}
4t & h_{\bf k} \\
h^*_{\bf k} & 4t 
\end{array}\right ),
\end{eqnarray}
where $h_{\bf k} = -2 t \left[ e^{4i\phi} \cos(k_x a) + e^{-4i\phi} \cos(k_y a) \right]$. The eigenvalues are 
\begin{eqnarray}
\epsilon_{\pm} (\mathbf{ k}) = 4t \pm |h_{\bf k}|.
\end{eqnarray}
For any given $\phi$ we can expand $\epsilon_{\pm}$ around small $k$, which gives 
\begin{eqnarray}\label{eq:eigentb}
\epsilon_{\pm} (\mathbf{k}) &\approx&  4 t \pm 2\sqrt{2+2\cos(8\phi)}t\\\nonumber
&& \mp \sqrt{\frac{1+\cos(8\phi)}{2}} (k_x^2 + k_y^2) a^2 t + O(k^3).
\end{eqnarray}
Thus when $\phi\rightarrow \pi/8$, the quadratic term approaches zero, i.e. the low-energy band for long wavelengths becomes flat. The magic value is therefore
\begin{eqnarray}\label{eq:mratioStb}
\frac{8\phi}{\pi} = \frac{\Phi}{\Phi_0} = 1,
\end{eqnarray}
or $\phi \approx 0.393$. At this value of $\phi$ the eigenenergies are
\begin{eqnarray}
\epsilon_{\pm} (\mathbf{ k}) = 4t \pm 2t |\cos(k_x a) - \cos(k_y a)|
\end{eqnarray}
where the 2nd term vanishes along $k_x = \pm k_y$. The density of states (at $\epsilon = 4t$) does not diverge at this exact point because of the linear band touching along $k_x = \pm k_y$. It will however diverge when $\phi$ is infinitely close to $\pi/2$. The band structure and DOS can be found in \cite{supp}. 

The magic value in Eq.~\eqref{eq:mratioStb} is smaller than the 1st one shown in Fig.~\ref{fig:fbSchr} (b). However, the tight-binding model above predicts a series of magic values
\begin{eqnarray}\label{eq:tbmagicphi}
\phi = \frac{(2n+1)\pi}{8},\,\, n\in \mathbb{Z},
\end{eqnarray}
with the periodicity $\Delta \phi = \pi/4 \approx 0.785$, which is close to the period of the oscillation in Fig.~\ref{fig:fbSchr} (b). We thus believe that the recurring magic values in the original problem of Schr\"{o}dinger electrons should be due to the same reason as the magicness in the minimal model. Moreover, the latter can help us make connections with many early examples of flat band lattice models ~\cite{3_sutherland_1986,3_lieb_1989,3_AME1,3_AME2,3_HTI1,3_HTI2}, where the origin of flat bands can be understood in terms of destructive interference. In the present case, the destructive interference comes from the values of $\phi$ in Eq.~\eqref{eq:tbmagicphi}, at which $t_{ij} = - t_{ji}$ for nearest neighbors $i$ and $j$. Specifically, for some local wavefunction having equal weights on two diagonal sites of a plaquette, which belong to the same sublattice, hopping to their common nearest neighbors will cancel out. This is the reason for the complete flatness of the bands along $k_x = \pm k_y$. At distances much larger than the lattice period, such cancellation leads to strong suppression of hopping along almost all directions, which is the reason for the vanishing inverse effective mass near $\mathbf{k}=0$. 

Plotting the inverse effective mass obtained from Eq.~\eqref{eq:eigentb} vs. $\phi$ gives Fig.~\ref{fig:tb} (b), which is similar to Fig.~\ref{fig:fbSchr} (b) in terms of the oscillation. It fails, however, to capture some fine features in the latter, e.g., the negative values of $m_{\rm eff}^{-1}$ near the magic values, the decreasing amplitudes of the oscillation with increasing $\phi$, etc., which is not surprising given the simplicity of the model. We do note that the decaying amplitude in Fig.~\ref{fig:fbSchr} (b) should be due to the general tendency of enhanced localization with increasing strength of the magnetic field. In the limit of strong magnetic field the eigenfunctions should be close to Landau orbits and all bands are expected to be very flat.

We finally comment on the Dirac case. Because it is sufficient to use a single Gaussian-like Wannier function to describe the lowest band (for a given spin), as shown in Sec.~\ref{sec:wannier}, the hopping parameters are real due to inversion symmetry. Thus a minimal model for it, more specifically for the squared Hamiltonian $(H^D)^2$, should be a nearest-neighbor hopping model on a square lattice with one site per unit cell. Such a trivial model obviously cannot describe the band flattening as it stands, unless one allows the hopping amplitude to depend on $\phi$ which is \emph{a posteriori}. Physically, the decreasing hopping with increasing $\phi$ should have two origins. The first is the Landau localization mentioned above. The second, which is unique to Dirac electrons, is the localization due to the Zeeman potential [last term in Eq.~\eqref{eq:HD2}], which has a Berry phase origin.

\section{Zeeman coupling and flat band Chern insulators}\label{sec:chern}

The Zeeman term in the squared Dirac Hamiltonian Eq.~\eqref{eq:HD2} motivates us to consider the Zeeman coupling between 2DEG and the periodic magnetic field, which always accompanies the orbital coupling. As mentioned in Sec.~\ref{sec:wannier} Dirac electrons in the present problem can be viewed as a special case of 2DEG plus Zeeman coupling with $gm/m_e = 2$. In common 2DEGs this ratio can vary significantly depending on materials realization \cite{Adachi_1982,Taskin_2011} and may even be tunable in a given system \cite{CBT,Giorgioni2016,Wang2014}. In this section we take the Zeeman coupling strength $gm/m_e$ as a variable and study how the flat band behaviors of Schr\"{o}dinger and Dirac electrons can be smoothly bridged by changing it between 0 and 2. More interestingly, we find that for $gm/m_e\in(0,2)$, not including the bounds, the band touching between the two lowest bands is removed, and the lowest band has a nonzero Chern number for each spin in almost all regions on the phase diagram plotted against $gm/m_e$ and $\phi$.

\begin{figure}[h]
	\subfloat[]{\includegraphics[width=1.7 in]{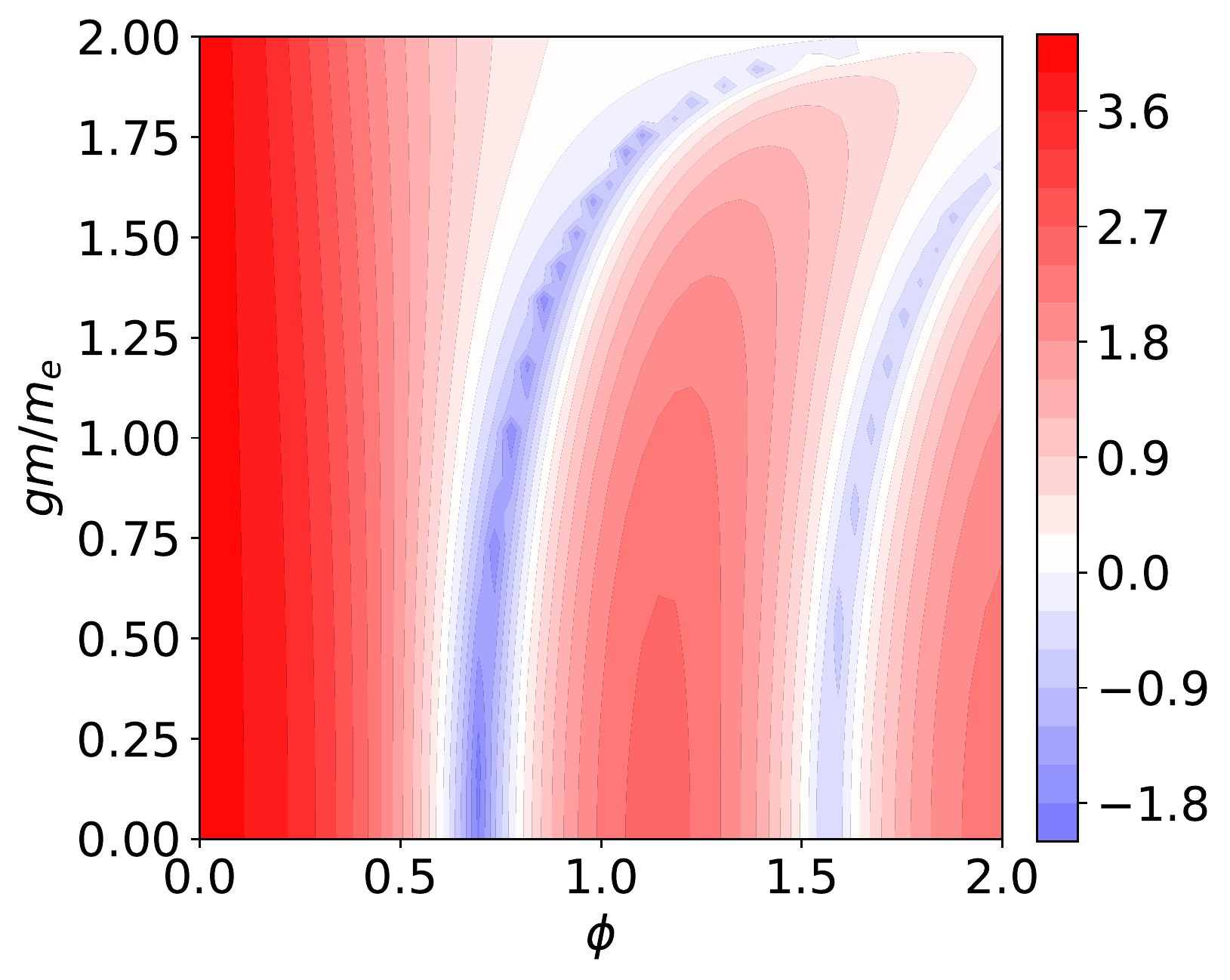}}
	\subfloat[]{\includegraphics[width=1.6 in]{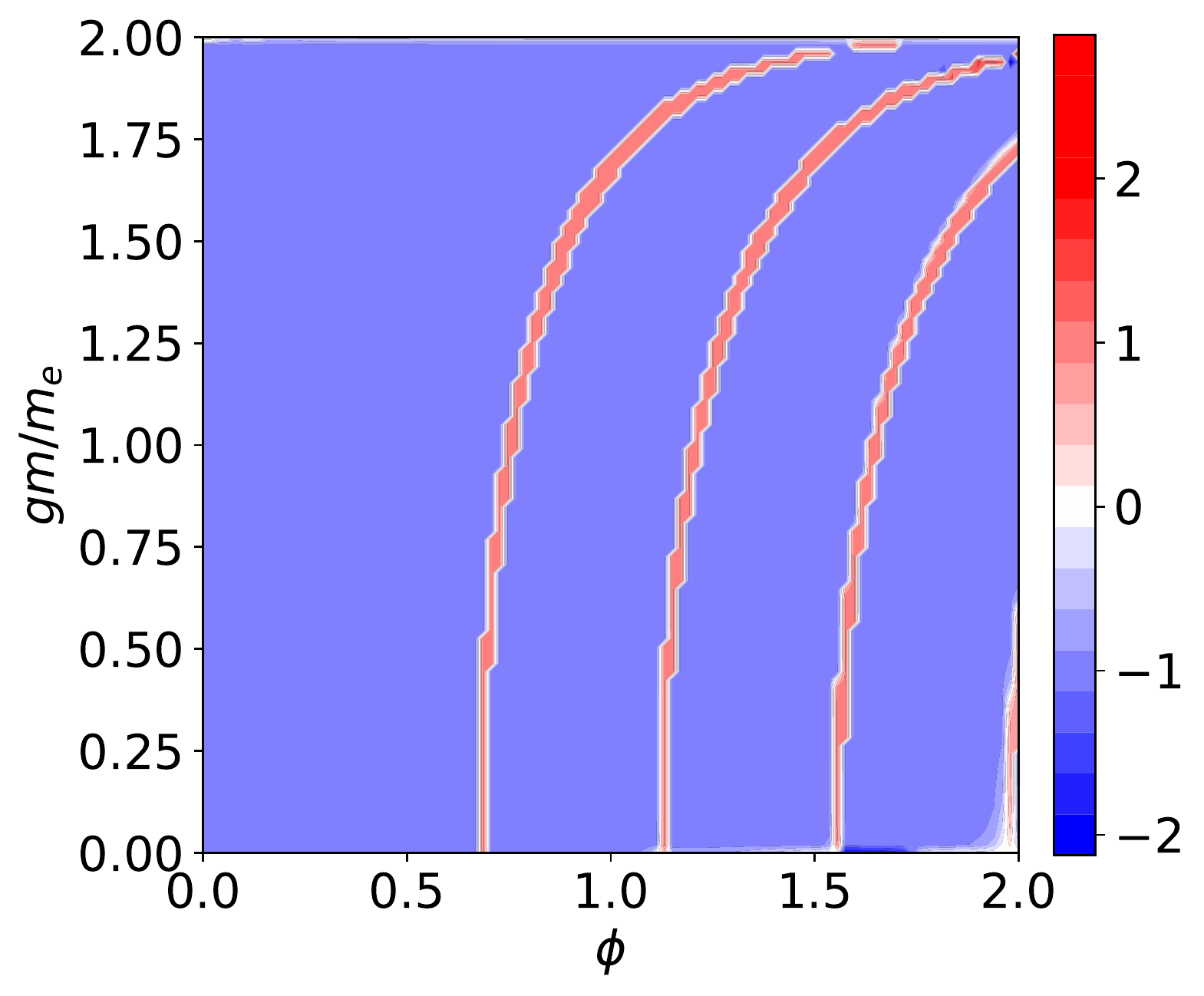}}
	\caption{(a) Inverse effective mass at $\mathbf{k} = 0$ and (b) Chern number of the up spin for the lowest band versus $gm/m_e$ and $\phi$. $K_c = 5K$. Brillouin zone discretization of $11\times 11$ and $10\times 10$ were used for calculating $m_{\rm eff}^{-1}$ and Chern number, respectively.}
	\label{fig:phasediag}
\end{figure}

Figure~\ref{fig:phasediag} (a) shows the phase diagram of the inverse effective mass $m_{\rm eff}^{-1}$ (in units of $m^{-1}$) at $\mathbf{k}=0$ versus $\phi$ and $gm/m_e$. One can see that along the horizontal line of $gm/m_e = 0$, i.e., pure Schr\"{o}dinger without Zeeman coupling, $m_{\rm eff}^{-1}$ oscillates between positive (red color) and negative (blue color) values, and reaches 0 (white color) at magic values of $\phi$. This is basically the same as Fig.~\ref{fig:fbSchr} (b). Similarly when $gm/m_e = 2$ the figure reproduces the monotonic decay of $m_{\rm eff}^{-1}$ for the Dirac case shown in Fig.~\ref{fig:fbDirac} (b). In between these two limits the regions with negative $m_{\rm eff}^{-1}$ form bands which start from being perpendicular to the $\phi$ axis when $gm/m_e = 0$, and gradually bend toward the horizontal $gm/m_e = 2$ line as $gm/m_e$ increases. Accordingly, the lines of magic values of $\phi$ and $gm/m_e$, defined by $m_{\rm eff}^{-1} = 0$, also bend to $gm/m_e = 2$ and disappear from the field of view. 

\begin{figure}[ht]
	\subfloat[]{\includegraphics[width=1.7 in]{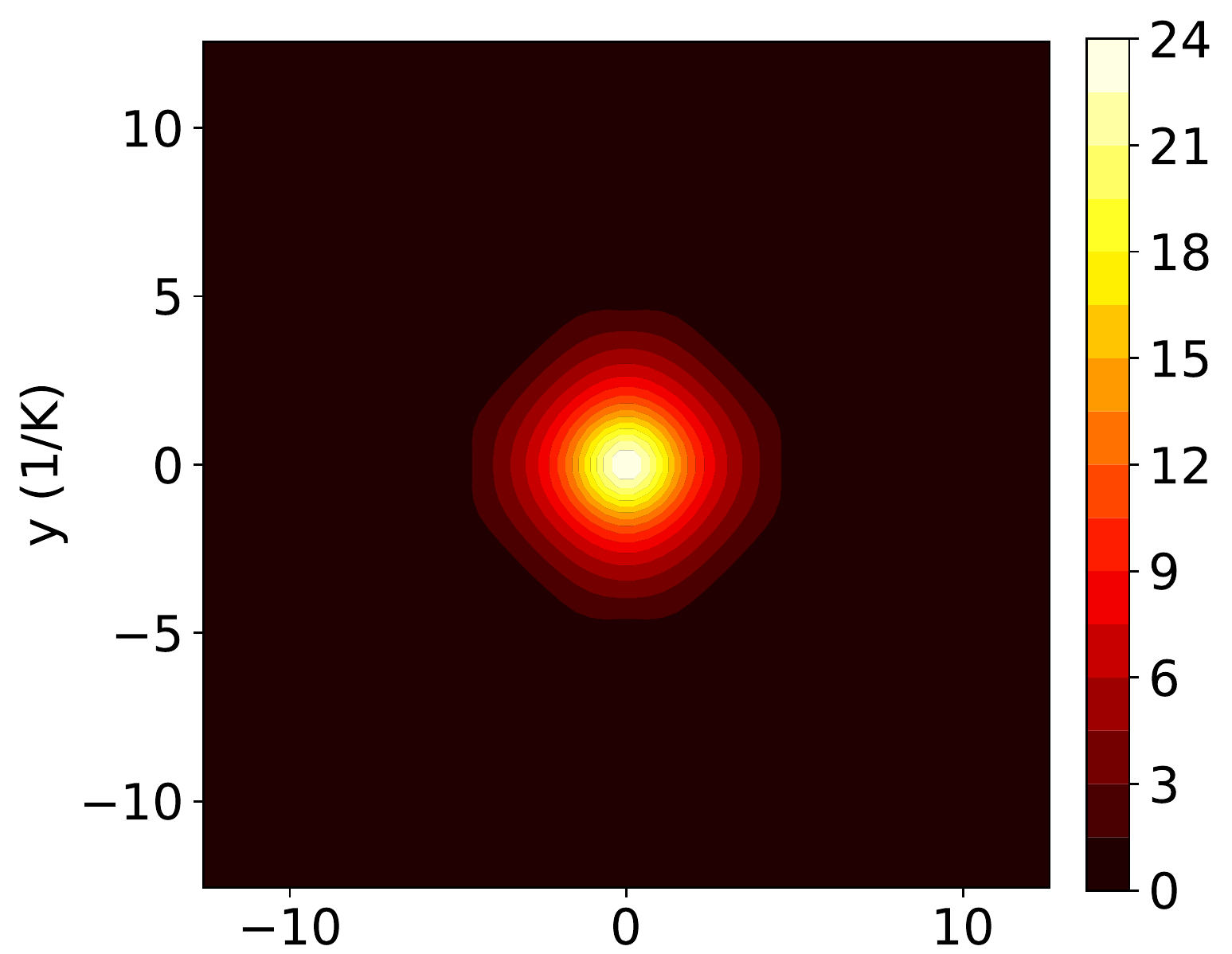}}
	\subfloat[]{\includegraphics[width=1.65 in]{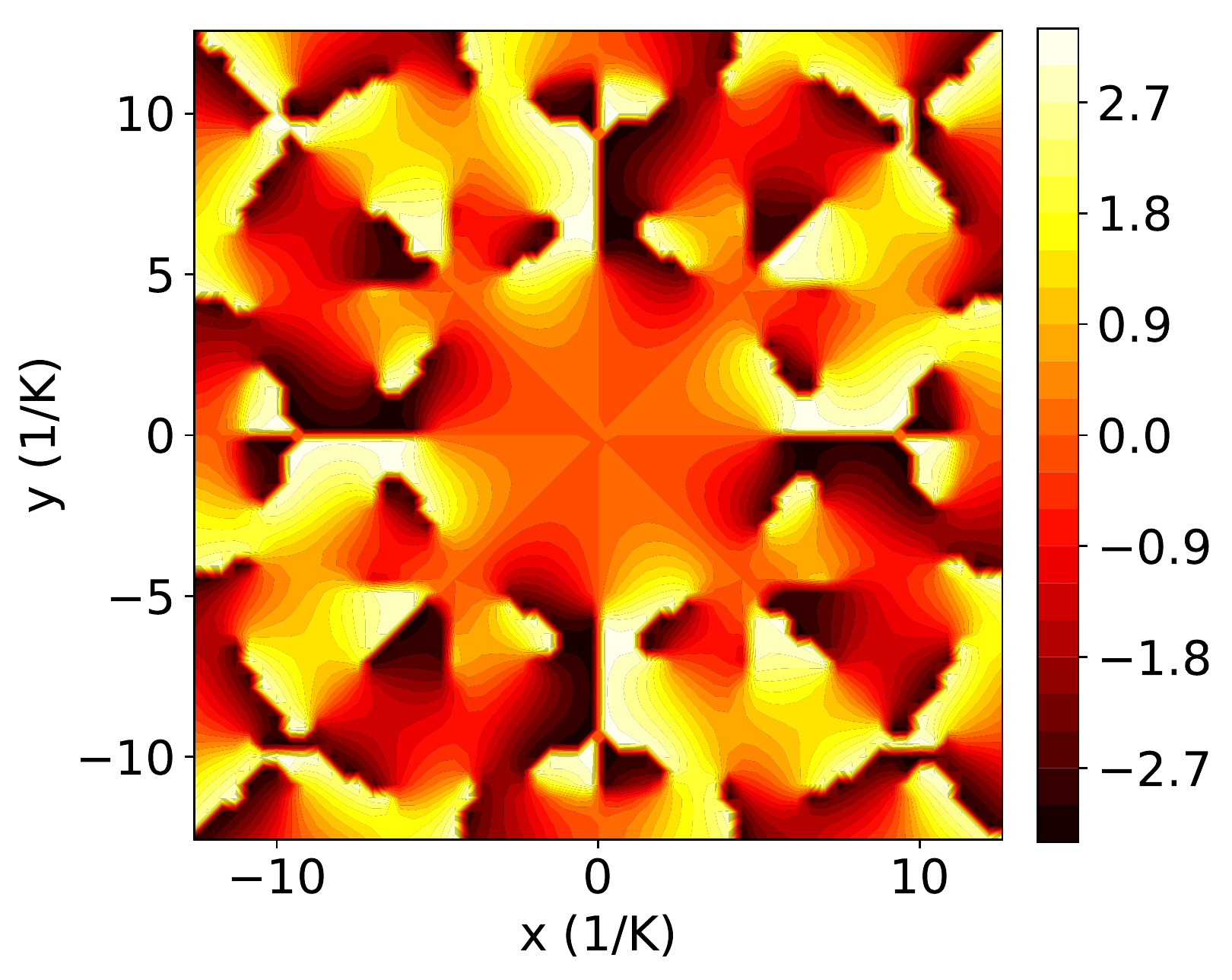}}\\
	\subfloat[]{\includegraphics[width=1.7 in]{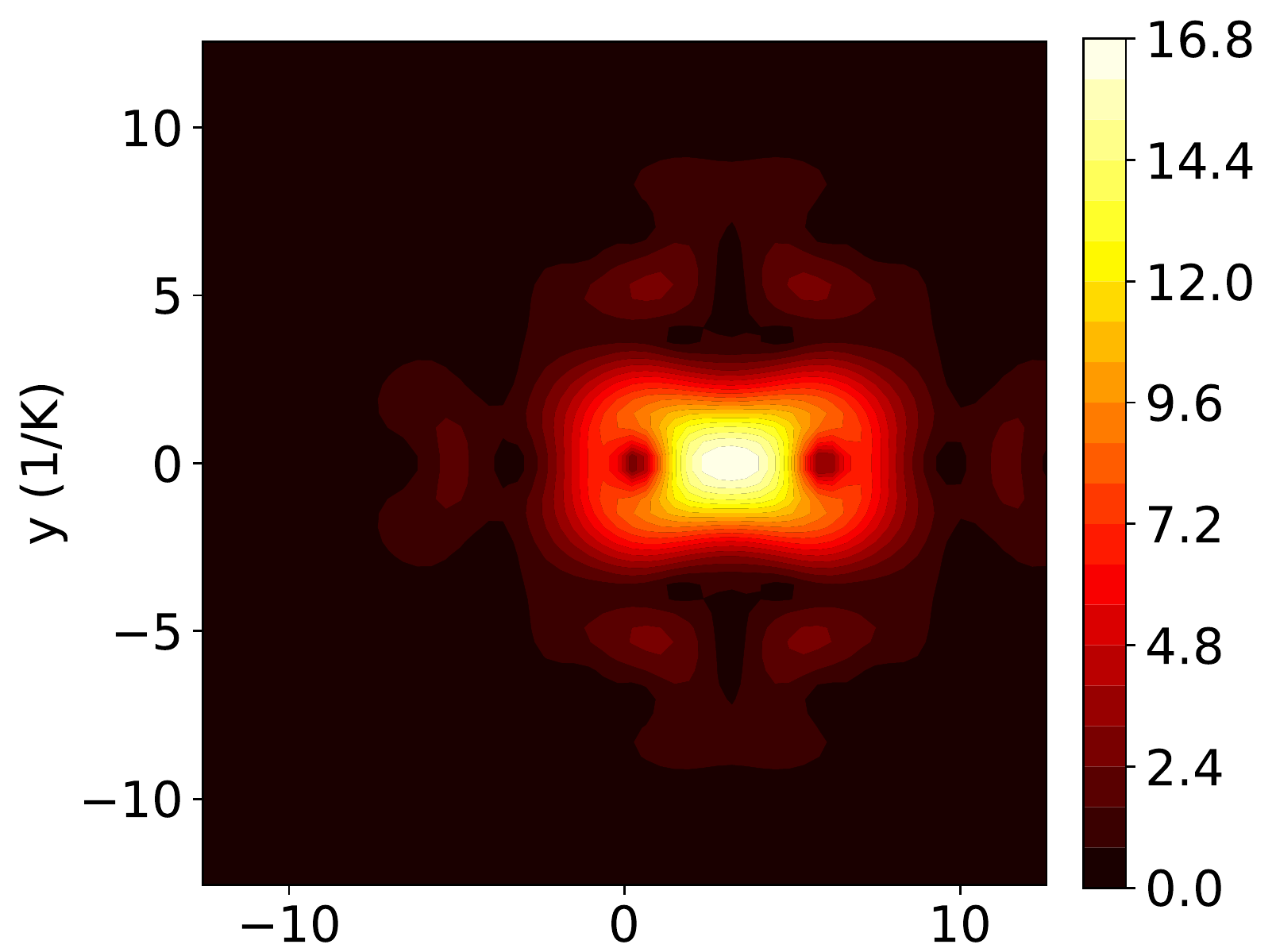}}
	\subfloat[]{\includegraphics[width=1.65 in]{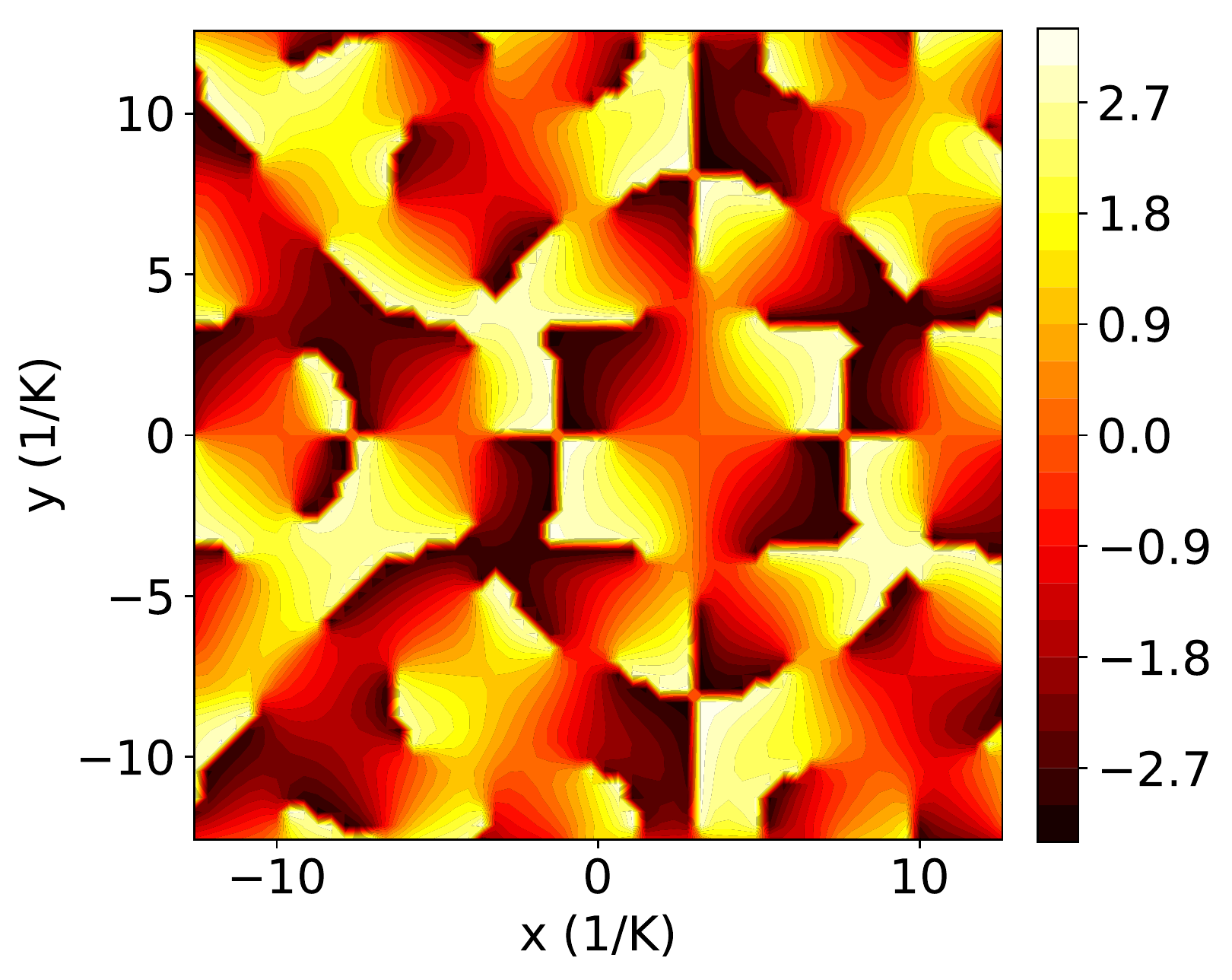}}\\
	\subfloat[]{\includegraphics[width=1.7 in]{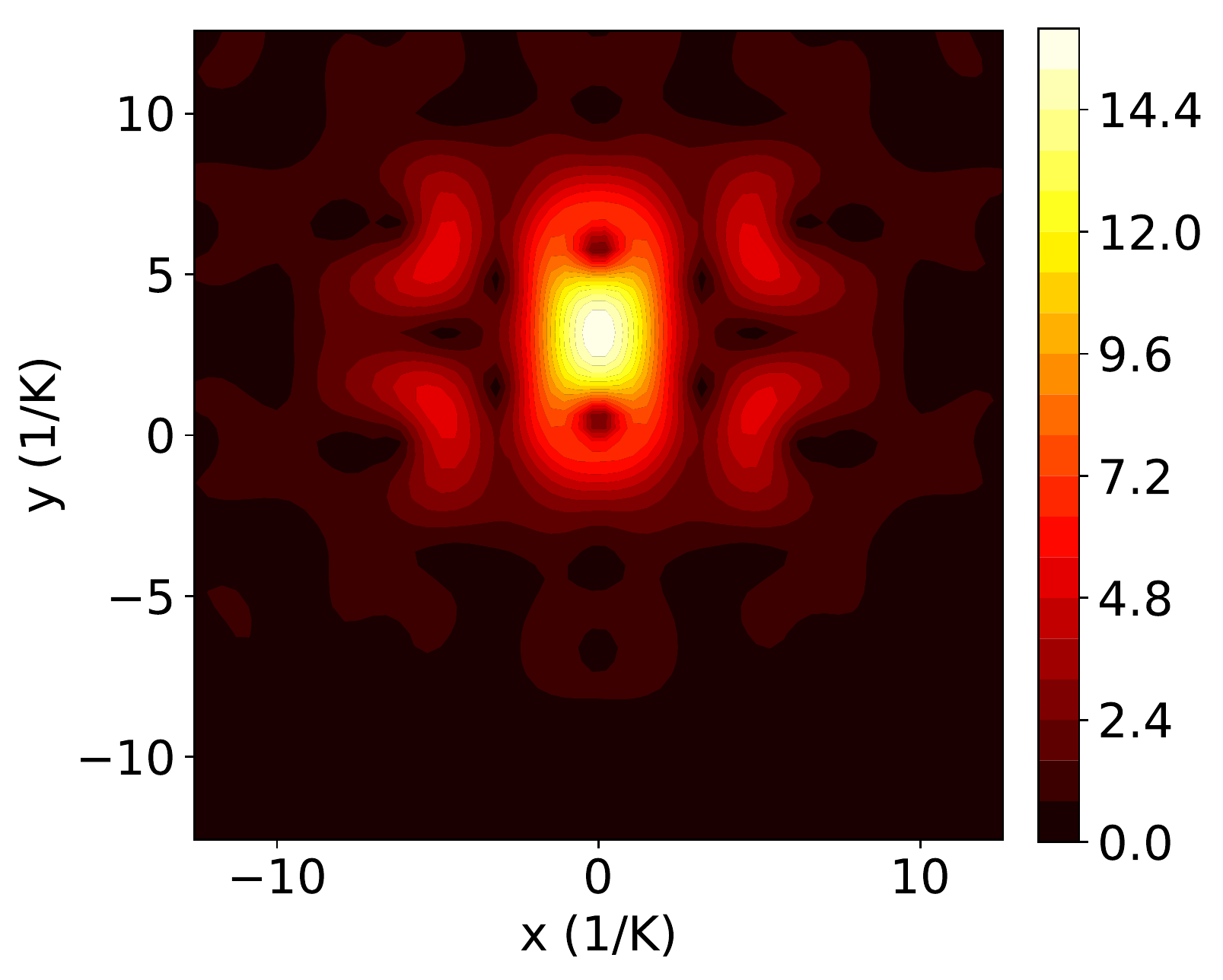}}
	\subfloat[]{\includegraphics[width=1.65 in]{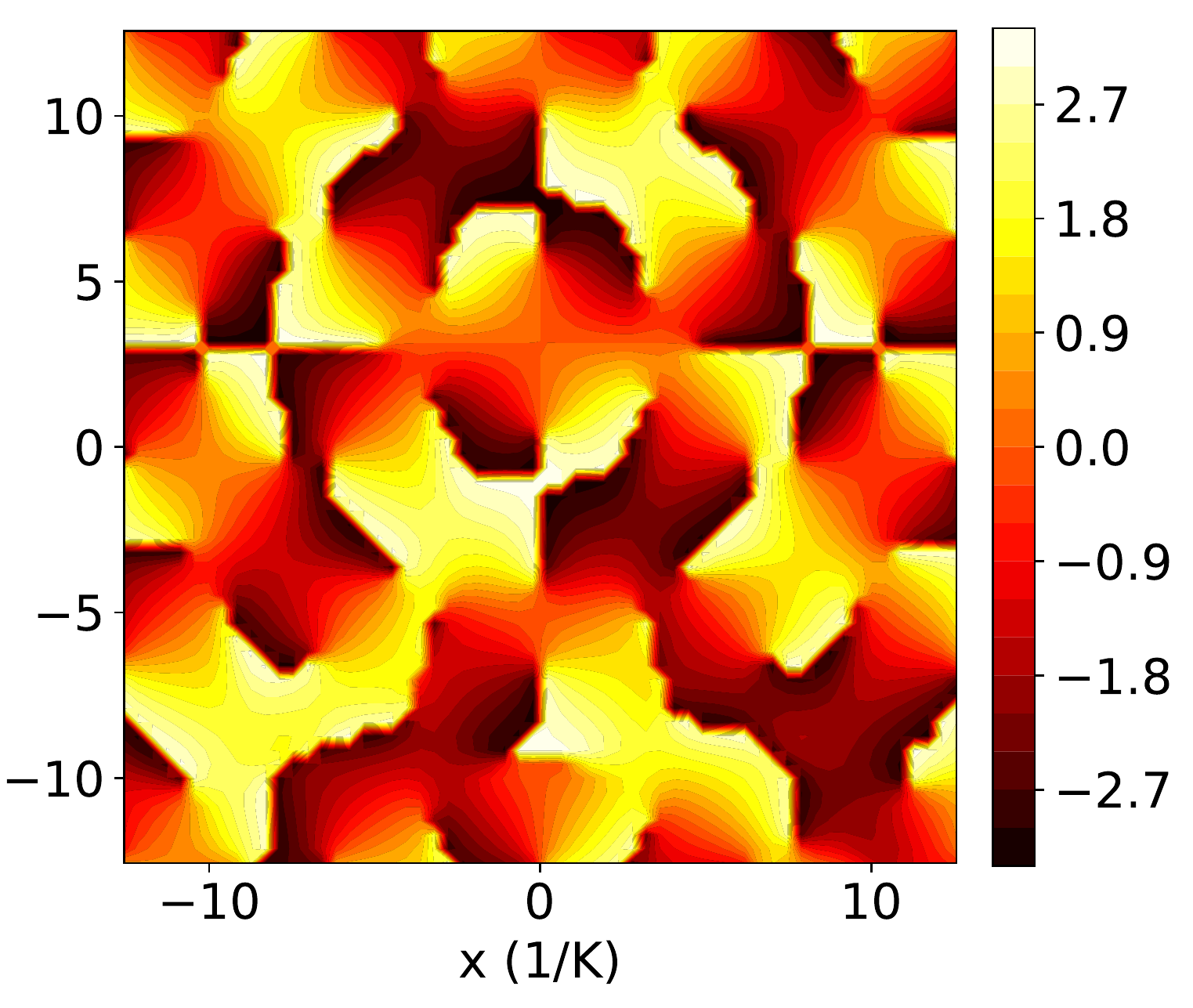}}\\
	\subfloat[]{\includegraphics[width=1.7 in]{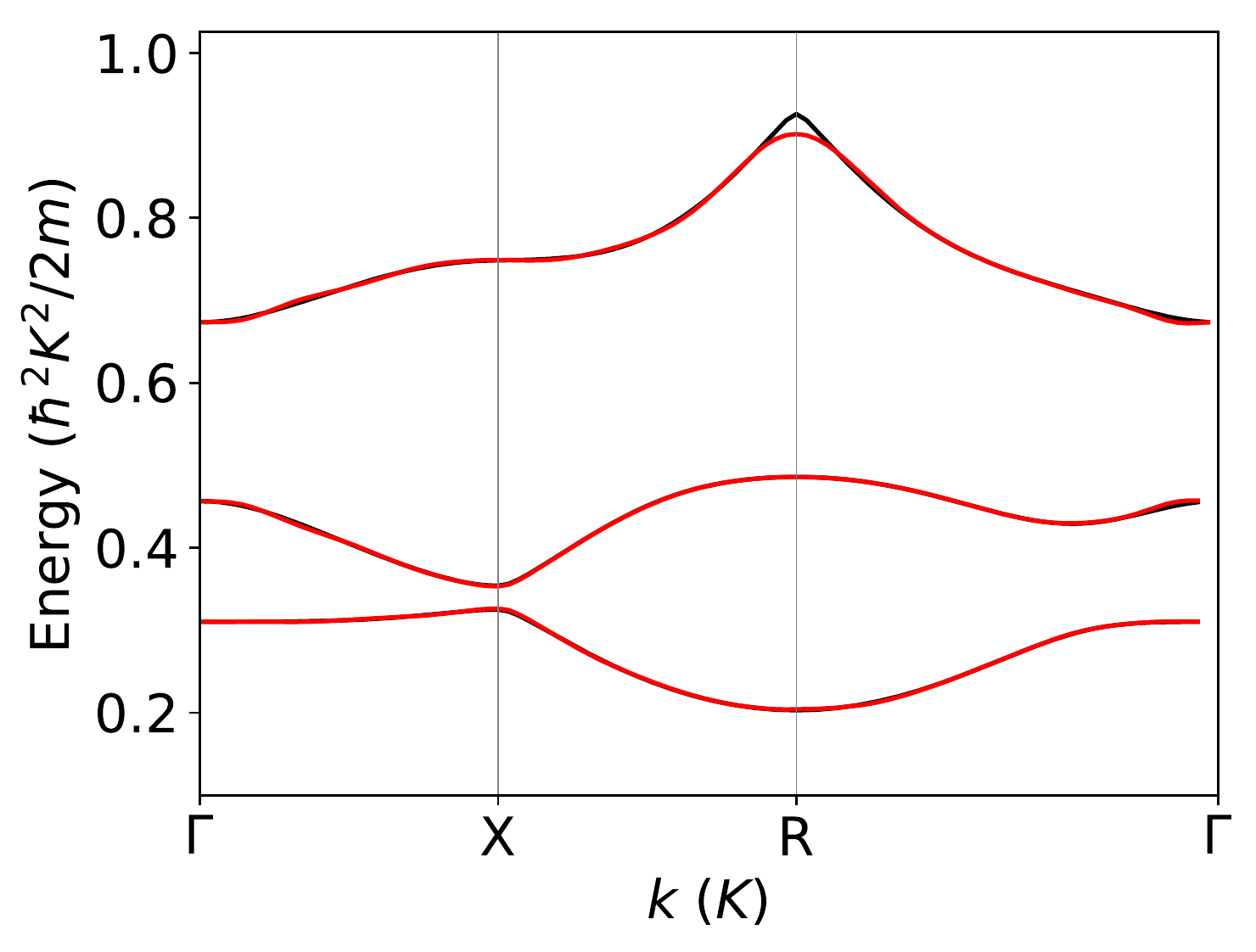}}
	\caption{\label{fig:WF3band} Norm (a, c, e) and phase (b, d, f) of the Wannier functions of the lowest 3 bands of a \emph{spin-down} Schr\"{o}dinger electron with $gm/m_e = 1.0$, $\phi = 0.67$. (g) Wannier-interpolated band structure (red solid lines) compared with the plane wave result (black solid lines). A plane wave cutoff of $K_c = 5K$ and a Brillouin zone discretization of $11\times 11$ are used. Width of the three Gaussians used for constructing the Wannier functions is set to $2/K$.}
\end{figure}

Since the two limiting cases of $gm/m_e = 0$ and $gm/m_e = 2$ can be respectively described by tight-binding models defined on different lattice sites, it is natural to ask if the cases with intermediate values of $gm/m_e$ can be described by a model with an enlarged basis. To see this we use the same method as explained in Sec.~\ref{sec:wannier} and project the lowest three bands, obtained at a set of magic values $\phi=0.67$ and $gm/m_e=1.0$, to three Gaussian functions located at A: $(0,0)$, B: $(\pi/K,0)$, and C: $(0,\pi/K)$ followed by orthonormalization. The resulting Wannier functions and the interpolated band structure are shown in Fig.~\ref{fig:WF3band}. Despite the different shape of the norm of the Wannier functions compared to Figs.~\ref{fig:WFSchrproj} and \ref{fig:MLWFDirac}, each of them still has a single peak at the expected location, and the phase distribution around the peak is qualitatively consistent with the Peierls phase. 

Looking into the real-space Hamiltonian in this Wannier basis, we find that although the hopping is still very short-ranged, the number of non-negligible hopping processes is larger than the 2-band model for pure Schr\"{o}dinger electrons. For example, the nearest neighbor hopping between same-sublattice sites is not small and has a nontrivial dependence on $\phi$. The on-site energies for the three sites are also different and depend on $\phi$. While it is possible to fine-tune the parameters of a minimal tight-binding model to fit the flat band behavior, it is not of our primary interest here since its predictive power is limited. Instead, we consider a 3-band tight-binding model with only the nearest-neighbor hoppings $t_{\rm BC}=t_1e^{\pm 4 i \phi}$, same as that in Sec.~\ref{sec:tb}, and $t_{\rm AB}=t_{\rm AC}=t_2$ which is real. While the $\mathbf{k}=0$ inverse effective mass of the lowest band of this model oscillates with $\phi$ in a similar manner as the 2-band model in Fig.~\ref{fig:tb} (b), a more interesting consequence of the extra orbital is that it removes the degeneracy of the two-band model Eq.~\eqref{eq:Htbk} at $(|k_x|,|k_y|) = (\pi/2a,\pi/2a)$, and the three bands do not touch one another in general. 

We find that the lowest band of the minimal 3-band model quite generally has a nonzero Chern number, making the model similar to the Haldane model of quantum anomalous Hall effect with zero net magnetic field \cite{haldane_1988}, but on the square lattice instead of the honeycomb lattice. Such an observation motivates us to calculate the Chern number of the lowest-band in a 2D parameter space spanned by $\phi$ and $gm/m_e$, and to see if in the original problem the flat bands can also be topologically nontrivial. The Chern number of the lowest band is calculated as
\begin{eqnarray}
\mathcal{C}_1 = \frac{1}{2\pi} \int_{\rm BZ} d^2\mathbf{k} \mathcal{F}^z_1,
\end{eqnarray}
where $\mathcal{F}^z_1 = (\nabla_{\mathbf{k}}\times \mathcal{A}_1)\cdot\hat{z}$ is the Berry curvature of the lowest band, and $\mathcal{A}_1 = i \langle u_{1\mathbf{k}} |\nabla_{\mathbf{k}}| u_{1\mathbf{k}} \rangle $ is the Berry connection of the lowest band. We make use of the algorithm proposed in \cite{fukui_2005} (with a different sign convention of the Chern number), which allows an accurate evaluation of the Chern number with a relatively coarse discretization of the Brillouin zone. 

The phase diagram of the Chern number, shown in Fig.~\ref{fig:phasediag} (b), is somewhat surprising since the Chern insulator phase is ubiquitous. Most regions have a $\mathcal{C}_1 = -1$ while on several narrow bands it is $+1$. These regions are separated by lines corresponding to band touching where the Chern number is ill-defined. Comparing Figs.~\ref{fig:phasediag} (a) and (b), one can see that the $\mathcal{C}_1 = 1$ regions coincide with places where $m_{\rm eff}^{-1} $ is extremal, indicating that there is band inversion near these values of $m_{\rm eff}^{-1} $. Most importantly, the regions with zero or vanishingly small $m_{\rm eff}^{-1} $ almost all have nonzero $\mathcal{C}_1$. Thus by tuning to the magic values of $\phi$ and $gm/m_e$ one could have flat bands and nontrivial topology simultaneously.

Above results have a caveat, however, due to spin degeneracy. The Schr\"{o}dinger Hamiltonian with the Zeeman term included has an emergent symmetry $T_{(\pi,\pi)}\otimes \mathcal{K}$, where $\mathcal{K}$ is complex conjugation and $T_{(\pi,\pi)}$ is a real space translation by $(\pi/K,\pi/K)$. Such a symmetry transforms the spin-up part of the Hamiltonian to the spin-down part and vice versa, and is the reason for the double degeneracy of the spinful bands. Since the Chern number changes sign under complex conjugation, the two spin species of a given band should always have opposite Chern numbers. This makes the net charge Chern number of a spinful band vanish, but not the spin Chern number, which is the difference between the Chern numbers of opposite spins. We note that the vanishing of the net Chern number of a spinful band is a consequence of the high symmetry of the present model, rather than a fundamental constraint. For example, adding a periodic scalar potential commensurate with the periodic magnetic field can have the same effect as the Zeeman potential for a single spin, and can thus make the net Chern number of the lowest band nonzero.

\section{Discussion and Conclusion}\label{sec:conclusion}

The magnetic field used in this work has a very simple form. In reality magnetic fields created by periodic arrays of bar magnets or superconducting wires will have more Fourier components, as well as finite in-plane magnetic fields. However, on the one hand the sinusoidal potential can be viewed as a legitimate first approximation if the spatial profile of the magnetic field is smooth. On the other hand, we expect the general low-energy behavior of Dirac electrons or 2DEG revealed in this work to qualitatively hold even with more realistic potential profiles. For example, Schr\"{o}dinger electrons will be likely to exhibit magicness since its low-energy Wannier orbitals should localize near zero-field lines, which will lead to complex hopping that periodically changes with field strength. 

The typical strength of fields needed to get flat bands should be such that the magnetic flux through each plaquette is on the order of $\Phi_0$. We emphasize that this is a rather modest requirement especially for large periods or small $K$. Since $\Phi_0\approx 4.136\times 10^{-3}$ T$\cdot\mu$m$^2$, a $\mu$m period field only needs to have an amplitude $\sim 10^2$ Gauss. In the case of graphene, such long wavelengths also mean the two valleys of graphene can be viewed as independent \cite{MY1,CHP}. Based on the lessons learned from the twisted multilayer graphene systems, for interaction-driven phases to appear the number of moir\'{e} unit cells in a given sample does not have to be macroscopically large--$10^2\times 10^2$ is sufficient. Artificial superlattices with such number of periods are not out of reach \cite{CBT,PDM,PDR,EAS,HAC,PDY}. Experimentally one can use either transport \cite{PDM, PDR, EAS, LAP, CRD, RKK} or spectroscopic \cite{MY1} methods to reveal the existence of the flat bands \cite{YCI, YCS, DPH, BGY} and in addition to look for exotic phases at very low temperatures. The complex hopping in the tight-binding models is reminiscent of the loop-current model for cuprates \cite{CMV1, CMV2}, thus suggesting potential new phases more proximate to high-temperature superconductors on a square lattice.

While our prescription works for the whole spectrum bridging Dirac materials and 2DEG, the former can take advantage of the various pseudo-magnetic fields through e.g. periodic strain or Zeeman field that may be easier to implement experimentally. Since the continuum description of graphene moir\'{e} also has the form of Dirac electrons subject to non-Abelian gauge potentials \cite{PJF,LJW,LGZ}, it is possible to use similar arguments to understand the origin of the moir\'{e} flat bands as well.

Although we have been focusing on periodic magnetic fields, band flattening as a general trend should be common for periodic potentials getting stronger and stronger. Even for Dirac electrons which are known to be difficult to confine with scalar potential wells, periodic scalar potentials can still lead to 1D flat bands \cite{MY1,CHP}. Finally, weak periodic electric potentials can be used together with a periodic magnetic field on 2DEG to get the ubiquitous Chern insulator phase as mentioned in Sec.~\ref{sec:chern}. 

In conclusion, we find that spatially periodic magnetic fields can be a practical and versatile approach to realizing emergent flat band lattices with different superlattice symmetries. The contrasting band-flattening behaviors of Dirac (no magicness) and Schr\"{o}dinger (with magicness) electrons can be understood through different minimal tight-binding models based on their respective Wannier functions localized by the periodic magnetic fields. In particular the magicness in the Schr\"{o}dinger case is due to a complex hopping amplitude along zero-field lines whose phase changes periodically with increasing field. The two limiting cases can be interpolated by considering the Zeeman coupling between the spin degrees of freedom of a 2DEG and the magnetic field and by varying the $g$-factor or the effective mass. The Zeeman coupling also quite generally leads to topologically nontrivial flat bands with nonzero Chern numbers for each spin. Future experimental and theoretical studies on this platform, which is a powerful alternative to moir\'{e} system, may reveal more exotic phases when interaction is taken into account.

\begin{acknowledgments}
MT and HC were supported by the start-up funding of CSU. OP is supported by NSF CAREER grant DMS-1452349. The authors are grateful to Allan MacDonald, Qian Niu, Di Xiao, Francois Peeters, and Pablo Jarillo-Herrero for helpful discussions. 
\end{acknowledgments}

\bibliography{refv2}

\begin{thebibliography}{99}%
\makeatletter
\providecommand \@ifxundefined [1]{%
 \@ifx{#1\undefined}
}%
\providecommand \@ifnum [1]{%
 \ifnum #1\expandafter \@firstoftwo
 \else \expandafter \@secondoftwo
 \fi
}%
\providecommand \@ifx [1]{%
 \ifx #1\expandafter \@firstoftwo
 \else \expandafter \@secondoftwo
 \fi
}%
\providecommand \natexlab [1]{#1}%
\providecommand \enquote  [1]{``#1''}%
\providecommand \bibnamefont  [1]{#1}%
\providecommand \bibfnamefont [1]{#1}%
\providecommand \citenamefont [1]{#1}%
\providecommand \href@noop [0]{\@secondoftwo}%
\providecommand \href [0]{\begingroup \@sanitize@url \@href}%
\providecommand \@href[1]{\@@startlink{#1}\@@href}%
\providecommand \@@href[1]{\endgroup#1\@@endlink}%
\providecommand \@sanitize@url [0]{\catcode `\\12\catcode `\$12\catcode
  `\&12\catcode `\#12\catcode `\^12\catcode `\_12\catcode `\%12\relax}%
\providecommand \@@startlink[1]{}%
\providecommand \@@endlink[0]{}%
\providecommand \url  [0]{\begingroup\@sanitize@url \@url }%
\providecommand \@url [1]{\endgroup\@href {#1}{\urlprefix }}%
\providecommand \urlprefix  [0]{URL }%
\providecommand \Eprint [0]{\href }%
\providecommand \doibase [0]{http://dx.doi.org/}%
\providecommand \selectlanguage [0]{\@gobble}%
\providecommand \bibinfo  [0]{\@secondoftwo}%
\providecommand \bibfield  [0]{\@secondoftwo}%
\providecommand \translation [1]{[#1]}%
\providecommand \BibitemOpen [0]{}%
\providecommand \bibitemStop [0]{}%
\providecommand \bibitemNoStop [0]{.\EOS\space}%
\providecommand \EOS [0]{\spacefactor3000\relax}%
\providecommand \BibitemShut  [1]{\csname bibitem#1\endcsname}%
\let\auto@bib@innerbib\@empty
\bibitem [{\citenamefont {Hass}\ \emph {et~al.}(2008)\citenamefont {Hass},
  \citenamefont {Varchon}, \citenamefont {Mill\'an-Otoya}, \citenamefont
  {Sprinkle}, \citenamefont {Sharma}, \citenamefont {de~Heer}, \citenamefont
  {Berger}, \citenamefont {First}, \citenamefont {Magaud},\ and\ \citenamefont
  {Conrad}}]{1_hass_2008}%
  \BibitemOpen
  \bibfield  {author} {\bibinfo {author} {\bibfnamefont {J.}~\bibnamefont
  {Hass}}, \bibinfo {author} {\bibfnamefont {F.}~\bibnamefont {Varchon}},
  \bibinfo {author} {\bibfnamefont {J.~E.}\ \bibnamefont {Mill\'an-Otoya}},
  \bibinfo {author} {\bibfnamefont {M.}~\bibnamefont {Sprinkle}}, \bibinfo
  {author} {\bibfnamefont {N.}~\bibnamefont {Sharma}}, \bibinfo {author}
  {\bibfnamefont {W.~A.}\ \bibnamefont {de~Heer}}, \bibinfo {author}
  {\bibfnamefont {C.}~\bibnamefont {Berger}}, \bibinfo {author} {\bibfnamefont
  {P.~N.}\ \bibnamefont {First}}, \bibinfo {author} {\bibfnamefont
  {L.}~\bibnamefont {Magaud}}, \ and\ \bibinfo {author} {\bibfnamefont {E.~H.}\
  \bibnamefont {Conrad}},\ }\href {\doibase 10.1103/PhysRevLett.100.125504}
  {\bibfield  {journal} {\bibinfo  {journal} {Phys. Rev. Lett.}\ }\textbf
  {\bibinfo {volume} {100}},\ \bibinfo {pages} {125504} (\bibinfo {year}
  {2008})}\BibitemShut {NoStop}%
\bibitem [{\citenamefont {Dean}\ \emph {et~al.}(2010)\citenamefont {Dean},
  \citenamefont {Young}, \citenamefont {Meric}, \citenamefont {Lee},
  \citenamefont {Wang}, \citenamefont {Sorgenfrei}, \citenamefont {Watanabe},
  \citenamefont {Taniguchi}, \citenamefont {Kim}, \citenamefont {Shepard},\
  and\ \citenamefont {Hone}}]{1_dean_2010}%
  \BibitemOpen
  \bibfield  {author} {\bibinfo {author} {\bibfnamefont {C.~R.}\ \bibnamefont
  {Dean}}, \bibinfo {author} {\bibfnamefont {A.~F.}\ \bibnamefont {Young}},
  \bibinfo {author} {\bibfnamefont {I.}~\bibnamefont {Meric}}, \bibinfo
  {author} {\bibfnamefont {C.}~\bibnamefont {Lee}}, \bibinfo {author}
  {\bibfnamefont {L.}~\bibnamefont {Wang}}, \bibinfo {author} {\bibfnamefont
  {S.}~\bibnamefont {Sorgenfrei}}, \bibinfo {author} {\bibfnamefont
  {K.}~\bibnamefont {Watanabe}}, \bibinfo {author} {\bibfnamefont
  {T.}~\bibnamefont {Taniguchi}}, \bibinfo {author} {\bibfnamefont
  {P.}~\bibnamefont {Kim}}, \bibinfo {author} {\bibfnamefont {K.~L.~J.}\
  \bibnamefont {Shepard}}, \ and\ \bibinfo {author} {\bibfnamefont
  {J.}~\bibnamefont {Hone}},\ }\href
  {https://www.nature.com/articles/nnano.2010.172} {\bibfield  {journal}
  {\bibinfo  {journal} {Nature Nanotechnology}\ }\textbf {\bibinfo {volume}
  {5}},\ \bibinfo {pages} {722} (\bibinfo {year} {2010})}\BibitemShut {NoStop}%
\bibitem [{\citenamefont {Xue}\ \emph {et~al.}(2011)\citenamefont {Xue},
  \citenamefont {Sanchez-Yamagishi}, \citenamefont {Bulmash}, \citenamefont
  {Jacquod}, \citenamefont {Deshpande}, \citenamefont {Watanabe}, \citenamefont
  {Taniguchi}, \citenamefont {Jarillo-Herrero},\ and\ \citenamefont
  {J.~LeRoy}}]{1_xue_2011}%
  \BibitemOpen
  \bibfield  {author} {\bibinfo {author} {\bibfnamefont {J.}~\bibnamefont
  {Xue}}, \bibinfo {author} {\bibfnamefont {J.}~\bibnamefont
  {Sanchez-Yamagishi}}, \bibinfo {author} {\bibfnamefont {D.}~\bibnamefont
  {Bulmash}}, \bibinfo {author} {\bibfnamefont {P.}~\bibnamefont {Jacquod}},
  \bibinfo {author} {\bibfnamefont {A.}~\bibnamefont {Deshpande}}, \bibinfo
  {author} {\bibfnamefont {K.}~\bibnamefont {Watanabe}}, \bibinfo {author}
  {\bibfnamefont {T.}~\bibnamefont {Taniguchi}}, \bibinfo {author}
  {\bibfnamefont {P.}~\bibnamefont {Jarillo-Herrero}}, \ and\ \bibinfo {author}
  {\bibfnamefont {B.}~\bibnamefont {J.~LeRoy}},\ }\href
  {https://www.nature.com/articles/nmat2968} {\bibfield  {journal} {\bibinfo
  {journal} {Nature Materials}\ }\textbf {\bibinfo {volume} {10}},\ \bibinfo
  {pages} {282} (\bibinfo {year} {2011})}\BibitemShut {NoStop}%
\bibitem [{\citenamefont {Wang}\ \emph {et~al.}(2016)\citenamefont {Wang},
  \citenamefont {Lu}, \citenamefont {Ding}, \citenamefont {Yao}, \citenamefont
  {Yan}, \citenamefont {Wan}, \citenamefont {Deng}, \citenamefont {Wang},
  \citenamefont {Chen}, \citenamefont {Ma}, \citenamefont {Jung}, \citenamefont
  {Fedorov}, \citenamefont {Zhang}, \citenamefont {Zhang},\ and\ \citenamefont
  {Zhou}}]{1_EWANG_2016}%
  \BibitemOpen
  \bibfield  {author} {\bibinfo {author} {\bibfnamefont {E.}~\bibnamefont
  {Wang}}, \bibinfo {author} {\bibfnamefont {X.}~\bibnamefont {Lu}}, \bibinfo
  {author} {\bibfnamefont {S.}~\bibnamefont {Ding}}, \bibinfo {author}
  {\bibfnamefont {W.}~\bibnamefont {Yao}}, \bibinfo {author} {\bibfnamefont
  {M.}~\bibnamefont {Yan}}, \bibinfo {author} {\bibfnamefont {G.}~\bibnamefont
  {Wan}}, \bibinfo {author} {\bibfnamefont {K.}~\bibnamefont {Deng}}, \bibinfo
  {author} {\bibfnamefont {S.}~\bibnamefont {Wang}}, \bibinfo {author}
  {\bibfnamefont {G.}~\bibnamefont {Chen}}, \bibinfo {author} {\bibfnamefont
  {L.}~\bibnamefont {Ma}}, \bibinfo {author} {\bibfnamefont {J.}~\bibnamefont
  {Jung}}, \bibinfo {author} {\bibfnamefont {A.~V.}\ \bibnamefont {Fedorov}},
  \bibinfo {author} {\bibfnamefont {Y.}~\bibnamefont {Zhang}}, \bibinfo
  {author} {\bibfnamefont {G.}~\bibnamefont {Zhang}}, \ and\ \bibinfo {author}
  {\bibfnamefont {S.}~\bibnamefont {Zhou}},\ }\href
  {https://www.nature.com/articles/nphys3856} {\bibfield  {journal} {\bibinfo
  {journal} {Nature Physics}\ }\textbf {\bibinfo {volume} {12}},\ \bibinfo
  {pages} {1111} (\bibinfo {year} {2016})}\BibitemShut {NoStop}%
\bibitem [{\citenamefont {Zhang}\ \emph {et~al.}(2017)\citenamefont {Zhang},
  \citenamefont {Chuu}, \citenamefont {Ren}, \citenamefont {Li}, \citenamefont
  {Li}, \citenamefont {Jin}, \citenamefont {Chou},\ and\ \citenamefont
  {Shih}}]{1_CZHANG_2017}%
  \BibitemOpen
  \bibfield  {author} {\bibinfo {author} {\bibfnamefont {C.}~\bibnamefont
  {Zhang}}, \bibinfo {author} {\bibfnamefont {C.-P.}\ \bibnamefont {Chuu}},
  \bibinfo {author} {\bibfnamefont {X.}~\bibnamefont {Ren}}, \bibinfo {author}
  {\bibfnamefont {M.-Y.}\ \bibnamefont {Li}}, \bibinfo {author} {\bibfnamefont
  {L.-J.}\ \bibnamefont {Li}}, \bibinfo {author} {\bibfnamefont
  {C.}~\bibnamefont {Jin}}, \bibinfo {author} {\bibfnamefont {M.-Y.}\
  \bibnamefont {Chou}}, \ and\ \bibinfo {author} {\bibfnamefont {C.-K.}\
  \bibnamefont {Shih}},\ }\href {\doibase 10.1126/sciadv.1601459} {\bibfield
  {journal} {\bibinfo  {journal} {Science Advances}\ }\textbf {\bibinfo
  {volume} {3}},\ \bibinfo {pages} {e1601459} (\bibinfo {year}
  {2017})}\BibitemShut {NoStop}%
\bibitem [{\citenamefont {Lopes~dos Santos}\ \emph {et~al.}(2007)\citenamefont
  {Lopes~dos Santos}, \citenamefont {Peres},\ and\ \citenamefont
  {Castro~Neto}}]{santos_2007}%
  \BibitemOpen
  \bibfield  {author} {\bibinfo {author} {\bibfnamefont {J.~M.~B.}\
  \bibnamefont {Lopes~dos Santos}}, \bibinfo {author} {\bibfnamefont
  {N.~M.~R.}\ \bibnamefont {Peres}}, \ and\ \bibinfo {author} {\bibfnamefont
  {A.~H.}\ \bibnamefont {Castro~Neto}},\ }\href {\doibase
  10.1103/PhysRevLett.99.256802} {\bibfield  {journal} {\bibinfo  {journal}
  {Phys. Rev. Lett.}\ }\textbf {\bibinfo {volume} {99}},\ \bibinfo {pages}
  {256802} (\bibinfo {year} {2007})}\BibitemShut {NoStop}%
\bibitem [{\citenamefont {Mele}(2010)}]{mele_2010}%
  \BibitemOpen
  \bibfield  {author} {\bibinfo {author} {\bibfnamefont {E.~J.}\ \bibnamefont
  {Mele}},\ }\href {\doibase 10.1103/PhysRevB.81.161405} {\bibfield  {journal}
  {\bibinfo  {journal} {Phys. Rev. B}\ }\textbf {\bibinfo {volume} {81}},\
  \bibinfo {pages} {161405} (\bibinfo {year} {2010})}\BibitemShut {NoStop}%
\bibitem [{\citenamefont {Bistritzer}\ and\ \citenamefont
  {MacDonald}(2011)}]{MD}%
  \BibitemOpen
  \bibfield  {author} {\bibinfo {author} {\bibfnamefont {R.}~\bibnamefont
  {Bistritzer}}\ and\ \bibinfo {author} {\bibfnamefont {A.~H.}\ \bibnamefont
  {MacDonald}},\ }\href {\doibase 10.1073/pnas.1108174108} {\bibfield
  {journal} {\bibinfo  {journal} {PNAS}\ }\textbf {\bibinfo {volume} {108}},\
  \bibinfo {pages} {12233} (\bibinfo {year} {2011})}\BibitemShut {NoStop}%
\bibitem [{\citenamefont {Cao}\ \emph {et~al.}(2018{\natexlab{a}})\citenamefont
  {Cao}, \citenamefont {Fatemi}, \citenamefont {Demir}, \citenamefont {Fang},
  \citenamefont {Tomarken}, \citenamefont {Luo}, \citenamefont
  {Sanchez-Yamagishi}, \citenamefont {Watanabe}, \citenamefont {Taniguchi},
  \citenamefont {Kaxiras}, \citenamefont {Ashoori},\ and\ \citenamefont
  {Jarillo-Herrero}}]{YCI}%
  \BibitemOpen
  \bibfield  {author} {\bibinfo {author} {\bibfnamefont {Y.}~\bibnamefont
  {Cao}}, \bibinfo {author} {\bibfnamefont {V.}~\bibnamefont {Fatemi}},
  \bibinfo {author} {\bibfnamefont {A.}~\bibnamefont {Demir}}, \bibinfo
  {author} {\bibfnamefont {S.}~\bibnamefont {Fang}}, \bibinfo {author}
  {\bibfnamefont {S.~L.}\ \bibnamefont {Tomarken}}, \bibinfo {author}
  {\bibfnamefont {J.~Y.}\ \bibnamefont {Luo}}, \bibinfo {author} {\bibfnamefont
  {J.~D.}\ \bibnamefont {Sanchez-Yamagishi}}, \bibinfo {author} {\bibfnamefont
  {K.}~\bibnamefont {Watanabe}}, \bibinfo {author} {\bibfnamefont
  {T.}~\bibnamefont {Taniguchi}}, \bibinfo {author} {\bibfnamefont
  {E.}~\bibnamefont {Kaxiras}}, \bibinfo {author} {\bibfnamefont {R.~C.}\
  \bibnamefont {Ashoori}}, \ and\ \bibinfo {author} {\bibfnamefont
  {P.}~\bibnamefont {Jarillo-Herrero}},\ }\href
  {https://www.nature.com/articles/nature26154} {\bibfield  {journal} {\bibinfo
   {journal} {Nature}\ }\textbf {\bibinfo {volume} {556}},\ \bibinfo {pages}
  {80} (\bibinfo {year} {2018}{\natexlab{a}})}\BibitemShut {NoStop}%
\bibitem [{\citenamefont {Cao}\ \emph {et~al.}(2018{\natexlab{b}})\citenamefont
  {Cao}, \citenamefont {Fatemi}, \citenamefont {Fang}, \citenamefont
  {Watanabe}, \citenamefont {Taniguchi}, \citenamefont {Kaxiras},\ and\
  \citenamefont {Jarillo-Herrero}}]{YCS}%
  \BibitemOpen
  \bibfield  {author} {\bibinfo {author} {\bibfnamefont {Y.}~\bibnamefont
  {Cao}}, \bibinfo {author} {\bibfnamefont {V.}~\bibnamefont {Fatemi}},
  \bibinfo {author} {\bibfnamefont {S.}~\bibnamefont {Fang}}, \bibinfo {author}
  {\bibfnamefont {K.}~\bibnamefont {Watanabe}}, \bibinfo {author}
  {\bibfnamefont {T.}~\bibnamefont {Taniguchi}}, \bibinfo {author}
  {\bibfnamefont {E.}~\bibnamefont {Kaxiras}}, \ and\ \bibinfo {author}
  {\bibfnamefont {P.}~\bibnamefont {Jarillo-Herrero}},\ }\href
  {https://www.nature.com/articles/nature26160} {\bibfield  {journal} {\bibinfo
   {journal} {Nature}\ }\textbf {\bibinfo {volume} {556}},\ \bibinfo {pages}
  {43} (\bibinfo {year} {2018}{\natexlab{b}})}\BibitemShut {NoStop}%
\bibitem [{\citenamefont {Yankowitz}\ \emph {et~al.}(2019)\citenamefont
  {Yankowitz}, \citenamefont {Chen}, \citenamefont {Polshyn}, \citenamefont
  {Zhang}, \citenamefont {Watanabe}, \citenamefont {Taniguchi}, \citenamefont
  {Graf}, \citenamefont {Young},\ and\ \citenamefont {Dean}}]{yankowitz_2018}%
  \BibitemOpen
  \bibfield  {author} {\bibinfo {author} {\bibfnamefont {M.}~\bibnamefont
  {Yankowitz}}, \bibinfo {author} {\bibfnamefont {S.}~\bibnamefont {Chen}},
  \bibinfo {author} {\bibfnamefont {H.}~\bibnamefont {Polshyn}}, \bibinfo
  {author} {\bibfnamefont {Y.}~\bibnamefont {Zhang}}, \bibinfo {author}
  {\bibfnamefont {K.}~\bibnamefont {Watanabe}}, \bibinfo {author}
  {\bibfnamefont {T.}~\bibnamefont {Taniguchi}}, \bibinfo {author}
  {\bibfnamefont {D.}~\bibnamefont {Graf}}, \bibinfo {author} {\bibfnamefont
  {A.~F.}\ \bibnamefont {Young}}, \ and\ \bibinfo {author} {\bibfnamefont
  {C.~R.}\ \bibnamefont {Dean}},\ }\href {\doibase 10.1126/science.aav1910}
  {\bibfield  {journal} {\bibinfo  {journal} {Science}\ }\textbf {\bibinfo
  {volume} {363}},\ \bibinfo {pages} {1059} (\bibinfo {year}
  {2019})}\BibitemShut {NoStop}%
\bibitem [{\citenamefont {Pierucci}\ \emph {et~al.}(2015)\citenamefont
  {Pierucci}, \citenamefont {Sediri}, \citenamefont {Hajlaoui}, \citenamefont
  {Girard}, \citenamefont {Brumme}, \citenamefont {Calandra}, \citenamefont
  {Velez-Fort}, \citenamefont {Patriarche}, \citenamefont {Silly},
  \citenamefont {Ferro}, \citenamefont {Soulière}, \citenamefont {Marangolo},
  \citenamefont {Sirotti}, \citenamefont {Mauri},\ and\ \citenamefont
  {Ouerghi}}]{DPH}%
  \BibitemOpen
  \bibfield  {author} {\bibinfo {author} {\bibfnamefont {D.}~\bibnamefont
  {Pierucci}}, \bibinfo {author} {\bibfnamefont {H.}~\bibnamefont {Sediri}},
  \bibinfo {author} {\bibfnamefont {M.}~\bibnamefont {Hajlaoui}}, \bibinfo
  {author} {\bibfnamefont {J.-C.}\ \bibnamefont {Girard}}, \bibinfo {author}
  {\bibfnamefont {T.}~\bibnamefont {Brumme}}, \bibinfo {author} {\bibfnamefont
  {M.}~\bibnamefont {Calandra}}, \bibinfo {author} {\bibfnamefont
  {E.}~\bibnamefont {Velez-Fort}}, \bibinfo {author} {\bibfnamefont
  {G.}~\bibnamefont {Patriarche}}, \bibinfo {author} {\bibfnamefont {M.~G.}\
  \bibnamefont {Silly}}, \bibinfo {author} {\bibfnamefont {G.}~\bibnamefont
  {Ferro}}, \bibinfo {author} {\bibfnamefont {V.}~\bibnamefont {Soulière}},
  \bibinfo {author} {\bibfnamefont {M.}~\bibnamefont {Marangolo}}, \bibinfo
  {author} {\bibfnamefont {F.}~\bibnamefont {Sirotti}}, \bibinfo {author}
  {\bibfnamefont {F.}~\bibnamefont {Mauri}}, \ and\ \bibinfo {author}
  {\bibfnamefont {A.}~\bibnamefont {Ouerghi}},\ }\href {\doibase
  10.1021/acsnano.5b01239} {\bibfield  {journal} {\bibinfo  {journal} {ACS
  Nano}\ }\textbf {\bibinfo {volume} {9}},\ \bibinfo {pages} {5432} (\bibinfo
  {year} {2015})}\BibitemShut {NoStop}%
\bibitem [{\citenamefont {Chittari}\ \emph {et~al.}(2019)\citenamefont
  {Chittari}, \citenamefont {Chen}, \citenamefont {Zhang}, \citenamefont
  {Wang},\ and\ \citenamefont {Jung}}]{BGY}%
  \BibitemOpen
  \bibfield  {author} {\bibinfo {author} {\bibfnamefont {B.~L.}\ \bibnamefont
  {Chittari}}, \bibinfo {author} {\bibfnamefont {G.}~\bibnamefont {Chen}},
  \bibinfo {author} {\bibfnamefont {Y.}~\bibnamefont {Zhang}}, \bibinfo
  {author} {\bibfnamefont {F.}~\bibnamefont {Wang}}, \ and\ \bibinfo {author}
  {\bibfnamefont {J.}~\bibnamefont {Jung}},\ }\href {\doibase
  10.1103/PhysRevLett.122.016401} {\bibfield  {journal} {\bibinfo  {journal}
  {Phys. Rev. Lett.}\ }\textbf {\bibinfo {volume} {122}},\ \bibinfo {pages}
  {016401} (\bibinfo {year} {2019})}\BibitemShut {NoStop}%
\bibitem [{\citenamefont {Volovik}(2018)}]{volovik_2018}%
  \BibitemOpen
  \bibfield  {author} {\bibinfo {author} {\bibfnamefont {G.~E.}\ \bibnamefont
  {Volovik}},\ }\href {\doibase 10.1134/S0021364018080052} {\bibfield
  {journal} {\bibinfo  {journal} {JETP Letters}\ }\textbf {\bibinfo {volume}
  {107}},\ \bibinfo {pages} {516} (\bibinfo {year} {2018})}\BibitemShut
  {NoStop}%
\bibitem [{\citenamefont {Xu}\ and\ \citenamefont
  {Balents}(2018)}]{22_xu_2018}%
  \BibitemOpen
  \bibfield  {author} {\bibinfo {author} {\bibfnamefont {C.}~\bibnamefont
  {Xu}}\ and\ \bibinfo {author} {\bibfnamefont {L.}~\bibnamefont {Balents}},\
  }\href {\doibase 10.1103/PhysRevLett.121.087001} {\bibfield  {journal}
  {\bibinfo  {journal} {Phys. Rev. Lett.}\ }\textbf {\bibinfo {volume} {121}},\
  \bibinfo {pages} {087001} (\bibinfo {year} {2018})}\BibitemShut {NoStop}%
\bibitem [{\citenamefont {DaSilva}\ \emph {et~al.}(2016)\citenamefont
  {DaSilva}, \citenamefont {Jung},\ and\ \citenamefont
  {MacDonald}}]{2_dasilva_2016}%
  \BibitemOpen
  \bibfield  {author} {\bibinfo {author} {\bibfnamefont {A.~M.}\ \bibnamefont
  {DaSilva}}, \bibinfo {author} {\bibfnamefont {J.}~\bibnamefont {Jung}}, \
  and\ \bibinfo {author} {\bibfnamefont {A.~H.}\ \bibnamefont {MacDonald}},\
  }\href {\doibase 10.1103/PhysRevLett.117.036802} {\bibfield  {journal}
  {\bibinfo  {journal} {Phys. Rev. Lett.}\ }\textbf {\bibinfo {volume} {117}},\
  \bibinfo {pages} {036802} (\bibinfo {year} {2016})}\BibitemShut {NoStop}%
\bibitem [{\citenamefont {Baskaran}(2018)}]{2_baskaran_2018}%
  \BibitemOpen
  \bibfield  {author} {\bibinfo {author} {\bibfnamefont {G.}~\bibnamefont
  {Baskaran}},\ }\href {https://arxiv.org/abs/1804.00627} {\bibfield  {journal}
  {\bibinfo  {journal} {arXiv:1804.00627}\ } (\bibinfo {year}
  {2018})}\BibitemShut {NoStop}%
\bibitem [{\citenamefont {Dodaro}\ \emph {et~al.}(2018)\citenamefont {Dodaro},
  \citenamefont {Kivelson}, \citenamefont {Schattner}, \citenamefont {Sun},\
  and\ \citenamefont {Wang}}]{2_dodaro_2018}%
  \BibitemOpen
  \bibfield  {author} {\bibinfo {author} {\bibfnamefont {J.~F.}\ \bibnamefont
  {Dodaro}}, \bibinfo {author} {\bibfnamefont {S.~A.}\ \bibnamefont
  {Kivelson}}, \bibinfo {author} {\bibfnamefont {Y.}~\bibnamefont {Schattner}},
  \bibinfo {author} {\bibfnamefont {X.~Q.}\ \bibnamefont {Sun}}, \ and\
  \bibinfo {author} {\bibfnamefont {C.}~\bibnamefont {Wang}},\ }\href {\doibase
  10.1103/PhysRevB.98.075154} {\bibfield  {journal} {\bibinfo  {journal} {Phys.
  Rev. B}\ }\textbf {\bibinfo {volume} {98}},\ \bibinfo {pages} {075154}
  (\bibinfo {year} {2018})}\BibitemShut {NoStop}%
\bibitem [{\citenamefont {Gonz\'alez}\ and\ \citenamefont
  {Stauber}(2019)}]{2_gonzalez_2018}%
  \BibitemOpen
  \bibfield  {author} {\bibinfo {author} {\bibfnamefont {J.}~\bibnamefont
  {Gonz\'alez}}\ and\ \bibinfo {author} {\bibfnamefont {T.}~\bibnamefont
  {Stauber}},\ }\href {\doibase 10.1103/PhysRevLett.122.026801} {\bibfield
  {journal} {\bibinfo  {journal} {Phys. Rev. Lett.}\ }\textbf {\bibinfo
  {volume} {122}},\ \bibinfo {pages} {026801} (\bibinfo {year}
  {2019})}\BibitemShut {NoStop}%
\bibitem [{\citenamefont {Guo}\ \emph {et~al.}(2018)\citenamefont {Guo},
  \citenamefont {Zhu}, \citenamefont {Feng},\ and\ \citenamefont
  {Scalettar}}]{2_guo_2018}%
  \BibitemOpen
  \bibfield  {author} {\bibinfo {author} {\bibfnamefont {H.}~\bibnamefont
  {Guo}}, \bibinfo {author} {\bibfnamefont {X.}~\bibnamefont {Zhu}}, \bibinfo
  {author} {\bibfnamefont {S.}~\bibnamefont {Feng}}, \ and\ \bibinfo {author}
  {\bibfnamefont {R.~T.}\ \bibnamefont {Scalettar}},\ }\href {\doibase
  10.1103/PhysRevB.97.235453} {\bibfield  {journal} {\bibinfo  {journal} {Phys.
  Rev. B}\ }\textbf {\bibinfo {volume} {97}},\ \bibinfo {pages} {235453}
  (\bibinfo {year} {2018})}\BibitemShut {NoStop}%
\bibitem [{\citenamefont {Laksonoa}\ \emph {et~al.}(2018)\citenamefont
  {Laksonoa}, \citenamefont {Leawa}, \citenamefont {Reavesc}, \citenamefont
  {Singhc}, \citenamefont {Wanga}, \citenamefont {Adama},\ and\ \citenamefont
  {Gu}}]{2_laksono_2018}%
  \BibitemOpen
  \bibfield  {author} {\bibinfo {author} {\bibfnamefont {E.}~\bibnamefont
  {Laksonoa}}, \bibinfo {author} {\bibfnamefont {J.~N.}\ \bibnamefont {Leawa}},
  \bibinfo {author} {\bibfnamefont {A.}~\bibnamefont {Reavesc}}, \bibinfo
  {author} {\bibfnamefont {M.}~\bibnamefont {Singhc}}, \bibinfo {author}
  {\bibfnamefont {X.}~\bibnamefont {Wanga}}, \bibinfo {author} {\bibfnamefont
  {S.}~\bibnamefont {Adama}}, \ and\ \bibinfo {author} {\bibfnamefont
  {X.}~\bibnamefont {Gu}},\ }\href {\doibase 10.1016/j.ssc.2018.07.013}
  {\bibfield  {journal} {\bibinfo  {journal} {Solid State Communications}\
  }\textbf {\bibinfo {volume} {282}},\ \bibinfo {pages} {38} (\bibinfo {year}
  {2018})}\BibitemShut {NoStop}%
\bibitem [{\citenamefont {Liu}\ \emph {et~al.}(2018)\citenamefont {Liu},
  \citenamefont {Zhang}, \citenamefont {Chen},\ and\ \citenamefont
  {Yang}}]{2_liu_2018}%
  \BibitemOpen
  \bibfield  {author} {\bibinfo {author} {\bibfnamefont {C.-C.}\ \bibnamefont
  {Liu}}, \bibinfo {author} {\bibfnamefont {L.-D.}\ \bibnamefont {Zhang}},
  \bibinfo {author} {\bibfnamefont {W.-Q.}\ \bibnamefont {Chen}}, \ and\
  \bibinfo {author} {\bibfnamefont {F.}~\bibnamefont {Yang}},\ }\href {\doibase
  10.1103/PhysRevLett.121.217001} {\bibfield  {journal} {\bibinfo  {journal}
  {Phys. Rev. Lett.}\ }\textbf {\bibinfo {volume} {121}},\ \bibinfo {pages}
  {217001} (\bibinfo {year} {2018})}\BibitemShut {NoStop}%
\bibitem [{\citenamefont {Po}\ \emph {et~al.}(2018)\citenamefont {Po},
  \citenamefont {Zou}, \citenamefont {Vishwanath},\ and\ \citenamefont
  {Senthil}}]{2_po_2018}%
  \BibitemOpen
  \bibfield  {author} {\bibinfo {author} {\bibfnamefont {H.~C.}\ \bibnamefont
  {Po}}, \bibinfo {author} {\bibfnamefont {L.}~\bibnamefont {Zou}}, \bibinfo
  {author} {\bibfnamefont {A.}~\bibnamefont {Vishwanath}}, \ and\ \bibinfo
  {author} {\bibfnamefont {T.}~\bibnamefont {Senthil}},\ }\href {\doibase
  10.1103/PhysRevX.8.031089} {\bibfield  {journal} {\bibinfo  {journal} {Phys.
  Rev. X}\ }\textbf {\bibinfo {volume} {8}},\ \bibinfo {pages} {031089}
  (\bibinfo {year} {2018})}\BibitemShut {NoStop}%
\bibitem [{\citenamefont {Ray}\ \emph {et~al.}(2019)\citenamefont {Ray},
  \citenamefont {Jung},\ and\ \citenamefont {Das}}]{2_ray_2018}%
  \BibitemOpen
  \bibfield  {author} {\bibinfo {author} {\bibfnamefont {S.}~\bibnamefont
  {Ray}}, \bibinfo {author} {\bibfnamefont {J.}~\bibnamefont {Jung}}, \ and\
  \bibinfo {author} {\bibfnamefont {T.}~\bibnamefont {Das}},\ }\href {\doibase
  10.1103/PhysRevB.99.134515} {\bibfield  {journal} {\bibinfo  {journal} {Phys.
  Rev. B}\ }\textbf {\bibinfo {volume} {99}},\ \bibinfo {pages} {134515}
  (\bibinfo {year} {2019})}\BibitemShut {NoStop}%
\bibitem [{\citenamefont {Thomson}\ \emph {et~al.}(2018)\citenamefont
  {Thomson}, \citenamefont {Chatterjee}, \citenamefont {Sachdev},\ and\
  \citenamefont {Scheurer}}]{2_thomson_2018}%
  \BibitemOpen
  \bibfield  {author} {\bibinfo {author} {\bibfnamefont {A.}~\bibnamefont
  {Thomson}}, \bibinfo {author} {\bibfnamefont {S.}~\bibnamefont {Chatterjee}},
  \bibinfo {author} {\bibfnamefont {S.}~\bibnamefont {Sachdev}}, \ and\
  \bibinfo {author} {\bibfnamefont {M.~S.}\ \bibnamefont {Scheurer}},\ }\href
  {\doibase 10.1103/PhysRevB.98.075109} {\bibfield  {journal} {\bibinfo
  {journal} {Phys. Rev. B}\ }\textbf {\bibinfo {volume} {98}},\ \bibinfo
  {pages} {075109} (\bibinfo {year} {2018})}\BibitemShut {NoStop}%
\bibitem [{\citenamefont {Wu}\ \emph {et~al.}(2018)\citenamefont {Wu},
  \citenamefont {MacDonald},\ and\ \citenamefont {Martin}}]{2_wu_2018}%
  \BibitemOpen
  \bibfield  {author} {\bibinfo {author} {\bibfnamefont {F.}~\bibnamefont
  {Wu}}, \bibinfo {author} {\bibfnamefont {A.~H.}\ \bibnamefont {MacDonald}}, \
  and\ \bibinfo {author} {\bibfnamefont {I.}~\bibnamefont {Martin}},\ }\href
  {\doibase 10.1103/PhysRevLett.121.257001} {\bibfield  {journal} {\bibinfo
  {journal} {Phys. Rev. Lett.}\ }\textbf {\bibinfo {volume} {121}},\ \bibinfo
  {pages} {257001} (\bibinfo {year} {2018})}\BibitemShut {NoStop}%
\bibitem [{\citenamefont {Xu}\ \emph {et~al.}(2018)\citenamefont {Xu},
  \citenamefont {Law},\ and\ \citenamefont {Lee}}]{2_xu_2018}%
  \BibitemOpen
  \bibfield  {author} {\bibinfo {author} {\bibfnamefont {X.~Y.}\ \bibnamefont
  {Xu}}, \bibinfo {author} {\bibfnamefont {K.~T.}\ \bibnamefont {Law}}, \ and\
  \bibinfo {author} {\bibfnamefont {P.~A.}\ \bibnamefont {Lee}},\ }\href
  {\doibase 10.1103/PhysRevB.98.121406} {\bibfield  {journal} {\bibinfo
  {journal} {Phys. Rev. B}\ }\textbf {\bibinfo {volume} {98}},\ \bibinfo
  {pages} {121406} (\bibinfo {year} {2018})}\BibitemShut {NoStop}%
\bibitem [{\citenamefont {Yuan}\ and\ \citenamefont {Fu}(2018)}]{2_yuan_2018}%
  \BibitemOpen
  \bibfield  {author} {\bibinfo {author} {\bibfnamefont {N.~F.~Q.}\
  \bibnamefont {Yuan}}\ and\ \bibinfo {author} {\bibfnamefont {L.}~\bibnamefont
  {Fu}},\ }\href {\doibase 10.1103/PhysRevB.98.045103} {\bibfield  {journal}
  {\bibinfo  {journal} {Phys. Rev. B}\ }\textbf {\bibinfo {volume} {98}},\
  \bibinfo {pages} {045103} (\bibinfo {year} {2018})}\BibitemShut {NoStop}%
\bibitem [{\citenamefont {San-Jose}\ \emph {et~al.}(2012)\citenamefont
  {San-Jose}, \citenamefont {Gonz\'alez},\ and\ \citenamefont {Guinea}}]{PJF}%
  \BibitemOpen
  \bibfield  {author} {\bibinfo {author} {\bibfnamefont {P.}~\bibnamefont
  {San-Jose}}, \bibinfo {author} {\bibfnamefont {J.}~\bibnamefont
  {Gonz\'alez}}, \ and\ \bibinfo {author} {\bibfnamefont {F.}~\bibnamefont
  {Guinea}},\ }\href {\doibase 10.1103/PhysRevLett.108.216802} {\bibfield
  {journal} {\bibinfo  {journal} {Phys. Rev. Lett.}\ }\textbf {\bibinfo
  {volume} {108}},\ \bibinfo {pages} {216802} (\bibinfo {year}
  {2012})}\BibitemShut {NoStop}%
\bibitem [{\citenamefont {Yin}\ \emph {et~al.}(2015)\citenamefont {Yin},
  \citenamefont {Qiao}, \citenamefont {Zuo}, \citenamefont {Li},\ and\
  \citenamefont {He}}]{LJW}%
  \BibitemOpen
  \bibfield  {author} {\bibinfo {author} {\bibfnamefont {L.-J.}\ \bibnamefont
  {Yin}}, \bibinfo {author} {\bibfnamefont {J.-B.}\ \bibnamefont {Qiao}},
  \bibinfo {author} {\bibfnamefont {W.-J.}\ \bibnamefont {Zuo}}, \bibinfo
  {author} {\bibfnamefont {W.-T.}\ \bibnamefont {Li}}, \ and\ \bibinfo {author}
  {\bibfnamefont {L.}~\bibnamefont {He}},\ }\href {\doibase
  10.1103/PhysRevB.92.081406} {\bibfield  {journal} {\bibinfo  {journal} {Phys.
  Rev. B}\ }\textbf {\bibinfo {volume} {92}},\ \bibinfo {pages} {081406}
  (\bibinfo {year} {2015})}\BibitemShut {NoStop}%
\bibitem [{\citenamefont {Gonz\'alez}(2016)}]{LGZ}%
  \BibitemOpen
  \bibfield  {author} {\bibinfo {author} {\bibfnamefont {J.}~\bibnamefont
  {Gonz\'alez}},\ }\href {\doibase 10.1103/PhysRevB.94.165401} {\bibfield
  {journal} {\bibinfo  {journal} {Phys. Rev. B}\ }\textbf {\bibinfo {volume}
  {94}},\ \bibinfo {pages} {165401} (\bibinfo {year} {2016})}\BibitemShut
  {NoStop}%
\bibitem [{\citenamefont {Levy}\ \emph {et~al.}(2010)\citenamefont {Levy},
  \citenamefont {Burke}, \citenamefont {Meaker}, \citenamefont {Panlasigui},
  \citenamefont {Zettl}, \citenamefont {Guinea}, \citenamefont {Neto},\ and\
  \citenamefont {Crommie}}]{4_levy_2010}%
  \BibitemOpen
  \bibfield  {author} {\bibinfo {author} {\bibfnamefont {N.}~\bibnamefont
  {Levy}}, \bibinfo {author} {\bibfnamefont {S.~A.}\ \bibnamefont {Burke}},
  \bibinfo {author} {\bibfnamefont {K.~L.}\ \bibnamefont {Meaker}}, \bibinfo
  {author} {\bibfnamefont {M.}~\bibnamefont {Panlasigui}}, \bibinfo {author}
  {\bibfnamefont {A.}~\bibnamefont {Zettl}}, \bibinfo {author} {\bibfnamefont
  {F.}~\bibnamefont {Guinea}}, \bibinfo {author} {\bibfnamefont {A.~H.~C.}\
  \bibnamefont {Neto}}, \ and\ \bibinfo {author} {\bibfnamefont {M.~F.}\
  \bibnamefont {Crommie}},\ }\href {\doibase 10.1126/science.1191700}
  {\bibfield  {journal} {\bibinfo  {journal} {Science}\ }\textbf {\bibinfo
  {volume} {329}},\ \bibinfo {pages} {544} (\bibinfo {year}
  {2010})}\BibitemShut {NoStop}%
\bibitem [{\citenamefont {Tang}\ and\ \citenamefont {Fu}(2014)}]{4_tang_2014}%
  \BibitemOpen
  \bibfield  {author} {\bibinfo {author} {\bibfnamefont {E.}~\bibnamefont
  {Tang}}\ and\ \bibinfo {author} {\bibfnamefont {L.}~\bibnamefont {Fu}},\
  }\href {https://www.nature.com/articles/nphys3109} {\bibfield  {journal}
  {\bibinfo  {journal} {Nature Physics}\ }\textbf {\bibinfo {volume} {10}},\
  \bibinfo {pages} {964} (\bibinfo {year} {2014})}\BibitemShut {NoStop}%
\bibitem [{\citenamefont {Mueed}\ \emph {et~al.}(2018)\citenamefont {Mueed},
  \citenamefont {Hossain}, \citenamefont {Jo}, \citenamefont {Pfeiffer},
  \citenamefont {West}, \citenamefont {Baldwin},\ and\ \citenamefont
  {Shayegan}}]{4_mueed_2018}%
  \BibitemOpen
  \bibfield  {author} {\bibinfo {author} {\bibfnamefont {M.~A.}\ \bibnamefont
  {Mueed}}, \bibinfo {author} {\bibfnamefont {M.~S.}\ \bibnamefont {Hossain}},
  \bibinfo {author} {\bibfnamefont {I.}~\bibnamefont {Jo}}, \bibinfo {author}
  {\bibfnamefont {L.~N.}\ \bibnamefont {Pfeiffer}}, \bibinfo {author}
  {\bibfnamefont {K.~W.}\ \bibnamefont {West}}, \bibinfo {author}
  {\bibfnamefont {K.~W.}\ \bibnamefont {Baldwin}}, \ and\ \bibinfo {author}
  {\bibfnamefont {M.}~\bibnamefont {Shayegan}},\ }\href {\doibase
  10.1103/PhysRevLett.121.036802} {\bibfield  {journal} {\bibinfo  {journal}
  {Phys. Rev. Lett.}\ }\textbf {\bibinfo {volume} {121}},\ \bibinfo {pages}
  {036802} (\bibinfo {year} {2018})}\BibitemShut {NoStop}%
\bibitem [{\citenamefont {Jiang}\ \emph {et~al.}(2019)\citenamefont {Jiang},
  \citenamefont {Andelkovic}, \citenamefont {Milovanovic}, \citenamefont
  {Covaci}, \citenamefont {Lai}, \citenamefont {Cao}, \citenamefont {Watanabe},
  \citenamefont {Taniguchi}, \citenamefont {Peeters}, \citenamefont {Geim},\
  and\ \citenamefont {Andrei}}]{Yuhang_Jiang_2019}%
  \BibitemOpen
  \bibfield  {author} {\bibinfo {author} {\bibfnamefont {Y.}~\bibnamefont
  {Jiang}}, \bibinfo {author} {\bibfnamefont {M.}~\bibnamefont {Andelkovic}},
  \bibinfo {author} {\bibfnamefont {S.~P.}\ \bibnamefont {Milovanovic}},
  \bibinfo {author} {\bibfnamefont {L.}~\bibnamefont {Covaci}}, \bibinfo
  {author} {\bibfnamefont {X.}~\bibnamefont {Lai}}, \bibinfo {author}
  {\bibfnamefont {Y.}~\bibnamefont {Cao}}, \bibinfo {author} {\bibfnamefont
  {K.}~\bibnamefont {Watanabe}}, \bibinfo {author} {\bibfnamefont
  {T.}~\bibnamefont {Taniguchi}}, \bibinfo {author} {\bibfnamefont {F.~M.}\
  \bibnamefont {Peeters}}, \bibinfo {author} {\bibfnamefont {A.~K.}\
  \bibnamefont {Geim}}, \ and\ \bibinfo {author} {\bibfnamefont {E.~Y.}\
  \bibnamefont {Andrei}},\ }\href {https://arxiv.org/abs/1904.10147} {\bibfield
   {journal} {\bibinfo  {journal} {arXiv:1904.10147v1}\ } (\bibinfo {year}
  {2019})}\BibitemShut {NoStop}%
\bibitem [{\citenamefont {Weiss}\ \emph {et~al.}(1989)\citenamefont {Weiss},
  \citenamefont {Klitzing}, \citenamefont {Ploog},\ and\ \citenamefont
  {Weimann}}]{DW}%
  \BibitemOpen
  \bibfield  {author} {\bibinfo {author} {\bibfnamefont {D.}~\bibnamefont
  {Weiss}}, \bibinfo {author} {\bibfnamefont {K.~V.}\ \bibnamefont {Klitzing}},
  \bibinfo {author} {\bibfnamefont {K.}~\bibnamefont {Ploog}}, \ and\ \bibinfo
  {author} {\bibfnamefont {G.}~\bibnamefont {Weimann}},\ }\href {\doibase
  10.1209/0295-5075/8/2/012} {\bibfield  {journal} {\bibinfo  {journal}
  {Europhysics Letters}\ }\textbf {\bibinfo {volume} {8}},\ \bibinfo {pages}
  {179} (\bibinfo {year} {1989})}\BibitemShut {NoStop}%
\bibitem [{\citenamefont {Gerhardts}\ \emph {et~al.}(1989)\citenamefont
  {Gerhardts}, \citenamefont {Weiss},\ and\ \citenamefont {Klitzing}}]{RRG}%
  \BibitemOpen
  \bibfield  {author} {\bibinfo {author} {\bibfnamefont {R.~R.}\ \bibnamefont
  {Gerhardts}}, \bibinfo {author} {\bibfnamefont {D.}~\bibnamefont {Weiss}}, \
  and\ \bibinfo {author} {\bibfnamefont {K.~v.}\ \bibnamefont {Klitzing}},\
  }\href {\doibase 10.1103/PhysRevLett.62.1173} {\bibfield  {journal} {\bibinfo
   {journal} {Phys. Rev. Lett.}\ }\textbf {\bibinfo {volume} {62}},\ \bibinfo
  {pages} {1173} (\bibinfo {year} {1989})}\BibitemShut {NoStop}%
\bibitem [{\citenamefont {Winkler}\ \emph {et~al.}(1989)\citenamefont
  {Winkler}, \citenamefont {Kotthaus},\ and\ \citenamefont {Ploog}}]{RWW}%
  \BibitemOpen
  \bibfield  {author} {\bibinfo {author} {\bibfnamefont {R.~W.}\ \bibnamefont
  {Winkler}}, \bibinfo {author} {\bibfnamefont {J.~P.}\ \bibnamefont
  {Kotthaus}}, \ and\ \bibinfo {author} {\bibfnamefont {K.}~\bibnamefont
  {Ploog}},\ }\href {\doibase 10.1103/PhysRevLett.62.1177} {\bibfield
  {journal} {\bibinfo  {journal} {Phys. Rev. Lett.}\ }\textbf {\bibinfo
  {volume} {62}},\ \bibinfo {pages} {1177} (\bibinfo {year}
  {1989})}\BibitemShut {NoStop}%
\bibitem [{\citenamefont {Beenakker}(1989)}]{CWJB}%
  \BibitemOpen
  \bibfield  {author} {\bibinfo {author} {\bibfnamefont {C.~W.~J.}\
  \bibnamefont {Beenakker}},\ }\href {\doibase 10.1103/PhysRevLett.62.2020}
  {\bibfield  {journal} {\bibinfo  {journal} {Phys. Rev. Lett.}\ }\textbf
  {\bibinfo {volume} {62}},\ \bibinfo {pages} {2020} (\bibinfo {year}
  {1989})}\BibitemShut {NoStop}%
\bibitem [{\citenamefont {Vasilopoulos}\ and\ \citenamefont
  {Peeters}(1989)}]{PVF}%
  \BibitemOpen
  \bibfield  {author} {\bibinfo {author} {\bibfnamefont {P.}~\bibnamefont
  {Vasilopoulos}}\ and\ \bibinfo {author} {\bibfnamefont {F.~M.}\ \bibnamefont
  {Peeters}},\ }\href {\doibase 10.1103/PhysRevLett.63.2120} {\bibfield
  {journal} {\bibinfo  {journal} {Phys. Rev. Lett.}\ }\textbf {\bibinfo
  {volume} {63}},\ \bibinfo {pages} {2120} (\bibinfo {year}
  {1989})}\BibitemShut {NoStop}%
\bibitem [{\citenamefont {Gerhardts}\ \emph {et~al.}(1991)\citenamefont
  {Gerhardts}, \citenamefont {Weiss},\ and\ \citenamefont {Wulf}}]{RRD}%
  \BibitemOpen
  \bibfield  {author} {\bibinfo {author} {\bibfnamefont {R.~R.}\ \bibnamefont
  {Gerhardts}}, \bibinfo {author} {\bibfnamefont {D.}~\bibnamefont {Weiss}}, \
  and\ \bibinfo {author} {\bibfnamefont {U.}~\bibnamefont {Wulf}},\ }\href
  {\doibase 10.1103/PhysRevB.43.5192} {\bibfield  {journal} {\bibinfo
  {journal} {Phys. Rev. B}\ }\textbf {\bibinfo {volume} {43}},\ \bibinfo
  {pages} {5192} (\bibinfo {year} {1991})}\BibitemShut {NoStop}%
\bibitem [{\citenamefont {Albrecht}\ \emph {et~al.}(2001)\citenamefont
  {Albrecht}, \citenamefont {Smet}, \citenamefont {von Klitzing}, \citenamefont
  {Weiss}, \citenamefont {Umansky},\ and\ \citenamefont {Schweizer}}]{CAJ}%
  \BibitemOpen
  \bibfield  {author} {\bibinfo {author} {\bibfnamefont {C.}~\bibnamefont
  {Albrecht}}, \bibinfo {author} {\bibfnamefont {J.~H.}\ \bibnamefont {Smet}},
  \bibinfo {author} {\bibfnamefont {K.}~\bibnamefont {von Klitzing}}, \bibinfo
  {author} {\bibfnamefont {D.}~\bibnamefont {Weiss}}, \bibinfo {author}
  {\bibfnamefont {V.}~\bibnamefont {Umansky}}, \ and\ \bibinfo {author}
  {\bibfnamefont {H.}~\bibnamefont {Schweizer}},\ }\href {\doibase
  10.1103/PhysRevLett.86.147} {\bibfield  {journal} {\bibinfo  {journal} {Phys.
  Rev. Lett.}\ }\textbf {\bibinfo {volume} {86}},\ \bibinfo {pages} {147}
  (\bibinfo {year} {2001})}\BibitemShut {NoStop}%
\bibitem [{\citenamefont {Geisler}\ \emph {et~al.}(2004)\citenamefont
  {Geisler}, \citenamefont {Smet}, \citenamefont {Umansky}, \citenamefont {von
  Klitzing}, \citenamefont {Naundorf}, \citenamefont {Ketzmerick},\ and\
  \citenamefont {Schweizer}}]{MCG}%
  \BibitemOpen
  \bibfield  {author} {\bibinfo {author} {\bibfnamefont {M.~C.}\ \bibnamefont
  {Geisler}}, \bibinfo {author} {\bibfnamefont {J.~H.}\ \bibnamefont {Smet}},
  \bibinfo {author} {\bibfnamefont {V.}~\bibnamefont {Umansky}}, \bibinfo
  {author} {\bibfnamefont {K.}~\bibnamefont {von Klitzing}}, \bibinfo {author}
  {\bibfnamefont {B.}~\bibnamefont {Naundorf}}, \bibinfo {author}
  {\bibfnamefont {R.}~\bibnamefont {Ketzmerick}}, \ and\ \bibinfo {author}
  {\bibfnamefont {H.}~\bibnamefont {Schweizer}},\ }\href {\doibase
  10.1103/PhysRevLett.92.256801} {\bibfield  {journal} {\bibinfo  {journal}
  {Phys. Rev. Lett.}\ }\textbf {\bibinfo {volume} {92}},\ \bibinfo {pages}
  {256801} (\bibinfo {year} {2004})}\BibitemShut {NoStop}%
\bibitem [{\citenamefont {Wang}\ \emph {et~al.}(2004)\citenamefont {Wang},
  \citenamefont {Vasilopoulos},\ and\ \citenamefont {Peeters}}]{XFW}%
  \BibitemOpen
  \bibfield  {author} {\bibinfo {author} {\bibfnamefont {X.~F.}\ \bibnamefont
  {Wang}}, \bibinfo {author} {\bibfnamefont {P.}~\bibnamefont {Vasilopoulos}},
  \ and\ \bibinfo {author} {\bibfnamefont {F.~M.}\ \bibnamefont {Peeters}},\
  }\href {\doibase 10.1103/PhysRevB.69.035331} {\bibfield  {journal} {\bibinfo
  {journal} {Phys. Rev. B}\ }\textbf {\bibinfo {volume} {69}},\ \bibinfo
  {pages} {035331} (\bibinfo {year} {2004})}\BibitemShut {NoStop}%
\bibitem [{\citenamefont {Albrecht}\ \emph {et~al.}(1999)\citenamefont
  {Albrecht}, \citenamefont {Smet}, \citenamefont {Weiss}, \citenamefont {von
  Klitzing}, \citenamefont {Hennig}, \citenamefont {Langenbuch}, \citenamefont
  {Suhrke}, \citenamefont {R\"ossler}, \citenamefont {Umansky},\ and\
  \citenamefont {Schweizer}}]{CAT}%
  \BibitemOpen
  \bibfield  {author} {\bibinfo {author} {\bibfnamefont {C.}~\bibnamefont
  {Albrecht}}, \bibinfo {author} {\bibfnamefont {J.~H.}\ \bibnamefont {Smet}},
  \bibinfo {author} {\bibfnamefont {D.}~\bibnamefont {Weiss}}, \bibinfo
  {author} {\bibfnamefont {K.}~\bibnamefont {von Klitzing}}, \bibinfo {author}
  {\bibfnamefont {R.}~\bibnamefont {Hennig}}, \bibinfo {author} {\bibfnamefont
  {M.}~\bibnamefont {Langenbuch}}, \bibinfo {author} {\bibfnamefont
  {M.}~\bibnamefont {Suhrke}}, \bibinfo {author} {\bibfnamefont
  {U.}~\bibnamefont {R\"ossler}}, \bibinfo {author} {\bibfnamefont
  {V.}~\bibnamefont {Umansky}}, \ and\ \bibinfo {author} {\bibfnamefont
  {H.}~\bibnamefont {Schweizer}},\ }\href {\doibase
  10.1103/PhysRevLett.83.2234} {\bibfield  {journal} {\bibinfo  {journal}
  {Phys. Rev. Lett.}\ }\textbf {\bibinfo {volume} {83}},\ \bibinfo {pages}
  {2234} (\bibinfo {year} {1999})}\BibitemShut {NoStop}%
\bibitem [{\citenamefont {Chowdhury}\ \emph {et~al.}(2000)\citenamefont
  {Chowdhury}, \citenamefont {Emeleus}, \citenamefont {Milton}, \citenamefont
  {Skuras}, \citenamefont {Long}, \citenamefont {Davies}, \citenamefont
  {Pennelli},\ and\ \citenamefont {Stanley}}]{SCC}%
  \BibitemOpen
  \bibfield  {author} {\bibinfo {author} {\bibfnamefont {S.}~\bibnamefont
  {Chowdhury}}, \bibinfo {author} {\bibfnamefont {C.~J.}\ \bibnamefont
  {Emeleus}}, \bibinfo {author} {\bibfnamefont {B.}~\bibnamefont {Milton}},
  \bibinfo {author} {\bibfnamefont {E.}~\bibnamefont {Skuras}}, \bibinfo
  {author} {\bibfnamefont {A.~R.}\ \bibnamefont {Long}}, \bibinfo {author}
  {\bibfnamefont {J.~H.}\ \bibnamefont {Davies}}, \bibinfo {author}
  {\bibfnamefont {G.}~\bibnamefont {Pennelli}}, \ and\ \bibinfo {author}
  {\bibfnamefont {C.~R.}\ \bibnamefont {Stanley}},\ }\href {\doibase
  10.1103/PhysRevB.62.R4821} {\bibfield  {journal} {\bibinfo  {journal} {Phys.
  Rev. B}\ }\textbf {\bibinfo {volume} {62}},\ \bibinfo {pages} {R4821}
  (\bibinfo {year} {2000})}\BibitemShut {NoStop}%
\bibitem [{\citenamefont {Chowdhury}\ \emph {et~al.}(2004)\citenamefont
  {Chowdhury}, \citenamefont {Long}, \citenamefont {Skuras}, \citenamefont
  {Davies}, \citenamefont {Lister}, \citenamefont {Pennelli},\ and\
  \citenamefont {Stanley}}]{SCA}%
  \BibitemOpen
  \bibfield  {author} {\bibinfo {author} {\bibfnamefont {S.}~\bibnamefont
  {Chowdhury}}, \bibinfo {author} {\bibfnamefont {A.~R.}\ \bibnamefont {Long}},
  \bibinfo {author} {\bibfnamefont {E.}~\bibnamefont {Skuras}}, \bibinfo
  {author} {\bibfnamefont {J.~H.}\ \bibnamefont {Davies}}, \bibinfo {author}
  {\bibfnamefont {K.}~\bibnamefont {Lister}}, \bibinfo {author} {\bibfnamefont
  {G.}~\bibnamefont {Pennelli}}, \ and\ \bibinfo {author} {\bibfnamefont
  {C.~R.}\ \bibnamefont {Stanley}},\ }\href {\doibase
  10.1103/PhysRevB.69.035330} {\bibfield  {journal} {\bibinfo  {journal} {Phys.
  Rev. B}\ }\textbf {\bibinfo {volume} {69}},\ \bibinfo {pages} {035330}
  (\bibinfo {year} {2004})}\BibitemShut {NoStop}%
\bibitem [{\citenamefont {Kato}\ \emph {et~al.}(2012)\citenamefont {Kato},
  \citenamefont {Endo}, \citenamefont {Katsumoto},\ and\ \citenamefont
  {Iye}}]{YKA}%
  \BibitemOpen
  \bibfield  {author} {\bibinfo {author} {\bibfnamefont {Y.}~\bibnamefont
  {Kato}}, \bibinfo {author} {\bibfnamefont {A.}~\bibnamefont {Endo}}, \bibinfo
  {author} {\bibfnamefont {S.}~\bibnamefont {Katsumoto}}, \ and\ \bibinfo
  {author} {\bibfnamefont {Y.}~\bibnamefont {Iye}},\ }\href {\doibase
  10.1103/PhysRevB.86.235315} {\bibfield  {journal} {\bibinfo  {journal} {Phys.
  Rev. B}\ }\textbf {\bibinfo {volume} {86}},\ \bibinfo {pages} {235315}
  (\bibinfo {year} {2012})}\BibitemShut {NoStop}%
\bibitem [{\citenamefont {Carmona}\ \emph {et~al.}(1995)\citenamefont
  {Carmona}, \citenamefont {Geim}, \citenamefont {Nogaret}, \citenamefont
  {Main}, \citenamefont {Foster}, \citenamefont {Henini}, \citenamefont
  {Beaumont},\ and\ \citenamefont {Blamire}}]{HAC}%
  \BibitemOpen
  \bibfield  {author} {\bibinfo {author} {\bibfnamefont {H.~A.}\ \bibnamefont
  {Carmona}}, \bibinfo {author} {\bibfnamefont {A.~K.}\ \bibnamefont {Geim}},
  \bibinfo {author} {\bibfnamefont {A.}~\bibnamefont {Nogaret}}, \bibinfo
  {author} {\bibfnamefont {P.~C.}\ \bibnamefont {Main}}, \bibinfo {author}
  {\bibfnamefont {T.~J.}\ \bibnamefont {Foster}}, \bibinfo {author}
  {\bibfnamefont {M.}~\bibnamefont {Henini}}, \bibinfo {author} {\bibfnamefont
  {S.~P.}\ \bibnamefont {Beaumont}}, \ and\ \bibinfo {author} {\bibfnamefont
  {M.~G.}\ \bibnamefont {Blamire}},\ }\href {\doibase
  10.1103/PhysRevLett.74.3009} {\bibfield  {journal} {\bibinfo  {journal}
  {Phys. Rev. Lett.}\ }\textbf {\bibinfo {volume} {74}},\ \bibinfo {pages}
  {3009} (\bibinfo {year} {1995})}\BibitemShut {NoStop}%
\bibitem [{\citenamefont {Ye}\ \emph {et~al.}(1995{\natexlab{a}})\citenamefont
  {Ye}, \citenamefont {Weiss}, \citenamefont {Gerhardts}, \citenamefont
  {Seeger}, \citenamefont {von Klitzing}, \citenamefont {Eberl},\ and\
  \citenamefont {Nickel}}]{PDY}%
  \BibitemOpen
  \bibfield  {author} {\bibinfo {author} {\bibfnamefont {P.~D.}\ \bibnamefont
  {Ye}}, \bibinfo {author} {\bibfnamefont {D.}~\bibnamefont {Weiss}}, \bibinfo
  {author} {\bibfnamefont {R.~R.}\ \bibnamefont {Gerhardts}}, \bibinfo {author}
  {\bibfnamefont {M.}~\bibnamefont {Seeger}}, \bibinfo {author} {\bibfnamefont
  {K.}~\bibnamefont {von Klitzing}}, \bibinfo {author} {\bibfnamefont
  {K.}~\bibnamefont {Eberl}}, \ and\ \bibinfo {author} {\bibfnamefont
  {H.}~\bibnamefont {Nickel}},\ }\href {\doibase 10.1103/PhysRevLett.74.3013}
  {\bibfield  {journal} {\bibinfo  {journal} {Phys. Rev. Lett.}\ }\textbf
  {\bibinfo {volume} {74}},\ \bibinfo {pages} {3013} (\bibinfo {year}
  {1995}{\natexlab{a}})}\BibitemShut {NoStop}%
\bibitem [{\citenamefont {Xue}\ and\ \citenamefont {Xiao}(1992)}]{DPX}%
  \BibitemOpen
  \bibfield  {author} {\bibinfo {author} {\bibfnamefont {D.~P.}\ \bibnamefont
  {Xue}}\ and\ \bibinfo {author} {\bibfnamefont {G.}~\bibnamefont {Xiao}},\
  }\href {\doibase 10.1103/PhysRevB.45.5986} {\bibfield  {journal} {\bibinfo
  {journal} {Phys. Rev. B}\ }\textbf {\bibinfo {volume} {45}},\ \bibinfo
  {pages} {5986} (\bibinfo {year} {1992})}\BibitemShut {NoStop}%
\bibitem [{\citenamefont {Peeters}\ and\ \citenamefont
  {Vasilopoulos}(1993)}]{FMP}%
  \BibitemOpen
  \bibfield  {author} {\bibinfo {author} {\bibfnamefont {F.~M.}\ \bibnamefont
  {Peeters}}\ and\ \bibinfo {author} {\bibfnamefont {P.}~\bibnamefont
  {Vasilopoulos}},\ }\href {\doibase 10.1103/PhysRevB.47.1466} {\bibfield
  {journal} {\bibinfo  {journal} {Phys. Rev. B}\ }\textbf {\bibinfo {volume}
  {47}},\ \bibinfo {pages} {1466} (\bibinfo {year} {1993})}\BibitemShut
  {NoStop}%
\bibitem [{\citenamefont {Chang}\ and\ \citenamefont {Niu}(1994)}]{MCQ}%
  \BibitemOpen
  \bibfield  {author} {\bibinfo {author} {\bibfnamefont {M.~C.}\ \bibnamefont
  {Chang}}\ and\ \bibinfo {author} {\bibfnamefont {Q.}~\bibnamefont {Niu}},\
  }\href {\doibase 10.1103/PhysRevB.50.10843} {\bibfield  {journal} {\bibinfo
  {journal} {Phys. Rev. B}\ }\textbf {\bibinfo {volume} {50}},\ \bibinfo
  {pages} {10843} (\bibinfo {year} {1994})}\BibitemShut {NoStop}%
\bibitem [{\citenamefont {Ye}\ \emph {et~al.}(1995{\natexlab{b}})\citenamefont
  {Ye}, \citenamefont {Weiss}, \citenamefont {Klitzing},\ and\ \citenamefont
  {Eberl}}]{PDM}%
  \BibitemOpen
  \bibfield  {author} {\bibinfo {author} {\bibfnamefont {P.~D.}\ \bibnamefont
  {Ye}}, \bibinfo {author} {\bibfnamefont {D.}~\bibnamefont {Weiss}}, \bibinfo
  {author} {\bibfnamefont {K.~v.}\ \bibnamefont {Klitzing}}, \ and\ \bibinfo
  {author} {\bibfnamefont {K.}~\bibnamefont {Eberl}},\ }\href {\doibase
  10.1063/1.114520} {\bibfield  {journal} {\bibinfo  {journal} {Appl. Phys.
  Lett.}\ }\textbf {\bibinfo {volume} {67}},\ \bibinfo {pages} {1441} (\bibinfo
  {year} {1995}{\natexlab{b}})}\BibitemShut {NoStop}%
\bibitem [{\citenamefont {Ye}\ \emph {et~al.}(1997)\citenamefont {Ye},
  \citenamefont {Weiss},\ and\ \citenamefont {Gerhardts}}]{PDR}%
  \BibitemOpen
  \bibfield  {author} {\bibinfo {author} {\bibfnamefont {P.~D.}\ \bibnamefont
  {Ye}}, \bibinfo {author} {\bibfnamefont {D.}~\bibnamefont {Weiss}}, \ and\
  \bibinfo {author} {\bibfnamefont {R.~R.}\ \bibnamefont {Gerhardts}},\ }\href
  {\doibase 10.1063/1.364565} {\bibfield  {journal} {\bibinfo  {journal}
  {Journal of Applied Physics}\ }\textbf {\bibinfo {volume} {81}},\ \bibinfo
  {pages} {5444} (\bibinfo {year} {1997})}\BibitemShut {NoStop}%
\bibitem [{\citenamefont {Skuras}\ \emph {et~al.}(2001)\citenamefont {Skuras},
  \citenamefont {Long}, \citenamefont {Chowdhury},\ and\ \citenamefont
  {Rahman}}]{EAS}%
  \BibitemOpen
  \bibfield  {author} {\bibinfo {author} {\bibfnamefont {E.}~\bibnamefont
  {Skuras}}, \bibinfo {author} {\bibfnamefont {A.~R.}\ \bibnamefont {Long}},
  \bibinfo {author} {\bibfnamefont {S.}~\bibnamefont {Chowdhury}}, \ and\
  \bibinfo {author} {\bibfnamefont {M.}~\bibnamefont {Rahman}},\ }\href
  {\doibase 10.1063/1.1388574} {\bibfield  {journal} {\bibinfo  {journal}
  {Journal of Applied Physics}\ }\textbf {\bibinfo {volume} {90}},\ \bibinfo
  {pages} {2623} (\bibinfo {year} {2001})}\BibitemShut {NoStop}%
\bibitem [{\citenamefont {Betthausen}\ \emph {et~al.}(2012)\citenamefont
  {Betthausen}, \citenamefont {Dollinger}, \citenamefont {Saarikoski},
  \citenamefont {Kolkovsky}, \citenamefont {Karczewski}, \citenamefont
  {Wojtowicz}, \citenamefont {Richter},\ and\ \citenamefont {Weiss}}]{CBT}%
  \BibitemOpen
  \bibfield  {author} {\bibinfo {author} {\bibfnamefont {C.}~\bibnamefont
  {Betthausen}}, \bibinfo {author} {\bibfnamefont {T.}~\bibnamefont
  {Dollinger}}, \bibinfo {author} {\bibfnamefont {H.}~\bibnamefont
  {Saarikoski}}, \bibinfo {author} {\bibfnamefont {V.}~\bibnamefont
  {Kolkovsky}}, \bibinfo {author} {\bibfnamefont {G.}~\bibnamefont
  {Karczewski}}, \bibinfo {author} {\bibfnamefont {T.}~\bibnamefont
  {Wojtowicz}}, \bibinfo {author} {\bibfnamefont {K.}~\bibnamefont {Richter}},
  \ and\ \bibinfo {author} {\bibfnamefont {D.}~\bibnamefont {Weiss}},\ }\href
  {\doibase 10.1126/science.1221350} {\bibfield  {journal} {\bibinfo  {journal}
  {Science}\ }\textbf {\bibinfo {volume} {337}},\ \bibinfo {pages} {324}
  (\bibinfo {year} {2012})}\BibitemShut {NoStop}%
\bibitem [{\citenamefont {Drienovsky}\ \emph {et~al.}(2018)\citenamefont
  {Drienovsky}, \citenamefont {Joachimsmeyer}, \citenamefont {Sandner},
  \citenamefont {Liu}, \citenamefont {Taniguchi}, \citenamefont {Watanabe},
  \citenamefont {Richter}, \citenamefont {Weiss},\ and\ \citenamefont
  {Eroms}}]{MDJ}%
  \BibitemOpen
  \bibfield  {author} {\bibinfo {author} {\bibfnamefont {M.}~\bibnamefont
  {Drienovsky}}, \bibinfo {author} {\bibfnamefont {J.}~\bibnamefont
  {Joachimsmeyer}}, \bibinfo {author} {\bibfnamefont {A.}~\bibnamefont
  {Sandner}}, \bibinfo {author} {\bibfnamefont {M.-H.}\ \bibnamefont {Liu}},
  \bibinfo {author} {\bibfnamefont {T.}~\bibnamefont {Taniguchi}}, \bibinfo
  {author} {\bibfnamefont {K.}~\bibnamefont {Watanabe}}, \bibinfo {author}
  {\bibfnamefont {K.}~\bibnamefont {Richter}}, \bibinfo {author} {\bibfnamefont
  {D.}~\bibnamefont {Weiss}}, \ and\ \bibinfo {author} {\bibfnamefont
  {J.}~\bibnamefont {Eroms}},\ }\href {\doibase 10.1103/PhysRevLett.121.026806}
  {\bibfield  {journal} {\bibinfo  {journal} {Phys. Rev. Lett.}\ }\textbf
  {\bibinfo {volume} {121}},\ \bibinfo {pages} {026806} (\bibinfo {year}
  {2018})}\BibitemShut {NoStop}%
\bibitem [{\citenamefont {Ponomarenko}\ \emph {et~al.}(2013)\citenamefont
  {Ponomarenko}, \citenamefont {Gorbachev}, \citenamefont {Yu}, \citenamefont
  {Elias}, \citenamefont {Jalil}, \citenamefont {Patel}, \citenamefont
  {Mishchenko}, \citenamefont {Mayorov}, \citenamefont {Woods}, \citenamefont
  {Wallbank}, \citenamefont {Mucha-Kruczynski}, \citenamefont {Piot},
  \citenamefont {Potemski}, \citenamefont {Grigorieva}, \citenamefont
  {Novoselov}, \citenamefont {Guinea}, \citenamefont {Falko},\ and\
  \citenamefont {Geim}}]{LAP}%
  \BibitemOpen
  \bibfield  {author} {\bibinfo {author} {\bibfnamefont {L.~A.}\ \bibnamefont
  {Ponomarenko}}, \bibinfo {author} {\bibfnamefont {R.~V.}\ \bibnamefont
  {Gorbachev}}, \bibinfo {author} {\bibfnamefont {G.~L.}\ \bibnamefont {Yu}},
  \bibinfo {author} {\bibfnamefont {D.~C.}\ \bibnamefont {Elias}}, \bibinfo
  {author} {\bibfnamefont {R.}~\bibnamefont {Jalil}}, \bibinfo {author}
  {\bibfnamefont {A.~A.}\ \bibnamefont {Patel}}, \bibinfo {author}
  {\bibfnamefont {A.}~\bibnamefont {Mishchenko}}, \bibinfo {author}
  {\bibfnamefont {A.~S.}\ \bibnamefont {Mayorov}}, \bibinfo {author}
  {\bibfnamefont {C.~R.}\ \bibnamefont {Woods}}, \bibinfo {author}
  {\bibfnamefont {J.~R.}\ \bibnamefont {Wallbank}}, \bibinfo {author}
  {\bibfnamefont {M.}~\bibnamefont {Mucha-Kruczynski}}, \bibinfo {author}
  {\bibfnamefont {B.~A.}\ \bibnamefont {Piot}}, \bibinfo {author}
  {\bibfnamefont {M.}~\bibnamefont {Potemski}}, \bibinfo {author}
  {\bibfnamefont {I.~V.}\ \bibnamefont {Grigorieva}}, \bibinfo {author}
  {\bibfnamefont {K.~S.}\ \bibnamefont {Novoselov}}, \bibinfo {author}
  {\bibfnamefont {F.}~\bibnamefont {Guinea}}, \bibinfo {author} {\bibfnamefont
  {V.~I.}\ \bibnamefont {Falko}}, \ and\ \bibinfo {author} {\bibfnamefont
  {A.~K.}\ \bibnamefont {Geim}},\ }\href
  {https://www.nature.com/articles/nature12187} {\bibfield  {journal} {\bibinfo
   {journal} {Nature}\ }\textbf {\bibinfo {volume} {497}},\ \bibinfo {pages}
  {594} (\bibinfo {year} {2013})}\BibitemShut {NoStop}%
\bibitem [{\citenamefont {Dean}\ \emph {et~al.}(2013)\citenamefont {Dean},
  \citenamefont {Wang}, \citenamefont {Maher}, \citenamefont {Forsythe},
  \citenamefont {Ghahari}, \citenamefont {Gao}, \citenamefont {Katoch},
  \citenamefont {Ishigami}, \citenamefont {Moon}, \citenamefont {Koshino},
  \citenamefont {Taniguchi}, \citenamefont {Watanabe}, \citenamefont {Shepard},
  \citenamefont {Hone},\ and\ \citenamefont {Kim}}]{CRD}%
  \BibitemOpen
  \bibfield  {author} {\bibinfo {author} {\bibfnamefont {C.~R.}\ \bibnamefont
  {Dean}}, \bibinfo {author} {\bibfnamefont {L.}~\bibnamefont {Wang}}, \bibinfo
  {author} {\bibfnamefont {P.}~\bibnamefont {Maher}}, \bibinfo {author}
  {\bibfnamefont {C.}~\bibnamefont {Forsythe}}, \bibinfo {author}
  {\bibfnamefont {F.}~\bibnamefont {Ghahari}}, \bibinfo {author} {\bibfnamefont
  {Y.}~\bibnamefont {Gao}}, \bibinfo {author} {\bibfnamefont {J.}~\bibnamefont
  {Katoch}}, \bibinfo {author} {\bibfnamefont {M.}~\bibnamefont {Ishigami}},
  \bibinfo {author} {\bibfnamefont {P.}~\bibnamefont {Moon}}, \bibinfo {author}
  {\bibfnamefont {M.}~\bibnamefont {Koshino}}, \bibinfo {author} {\bibfnamefont
  {T.}~\bibnamefont {Taniguchi}}, \bibinfo {author} {\bibfnamefont
  {K.}~\bibnamefont {Watanabe}}, \bibinfo {author} {\bibfnamefont {K.~L.}\
  \bibnamefont {Shepard}}, \bibinfo {author} {\bibfnamefont {J.}~\bibnamefont
  {Hone}}, \ and\ \bibinfo {author} {\bibfnamefont {P.}~\bibnamefont {Kim}},\
  }\href {https://www.nature.com/articles/nature12186} {\bibfield  {journal}
  {\bibinfo  {journal} {Nature}\ }\textbf {\bibinfo {volume} {497}},\ \bibinfo
  {pages} {598} (\bibinfo {year} {2013})}\BibitemShut {NoStop}%
\bibitem [{\citenamefont {Kumar}\ \emph {et~al.}(2018)\citenamefont {Kumar},
  \citenamefont {Mishchenko}, \citenamefont {Chen}, \citenamefont {Pezzini},
  \citenamefont {Auton}, \citenamefont {Ponomarenko}, \citenamefont {Zeitler},
  \citenamefont {Eaves}, \citenamefont {Falko},\ and\ \citenamefont
  {Geim}}]{RKK}%
  \BibitemOpen
  \bibfield  {author} {\bibinfo {author} {\bibfnamefont {R.~K.}\ \bibnamefont
  {Kumar}}, \bibinfo {author} {\bibfnamefont {A.}~\bibnamefont {Mishchenko}},
  \bibinfo {author} {\bibfnamefont {X.}~\bibnamefont {Chen}}, \bibinfo {author}
  {\bibfnamefont {S.}~\bibnamefont {Pezzini}}, \bibinfo {author} {\bibfnamefont
  {G.~H.}\ \bibnamefont {Auton}}, \bibinfo {author} {\bibfnamefont {L.~A.}\
  \bibnamefont {Ponomarenko}}, \bibinfo {author} {\bibfnamefont
  {U.}~\bibnamefont {Zeitler}}, \bibinfo {author} {\bibfnamefont
  {L.}~\bibnamefont {Eaves}}, \bibinfo {author} {\bibfnamefont {V.~I.}\
  \bibnamefont {Falko}}, \ and\ \bibinfo {author} {\bibfnamefont {A.~K.}\
  \bibnamefont {Geim}},\ }\href {\doibase 10.1073/pnas.1804572115} {\bibfield
  {journal} {\bibinfo  {journal} {PNAS}\ }\textbf {\bibinfo {volume} {115}},\
  \bibinfo {pages} {5135} (\bibinfo {year} {2018})}\BibitemShut {NoStop}%
\bibitem [{\citenamefont {Ibrahim}\ and\ \citenamefont
  {Peeters}(1995)}]{5_ibrahim_1995}%
  \BibitemOpen
  \bibfield  {author} {\bibinfo {author} {\bibfnamefont {I.~S.}\ \bibnamefont
  {Ibrahim}}\ and\ \bibinfo {author} {\bibfnamefont {F.~M.}\ \bibnamefont
  {Peeters}},\ }\href {\doibase 10.1103/PhysRevB.52.17321} {\bibfield
  {journal} {\bibinfo  {journal} {Phys. Rev. B}\ }\textbf {\bibinfo {volume}
  {52}},\ \bibinfo {pages} {17321} (\bibinfo {year} {1995})}\BibitemShut
  {NoStop}%
\bibitem [{\citenamefont {Chiu}\ \emph {et~al.}(2008)\citenamefont {Chiu},
  \citenamefont {Lai}, \citenamefont {Ho}, \citenamefont {Chuu},\ and\
  \citenamefont {Lin}}]{5_chiu_2008}%
  \BibitemOpen
  \bibfield  {author} {\bibinfo {author} {\bibfnamefont {Y.~H.}\ \bibnamefont
  {Chiu}}, \bibinfo {author} {\bibfnamefont {Y.~H.}\ \bibnamefont {Lai}},
  \bibinfo {author} {\bibfnamefont {J.~H.}\ \bibnamefont {Ho}}, \bibinfo
  {author} {\bibfnamefont {D.~S.}\ \bibnamefont {Chuu}}, \ and\ \bibinfo
  {author} {\bibfnamefont {M.~F.}\ \bibnamefont {Lin}},\ }\href {\doibase
  10.1103/PhysRevB.77.045407} {\bibfield  {journal} {\bibinfo  {journal} {Phys.
  Rev. B}\ }\textbf {\bibinfo {volume} {77}},\ \bibinfo {pages} {045407}
  (\bibinfo {year} {2008})}\BibitemShut {NoStop}%
\bibitem [{\citenamefont {Dell'Anna}\ and\ \citenamefont
  {De~Martino}(2009)}]{5_luca_2009}%
  \BibitemOpen
  \bibfield  {author} {\bibinfo {author} {\bibfnamefont {L.}~\bibnamefont
  {Dell'Anna}}\ and\ \bibinfo {author} {\bibfnamefont {A.}~\bibnamefont
  {De~Martino}},\ }\href {\doibase 10.1103/PhysRevB.79.045420} {\bibfield
  {journal} {\bibinfo  {journal} {Phys. Rev. B}\ }\textbf {\bibinfo {volume}
  {79}},\ \bibinfo {pages} {045420} (\bibinfo {year} {2009})}\BibitemShut
  {NoStop}%
\bibitem [{\citenamefont {Masir}\ \emph {et~al.}(2009)\citenamefont {Masir},
  \citenamefont {Vasilopoulos},\ and\ \citenamefont {Peeters}}]{5_masir_2009}%
  \BibitemOpen
  \bibfield  {author} {\bibinfo {author} {\bibfnamefont {M.~R.}\ \bibnamefont
  {Masir}}, \bibinfo {author} {\bibfnamefont {P.}~\bibnamefont {Vasilopoulos}},
  \ and\ \bibinfo {author} {\bibfnamefont {F.~M.}\ \bibnamefont {Peeters}},\
  }\href {\doibase 10.1088/1367-2630/11/9/095009} {\bibfield  {journal}
  {\bibinfo  {journal} {New Journal of Physics}\ }\textbf {\bibinfo {volume}
  {11}},\ \bibinfo {pages} {095009} (\bibinfo {year} {2009})}\BibitemShut
  {NoStop}%
\bibitem [{\citenamefont {Tan}\ \emph {et~al.}(2010)\citenamefont {Tan},
  \citenamefont {Park},\ and\ \citenamefont {Louie}}]{5_tan_2010}%
  \BibitemOpen
  \bibfield  {author} {\bibinfo {author} {\bibfnamefont {L.~Z.}\ \bibnamefont
  {Tan}}, \bibinfo {author} {\bibfnamefont {C.-H.}\ \bibnamefont {Park}}, \
  and\ \bibinfo {author} {\bibfnamefont {S.~G.}\ \bibnamefont {Louie}},\ }\href
  {\doibase 10.1103/PhysRevB.81.195426} {\bibfield  {journal} {\bibinfo
  {journal} {Phys. Rev. B}\ }\textbf {\bibinfo {volume} {81}},\ \bibinfo
  {pages} {195426} (\bibinfo {year} {2010})}\BibitemShut {NoStop}%
\bibitem [{\citenamefont {Taillefumier}\ \emph {et~al.}(2011)\citenamefont
  {Taillefumier}, \citenamefont {Dugaev}, \citenamefont {Canals}, \citenamefont
  {Lacroix},\ and\ \citenamefont {Bruno}}]{5_taillefumier_2011}%
  \BibitemOpen
  \bibfield  {author} {\bibinfo {author} {\bibfnamefont {M.}~\bibnamefont
  {Taillefumier}}, \bibinfo {author} {\bibfnamefont {V.~K.}\ \bibnamefont
  {Dugaev}}, \bibinfo {author} {\bibfnamefont {B.}~\bibnamefont {Canals}},
  \bibinfo {author} {\bibfnamefont {C.}~\bibnamefont {Lacroix}}, \ and\
  \bibinfo {author} {\bibfnamefont {P.}~\bibnamefont {Bruno}},\ }\href
  {\doibase 10.1103/PhysRevB.84.085427} {\bibfield  {journal} {\bibinfo
  {journal} {Phys. Rev. B}\ }\textbf {\bibinfo {volume} {84}},\ \bibinfo
  {pages} {085427} (\bibinfo {year} {2011})}\BibitemShut {NoStop}%
\bibitem [{\citenamefont {Sutherland}(1986)}]{3_sutherland_1986}%
  \BibitemOpen
  \bibfield  {author} {\bibinfo {author} {\bibfnamefont {B.}~\bibnamefont
  {Sutherland}},\ }\href {\doibase 10.1103/PhysRevB.34.5208} {\bibfield
  {journal} {\bibinfo  {journal} {Phys. Rev. B}\ }\textbf {\bibinfo {volume}
  {34}},\ \bibinfo {pages} {5208} (\bibinfo {year} {1986})}\BibitemShut
  {NoStop}%
\bibitem [{\citenamefont {Lieb}(1989)}]{3_lieb_1989}%
  \BibitemOpen
  \bibfield  {author} {\bibinfo {author} {\bibfnamefont {E.~H.}\ \bibnamefont
  {Lieb}},\ }\href {\doibase 10.1103/PhysRevLett.62.1201} {\bibfield  {journal}
  {\bibinfo  {journal} {Phys. Rev. Lett.}\ }\textbf {\bibinfo {volume} {62}},\
  \bibinfo {pages} {1201} (\bibinfo {year} {1989})}\BibitemShut {NoStop}%
\bibitem [{\citenamefont {Mielke}(1991{\natexlab{a}})}]{3_AME1}%
  \BibitemOpen
  \bibfield  {author} {\bibinfo {author} {\bibfnamefont {A.}~\bibnamefont
  {Mielke}},\ }\href {\doibase 10.1088/0305-4470/24/2/005} {\bibfield
  {journal} {\bibinfo  {journal} {Journal of Physics A: Mathematical and
  General}\ }\textbf {\bibinfo {volume} {24}},\ \bibinfo {pages} {L73}
  (\bibinfo {year} {1991}{\natexlab{a}})}\BibitemShut {NoStop}%
\bibitem [{\citenamefont {Mielke}(1991{\natexlab{b}})}]{3_AME2}%
  \BibitemOpen
  \bibfield  {author} {\bibinfo {author} {\bibfnamefont {A.}~\bibnamefont
  {Mielke}},\ }\href {\doibase 10.1088/0305-4470/24/14/018} {\bibfield
  {journal} {\bibinfo  {journal} {Journal of Physics A: Mathematical and
  General}\ }\textbf {\bibinfo {volume} {24}},\ \bibinfo {pages} {3311}
  (\bibinfo {year} {1991}{\natexlab{b}})}\BibitemShut {NoStop}%
\bibitem [{\citenamefont {Tasaki}(1992)}]{3_HTI1}%
  \BibitemOpen
  \bibfield  {author} {\bibinfo {author} {\bibfnamefont {H.}~\bibnamefont
  {Tasaki}},\ }\href {\doibase 10.1103/PhysRevLett.69.1608} {\bibfield
  {journal} {\bibinfo  {journal} {Phys. Rev. Lett.}\ }\textbf {\bibinfo
  {volume} {69}},\ \bibinfo {pages} {1608} (\bibinfo {year}
  {1992})}\BibitemShut {NoStop}%
\bibitem [{\citenamefont {Tasaki}(2008)}]{3_HTI2}%
  \BibitemOpen
  \bibfield  {author} {\bibinfo {author} {\bibfnamefont {H.}~\bibnamefont
  {Tasaki}},\ }\href
  {https://epjb.epj.org/articles/epjb/abs/2008/15/b07754/b07754.html}
  {\bibfield  {journal} {\bibinfo  {journal} {Eur. Phys. J. B}\ }\textbf
  {\bibinfo {volume} {64}},\ \bibinfo {pages} {365} (\bibinfo {year}
  {2008})}\BibitemShut {NoStop}%
\bibitem [{sup()}]{supp}%
  \BibitemOpen
  \href@noop {} {\bibinfo  {journal} {Supplemental Information}\ }\BibitemShut
  {NoStop}%
\bibitem [{\citenamefont {Lopes~dos Santos}\ \emph {et~al.}(2012)\citenamefont
  {Lopes~dos Santos}, \citenamefont {Peres},\ and\ \citenamefont
  {Castro~Neto}}]{JNA}%
  \BibitemOpen
\bibfield  {journal} {  }\bibfield  {author} {\bibinfo {author} {\bibfnamefont
  {J.~M.~B.}\ \bibnamefont {Lopes~dos Santos}}, \bibinfo {author}
  {\bibfnamefont {N.~M.~R.}\ \bibnamefont {Peres}}, \ and\ \bibinfo {author}
  {\bibfnamefont {A.~H.}\ \bibnamefont {Castro~Neto}},\ }\href {\doibase
  10.1103/PhysRevB.86.155449} {\bibfield  {journal} {\bibinfo  {journal} {Phys.
  Rev. B}\ }\textbf {\bibinfo {volume} {86}},\ \bibinfo {pages} {155449}
  (\bibinfo {year} {2012})}\BibitemShut {NoStop}%
\bibitem [{\citenamefont {Jackiw}(1984)}]{JACKIW}%
  \BibitemOpen
  \bibfield  {author} {\bibinfo {author} {\bibfnamefont {R.}~\bibnamefont
  {Jackiw}},\ }\href {\doibase 10.1103/PhysRevD.29.2375} {\bibfield  {journal}
  {\bibinfo  {journal} {Phys. Rev. D}\ }\textbf {\bibinfo {volume} {29}},\
  \bibinfo {pages} {2375} (\bibinfo {year} {1984})}\BibitemShut {NoStop}%
\bibitem [{\citenamefont {Snyman}(2009)}]{snyman_2009}%
  \BibitemOpen
  \bibfield  {author} {\bibinfo {author} {\bibfnamefont {I.}~\bibnamefont
  {Snyman}},\ }\href {\doibase 10.1103/PhysRevB.80.054303} {\bibfield
  {journal} {\bibinfo  {journal} {Phys. Rev. B}\ }\textbf {\bibinfo {volume}
  {80}},\ \bibinfo {pages} {054303} (\bibinfo {year} {2009})}\BibitemShut
  {NoStop}%
\bibitem [{\citenamefont {Allaire}\ and\ \citenamefont
  {Piatnitski}(2005)}]{allaire}%
  \BibitemOpen
  \bibfield  {author} {\bibinfo {author} {\bibfnamefont {G.}~\bibnamefont
  {Allaire}}\ and\ \bibinfo {author} {\bibfnamefont {A.}~\bibnamefont
  {Piatnitski}},\ }\href {\doibase 10.1007/s00220-005-1329-2} {\bibfield
  {journal} {\bibinfo  {journal} {Communications in Mathematical Physics}\
  }\textbf {\bibinfo {volume} {258}},\ \bibinfo {pages} {1} (\bibinfo {year}
  {2005})}\BibitemShut {NoStop}%
\bibitem [{\citenamefont {Barletti}\ and\ \citenamefont
  {Ben~Abdallah}(2011)}]{naoufel}%
  \BibitemOpen
  \bibfield  {author} {\bibinfo {author} {\bibfnamefont {L.}~\bibnamefont
  {Barletti}}\ and\ \bibinfo {author} {\bibfnamefont {N.}~\bibnamefont
  {Ben~Abdallah}},\ }\href {\doibase 10.1007/s00220-011-1344-4} {\bibfield
  {journal} {\bibinfo  {journal} {Communications in Mathematical Physics}\
  }\textbf {\bibinfo {volume} {307}},\ \bibinfo {pages} {567} (\bibinfo {year}
  {2011})}\BibitemShut {NoStop}%
\bibitem [{\citenamefont {Chen}\ \emph {et~al.}()\citenamefont {Chen},
  \citenamefont {Pinaud},\ and\ \citenamefont {Tahir}}]{future}%
  \BibitemOpen
  \bibfield  {author} {\bibinfo {author} {\bibfnamefont {H.}~\bibnamefont
  {Chen}}, \bibinfo {author} {\bibfnamefont {O.}~\bibnamefont {Pinaud}}, \ and\
  \bibinfo {author} {\bibfnamefont {M.}~\bibnamefont {Tahir}},\ }\href@noop {}
  {\bibinfo  {journal} {in preparation}\ }\BibitemShut {NoStop}%
\bibitem [{\citenamefont {Hammer}\ \emph
  {et~al.}(2014{\natexlab{a}})\citenamefont {Hammer}, \citenamefont {Pötz},\
  and\ \citenamefont {Arnold}}]{hammer1}%
  \BibitemOpen
\bibfield  {journal} {  }\bibfield  {author} {\bibinfo {author} {\bibfnamefont
  {R.}~\bibnamefont {Hammer}}, \bibinfo {author} {\bibfnamefont
  {W.}~\bibnamefont {Pötz}}, \ and\ \bibinfo {author} {\bibfnamefont
  {A.}~\bibnamefont {Arnold}},\ }\href {\doibase
  https://doi.org/10.1016/j.jcp.2013.09.022} {\bibfield  {journal} {\bibinfo
  {journal} {Journal of Computational Physics}\ }\textbf {\bibinfo {volume}
  {256}},\ \bibinfo {pages} {728 } (\bibinfo {year}
  {2014}{\natexlab{a}})}\BibitemShut {NoStop}%
\bibitem [{\citenamefont {Hammer}\ \emph
  {et~al.}(2014{\natexlab{b}})\citenamefont {Hammer}, \citenamefont {Pötz},\
  and\ \citenamefont {Arnold}}]{hammer2}%
  \BibitemOpen
  \bibfield  {author} {\bibinfo {author} {\bibfnamefont {R.}~\bibnamefont
  {Hammer}}, \bibinfo {author} {\bibfnamefont {W.}~\bibnamefont {Pötz}}, \ and\
  \bibinfo {author} {\bibfnamefont {A.}~\bibnamefont {Arnold}},\ }\href
  {\doibase https://doi.org/10.1016/j.jcp.2014.01.028} {\bibfield  {journal}
  {\bibinfo  {journal} {Journal of Computational Physics}\ }\textbf {\bibinfo
  {volume} {265}},\ \bibinfo {pages} {50 } (\bibinfo {year}
  {2014}{\natexlab{b}})}\BibitemShut {NoStop}%
\bibitem [{\citenamefont {Luttinger}(1951)}]{luttinger_1951}%
  \BibitemOpen
  \bibfield  {author} {\bibinfo {author} {\bibfnamefont {J.~M.}\ \bibnamefont
  {Luttinger}},\ }\href {\doibase 10.1103/PhysRev.84.814} {\bibfield  {journal}
  {\bibinfo  {journal} {Phys. Rev.}\ }\textbf {\bibinfo {volume} {84}},\
  \bibinfo {pages} {814} (\bibinfo {year} {1951})}\BibitemShut {NoStop}%
\bibitem [{\citenamefont {Wannier}(1962)}]{wannier_1962}%
  \BibitemOpen
  \bibfield  {author} {\bibinfo {author} {\bibfnamefont {G.~H.}\ \bibnamefont
  {Wannier}},\ }\href {\doibase 10.1103/RevModPhys.34.645} {\bibfield
  {journal} {\bibinfo  {journal} {Rev. Mod. Phys.}\ }\textbf {\bibinfo {volume}
  {34}},\ \bibinfo {pages} {645} (\bibinfo {year} {1962})}\BibitemShut
  {NoStop}%
\bibitem [{\citenamefont {Blount}(1962)}]{blount_1962}%
  \BibitemOpen
  \bibfield  {author} {\bibinfo {author} {\bibfnamefont {E.~I.}\ \bibnamefont
  {Blount}},\ }\href {\doibase 10.1103/PhysRev.126.1636} {\bibfield  {journal}
  {\bibinfo  {journal} {Phys. Rev.}\ }\textbf {\bibinfo {volume} {126}},\
  \bibinfo {pages} {1636} (\bibinfo {year} {1962})}\BibitemShut {NoStop}%
\bibitem [{\citenamefont {Kohn}(1959)}]{kohn_1959}%
  \BibitemOpen
  \bibfield  {author} {\bibinfo {author} {\bibfnamefont {W.}~\bibnamefont
  {Kohn}},\ }\href {\doibase 10.1103/PhysRev.115.809} {\bibfield  {journal}
  {\bibinfo  {journal} {Phys. Rev.}\ }\textbf {\bibinfo {volume} {115}},\
  \bibinfo {pages} {809} (\bibinfo {year} {1959})}\BibitemShut {NoStop}%
\bibitem [{\citenamefont {Nenciu}(1983)}]{nenciu_1983}%
  \BibitemOpen
  \bibfield  {author} {\bibinfo {author} {\bibfnamefont {G.}~\bibnamefont
  {Nenciu}},\ }\href {https://projecteuclid.org/euclid.cmp/1103940475}
  {\bibfield  {journal} {\bibinfo  {journal} {Communications in Mathematical
  Physics}\ }\textbf {\bibinfo {volume} {91}},\ \bibinfo {pages} {81} (\bibinfo
  {year} {1983})}\BibitemShut {NoStop}%
\bibitem [{\citenamefont {Brouder}\ \emph {et~al.}(2007)\citenamefont
  {Brouder}, \citenamefont {Panati}, \citenamefont {Calandra}, \citenamefont
  {Mourougane},\ and\ \citenamefont {Marzari}}]{brouder_2007}%
  \BibitemOpen
  \bibfield  {author} {\bibinfo {author} {\bibfnamefont {C.}~\bibnamefont
  {Brouder}}, \bibinfo {author} {\bibfnamefont {G.}~\bibnamefont {Panati}},
  \bibinfo {author} {\bibfnamefont {M.}~\bibnamefont {Calandra}}, \bibinfo
  {author} {\bibfnamefont {C.}~\bibnamefont {Mourougane}}, \ and\ \bibinfo
  {author} {\bibfnamefont {N.}~\bibnamefont {Marzari}},\ }\href {\doibase
  10.1103/PhysRevLett.98.046402} {\bibfield  {journal} {\bibinfo  {journal}
  {Phys. Rev. Lett.}\ }\textbf {\bibinfo {volume} {98}},\ \bibinfo {pages}
  {046402} (\bibinfo {year} {2007})}\BibitemShut {NoStop}%
\bibitem [{\citenamefont {Marzari}\ and\ \citenamefont
  {Vanderbilt}(1997)}]{marzari_1997}%
  \BibitemOpen
  \bibfield  {author} {\bibinfo {author} {\bibfnamefont {N.}~\bibnamefont
  {Marzari}}\ and\ \bibinfo {author} {\bibfnamefont {D.}~\bibnamefont
  {Vanderbilt}},\ }\href {\doibase 10.1103/PhysRevB.56.12847} {\bibfield
  {journal} {\bibinfo  {journal} {Phys. Rev. B}\ }\textbf {\bibinfo {volume}
  {56}},\ \bibinfo {pages} {12847} (\bibinfo {year} {1997})}\BibitemShut
  {NoStop}%
\bibitem [{\citenamefont {Haldane}(1988)}]{haldane_1988}%
  \BibitemOpen
  \bibfield  {author} {\bibinfo {author} {\bibfnamefont {F.~D.~M.}\
  \bibnamefont {Haldane}},\ }\href {\doibase 10.1103/PhysRevLett.61.2015}
  {\bibfield  {journal} {\bibinfo  {journal} {Phys. Rev. Lett.}\ }\textbf
  {\bibinfo {volume} {61}},\ \bibinfo {pages} {2015} (\bibinfo {year}
  {1988})}\BibitemShut {NoStop}%
\bibitem [{\citenamefont {Adachi}(1982)}]{Adachi_1982}%
  \BibitemOpen
  \bibfield  {author} {\bibinfo {author} {\bibfnamefont {S.}~\bibnamefont
  {Adachi}},\ }\href {\doibase 10.1063/1.330480} {\bibfield  {journal}
  {\bibinfo  {journal} {J. Appl. Phys.}\ }\textbf {\bibinfo {volume} {53}},\
  \bibinfo {pages} {8775} (\bibinfo {year} {1982})}\BibitemShut {NoStop}%
\bibitem [{\citenamefont {Taskin}\ and\ \citenamefont
  {Ando}(2011)}]{Taskin_2011}%
  \BibitemOpen
  \bibfield  {author} {\bibinfo {author} {\bibfnamefont {A.~A.}\ \bibnamefont
  {Taskin}}\ and\ \bibinfo {author} {\bibfnamefont {Y.}~\bibnamefont {Ando}},\
  }\href {\doibase 10.1103/PhysRevB.84.035301} {\bibfield  {journal} {\bibinfo
  {journal} {Phys. Rev. B}\ }\textbf {\bibinfo {volume} {84}},\ \bibinfo
  {pages} {035301} (\bibinfo {year} {2011})}\BibitemShut {NoStop}%
\bibitem [{\citenamefont {Giorgioni}\ \emph {et~al.}(2016)\citenamefont
  {Giorgioni}, \citenamefont {Paleari}, \citenamefont {Cecchi}, \citenamefont
  {Vitiello}, \citenamefont {Grilli}, \citenamefont {Isella}, \citenamefont
  {Jantsch}, \citenamefont {Fanciulli},\ and\ \citenamefont
  {Pezzoli}}]{Giorgioni2016}%
  \BibitemOpen
  \bibfield  {author} {\bibinfo {author} {\bibfnamefont {A.}~\bibnamefont
  {Giorgioni}}, \bibinfo {author} {\bibfnamefont {S.}~\bibnamefont {Paleari}},
  \bibinfo {author} {\bibfnamefont {S.}~\bibnamefont {Cecchi}}, \bibinfo
  {author} {\bibfnamefont {E.}~\bibnamefont {Vitiello}}, \bibinfo {author}
  {\bibfnamefont {E.}~\bibnamefont {Grilli}}, \bibinfo {author} {\bibfnamefont
  {G.}~\bibnamefont {Isella}}, \bibinfo {author} {\bibfnamefont
  {W.}~\bibnamefont {Jantsch}}, \bibinfo {author} {\bibfnamefont
  {M.}~\bibnamefont {Fanciulli}}, \ and\ \bibinfo {author} {\bibfnamefont
  {F.}~\bibnamefont {Pezzoli}},\ }\href@noop {} {\bibfield  {journal} {\bibinfo
   {journal} {Nature Communications}\ }\textbf {\bibinfo {volume} {7}},\
  \bibinfo {pages} {13886} (\bibinfo {year} {2016})}\BibitemShut {NoStop}%
\bibitem [{\citenamefont {Wang}\ \emph {et~al.}(2014)\citenamefont {Wang},
  \citenamefont {Zhong}, \citenamefont {Hao}, \citenamefont {Gerhold},
  \citenamefont {St{\"o}ger}, \citenamefont {Schmid}, \citenamefont
  {S{\'a}nchez-Barriga}, \citenamefont {Varykhalov}, \citenamefont {Franchini},
  \citenamefont {Held},\ and\ \citenamefont {Diebold}}]{Wang2014}%
  \BibitemOpen
  \bibfield  {author} {\bibinfo {author} {\bibfnamefont {Z.}~\bibnamefont
  {Wang}}, \bibinfo {author} {\bibfnamefont {Z.}~\bibnamefont {Zhong}},
  \bibinfo {author} {\bibfnamefont {X.}~\bibnamefont {Hao}}, \bibinfo {author}
  {\bibfnamefont {S.}~\bibnamefont {Gerhold}}, \bibinfo {author} {\bibfnamefont
  {B.}~\bibnamefont {St{\"o}ger}}, \bibinfo {author} {\bibfnamefont
  {M.}~\bibnamefont {Schmid}}, \bibinfo {author} {\bibfnamefont
  {J.}~\bibnamefont {S{\'a}nchez-Barriga}}, \bibinfo {author} {\bibfnamefont
  {A.}~\bibnamefont {Varykhalov}}, \bibinfo {author} {\bibfnamefont
  {C.}~\bibnamefont {Franchini}}, \bibinfo {author} {\bibfnamefont
  {K.}~\bibnamefont {Held}}, \ and\ \bibinfo {author} {\bibfnamefont
  {U.}~\bibnamefont {Diebold}},\ }\href@noop {} {\bibfield  {journal} {\bibinfo
   {journal} {Proc. Natl. Acad. Sci.}\ }\textbf {\bibinfo {volume} {111}},\
  \bibinfo {pages} {3933} (\bibinfo {year} {2014})}\BibitemShut {NoStop}%
\bibitem [{\citenamefont {Fukui}\ \emph {et~al.}(2005)\citenamefont {Fukui},
  \citenamefont {Hatsugai},\ and\ \citenamefont {Suzuki}}]{fukui_2005}%
  \BibitemOpen
  \bibfield  {author} {\bibinfo {author} {\bibfnamefont {T.}~\bibnamefont
  {Fukui}}, \bibinfo {author} {\bibfnamefont {Y.}~\bibnamefont {Hatsugai}}, \
  and\ \bibinfo {author} {\bibfnamefont {H.}~\bibnamefont {Suzuki}},\ }\href
  {\doibase 10.1143/JPSJ.74.1674} {\bibfield  {journal} {\bibinfo  {journal}
  {Journal of the Physical Society of Japan}\ }\textbf {\bibinfo {volume}
  {74}},\ \bibinfo {pages} {1674} (\bibinfo {year} {2005})}\BibitemShut
  {NoStop}%
\bibitem [{\citenamefont {Yankowitz}\ \emph {et~al.}(2012)\citenamefont
  {Yankowitz}, \citenamefont {Xue}, \citenamefont {Cormode}, \citenamefont
  {Sanchez-Yamagishi}, \citenamefont {Watanabe}, \citenamefont {Taniguchi},
  \citenamefont {Jarillo-Herrero}, \citenamefont {Jacquod},\ and\ \citenamefont
  {LeRoy}}]{MY1}%
  \BibitemOpen
  \bibfield  {author} {\bibinfo {author} {\bibfnamefont {M.}~\bibnamefont
  {Yankowitz}}, \bibinfo {author} {\bibfnamefont {J.}~\bibnamefont {Xue}},
  \bibinfo {author} {\bibfnamefont {D.}~\bibnamefont {Cormode}}, \bibinfo
  {author} {\bibfnamefont {J.~D.}\ \bibnamefont {Sanchez-Yamagishi}}, \bibinfo
  {author} {\bibfnamefont {K.}~\bibnamefont {Watanabe}}, \bibinfo {author}
  {\bibfnamefont {T.}~\bibnamefont {Taniguchi}}, \bibinfo {author}
  {\bibfnamefont {P.}~\bibnamefont {Jarillo-Herrero}}, \bibinfo {author}
  {\bibfnamefont {P.}~\bibnamefont {Jacquod}}, \ and\ \bibinfo {author}
  {\bibfnamefont {B.~J.}\ \bibnamefont {LeRoy}},\ }\href
  {https://www.nature.com/articles/nphys2272} {\bibfield  {journal} {\bibinfo
  {journal} {Nature Physics}\ }\textbf {\bibinfo {volume} {8}},\ \bibinfo
  {pages} {382} (\bibinfo {year} {2012})}\BibitemShut {NoStop}%
\bibitem [{\citenamefont {Park}\ \emph {et~al.}(2008)\citenamefont {Park},
  \citenamefont {Yang}, \citenamefont {Son}, \citenamefont {Cohen},\ and\
  \citenamefont {Louie}}]{CHP}%
  \BibitemOpen
  \bibfield  {author} {\bibinfo {author} {\bibfnamefont {C.-H.}\ \bibnamefont
  {Park}}, \bibinfo {author} {\bibfnamefont {L.}~\bibnamefont {Yang}}, \bibinfo
  {author} {\bibfnamefont {Y.-W.}\ \bibnamefont {Son}}, \bibinfo {author}
  {\bibfnamefont {M.~L.}\ \bibnamefont {Cohen}}, \ and\ \bibinfo {author}
  {\bibfnamefont {S.~G.}\ \bibnamefont {Louie}},\ }\href
  {https://www.nature.com/articles/nphys890} {\bibfield  {journal} {\bibinfo
  {journal} {Nature Physics}\ }\textbf {\bibinfo {volume} {4}},\ \bibinfo
  {pages} {213} (\bibinfo {year} {2008})}\BibitemShut {NoStop}%
\bibitem [{\citenamefont {Varma}(1997)}]{CMV1}%
  \BibitemOpen
  \bibfield  {author} {\bibinfo {author} {\bibfnamefont {C.~M.}\ \bibnamefont
  {Varma}},\ }\href {\doibase 10.1103/PhysRevB.55.14554} {\bibfield  {journal}
  {\bibinfo  {journal} {Phys. Rev. B}\ }\textbf {\bibinfo {volume} {55}},\
  \bibinfo {pages} {14554} (\bibinfo {year} {1997})}\BibitemShut {NoStop}%
\bibitem [{\citenamefont {Varma}(2006)}]{CMV2}%
  \BibitemOpen
  \bibfield  {author} {\bibinfo {author} {\bibfnamefont {C.~M.}\ \bibnamefont
  {Varma}},\ }\href {\doibase 10.1103/PhysRevB.73.155113} {\bibfield  {journal}
  {\bibinfo  {journal} {Phys. Rev. B}\ }\textbf {\bibinfo {volume} {73}},\
  \bibinfo {pages} {155113} (\bibinfo {year} {2006})}\BibitemShut {NoStop}%
\end{thebibliography}%

\end{document}